\documentclass[11pt,letterpaper]{article}
\pdfoutput=1
\usepackage{jheppub}
\usepackage{amsmath,amssymb,amsfonts,bm}
\usepackage{graphicx,wrapfig}
\usepackage{multirow}
\usepackage{verbatim}
\usepackage{appendix}
\usepackage{rotating}

\graphicspath{{./images/}}

\numberwithin{equation}{section}
\renewcommand\afterTocRuleSpace{\bigskip}

\newcommand{\x}{\mathbf{x}}
\newcommand{\y}{\mathbf{y}}
\newcommand{\m}{\mathbf{m}}
\newcommand{\n}{\mathbf{n}}
\newcommand{\bmu}{\bm{\mu}}
\newcommand{\bnu}{\bm{\nu}}
\newcommand{\p}{{\mathfrak p}}

\newcommand{\bea}{\begin{eqnarray}}
\newcommand{\eea}{\end{eqnarray}}
\newcommand{\be}{\begin{equation}}
\newcommand{\ee}{\end{equation}}
\newcommand{\bse}{\begin{subequations}}
\newcommand{\ese}{\end{subequations}}
\newcommand{\mb}{\mathbf}

\newcommand{\wt}{\widetilde}

\newcommand{\ol}{\overline}
\newcommand{\ds}{\displaystyle}

\newcommand{\eg}{\emph{e.g.}}
\newcommand{\ie}{\emph{i.e.}}
\newcommand{\cf}{\emph{cf.}}

\newcommand{\Z}{{\mathbb Z}}
\newcommand{\R}{{\mathbb R}}
\newcommand{\C}{{\mathbb C}}
\newcommand{\Q}{{\mathbb Q}}

\newcommand{\Tr}{{\rm Tr \,}}
\renewcommand{\Re}{{\rm Re}}
\renewcommand{\Im}{{\rm Im}}
\newcommand{\bs}{\backslash}
\newcommand{\pd}{\partial}

\newcommand{\CA}{\mathcal{A}}

\newcommand{\CF}{\mathcal{F}}
\newcommand{\CG}{\mathcal{G}}
\newcommand{\CH}{\mathcal{H}}
\newcommand{\CI}{\mathcal{I}}

\newcommand{\CL}{\mathcal{L}}

\newcommand{\CN}{\mathcal{N}}
\newcommand{\CO}{\mathcal{O}}
\newcommand{\CP}{\mathcal{P}}

\newcommand{\CS}{\mathcal{S}}

\newcommand{\CZ}{\mathcal{Z}}

\title{Complex Chern-Simons theory at level $k$ via the 3d-3d correspondence}

\author{Tudor Dimofte}

\affiliation{\hspace{-.3cm}Institute for Advanced Study, Einstein Dr., Princeton, NJ 08540}

\abstract{
We use the 3d-3d correspondence together with the DGG construction of theories $T_n[M]$ labelled by 3-manifolds $M$ to define a non-perturbative state-integral model for $SL(n,\C)$ Chern-Simons theory at any level $k$, based on ideal triangulations.
The resulting partition functions generalize a widely studied $k=1$ state-integral as well as the 3d index, which is $k=0$.
The Chern-Simons partition functions correspond to partition functions of $T_n[M]$ on squashed lens spaces $L(k,1)$.  At any $k$, they admit a holomorphic-antiholomorphic factorization, corresponding to the decomposition of $L(k,1)$ into two solid tori, and the associated holomorphic block decomposition of the partition functions of $T_n[M]$. A generalization to $L(k,p)$ is also presented.
Convergence of the state integrals, for any $k$, requires triangulations to admit a positive angle structure; we propose that this is also necessary for the DGG gauge theory $T_n[M]$ to flow to a desired IR SCFT.}

\begin{document}

\maketitle


\section{Introduction}
\label{sec:intro}

The 3d-3d correspondence associates a 3d $\CN=2$ superconformal field theory (SCFT) $T_{\mathfrak g}[M]$ to a 3-manifold $M$ and an ADE Lie algebra $\mathfrak g$\,:
\be M\qquad \overset{\CF_{3d3d}}{\leadsto} \qquad T_{\mathfrak g}[M]\,. \label{3d3d} \ee
The theory $T_{\mathfrak g}[M]$ can be defined as the IR limit of a twisted compactification of the 6d $(2,0)$ theory on $M$, or alternatively, for $\mathfrak g = A_{n-1}$, as the effective low-energy worldvolume theory of $n$ M5 branes wrapped on $M$ in the 11d geometry $T^*M\times \R^5$.%
\footnote{Various foundational aspects of the 3d-3d correspondence were uncovered in (\eg) \cite{DGH, Yamazaki-3d, DGG, CCV, DGV-hybrid, CJ-S3, CDGS}, though much is still being studied.
We also refer the reader to the review \cite{D-volume} and references therein.} %
A central feature of \eqref{3d3d} is that, conjecturally, the theory $T_{\mathfrak g}[M]$ only depends on the \emph{topology} of $M$ --- and on some choices of boundary conditions if $M$ has a boundary. Indeed, \eqref{3d3d} should be thought of as a 3d TQFT functor that is ``valued in 3d SCFT's.'' Any observable computed in the physical theory $T_{\mathfrak g}[M]$ then becomes a (classical, quantum, or categorical) topological invariant of $M$.

In the present paper, we will study the partition functions of $T_{n}[M] := T_{A_{n-1}}[M]$ on \emph{squashed lens spaces} $L(k,p)_b$ and use them to build state-integral models for $SL(n,\C)$ Chern-Simons theory on $M$ itself.%
\footnote{We will not be careful to differentiate between the integral forms $SL(n,\C)$, $PGL(n,\C)$, or other integral forms of the Lie algebra $A_{n-1}$ in this paper. All these forms can arise in the 3d-3d correspondence \cite{Witten-GL6d, Tachi-discrete}, and the distinction is subtle. In our state-integral models, the tetrahedron building blocks are always $PGL(n,\C)$, but (\eg) the state integral for knot complements actually lifts to an $SL(n,\C)$ partition function.} %
Chern-Simons theory with complex gauge group was introduced in \cite{Witten-gravCS, Witten-cx}. Perturbative aspects of the theory were studied in \cite{BW-cx, gukov-2003}, and the first (perturbative) state-integral models appeared in \cite{hikami-2006,DGLZ}.  In contrast, the state-integral models that we present here are \emph{non-perturbative} --- and provide some of the first examples of non-perturbative partition functions in complex Chern-Simons theory at general levels.

The squashed lens space $L(k,p)_b$, with $k,p$ relatively prime, is a quotient of a squashed 3-sphere (or simply an ellipsoid) $S^3_b$,
\be \label{lens}
L(k,p)_b \, = \, \big\{(z,w)\in \C^2\,\big|\, b^2|z|^2 + b^{-2}|w|^2 =1\,\big\}\big/\, \raisebox{-.1cm}{$(z,w)\sim (e^{\frac{2\pi i}{k}}z, e^{-\frac{2\pi ip}{k}}w)$}\,,
\ee 
whose metric is induced from the flat metric on $\C^2$.
The metric is important when considering compactification of the 3d $\CN=2$ theory $T_n[M]$ on $L(k,p)_b$, since $T_{\mathfrak g}[M]$ is a superconformal but not a topological theory. Moreover, since $L(k,p)_b$ admits an integrable almost contact structure, the compactification can be arranged to preserve some supersymmetry \cite{FS-curvedSUSY, CDFZ-3d, CDFZ-geom}.

The case $(k,p)=(1,1)$, \ie\ the partition function of $T_n[M]$ on the ellipsoid $S^3_b$, was studied in the first works on the 3d-3d correspondence \cite{Yamazaki-3d, DGG}, using field-theory computations from \cite{Kapustin-3dloc, HHL}, and is closely related to the AGT correspondence \cite{AGT, HLP-wall}. It was argued therein that this partition function captures a sector of $SL(n,\C)$ or an ``$SL(n,\R)$-like'' Chern-Simons theory on $M$. Later, C\'ordova and Jafferis \cite{CJ-S3} derived (from six dimensions) that $T_n[M]$ on $S^3_b$ is precisely equivalent to $SL(n,\C)$ Chern-Simons theory on $M$, with levels
\be   k = 1\,,\qquad \sigma = \frac{1-b^2}{1+b^2}\,.\ee
Here we may recall that complex Chern-Simons theory has \emph{two} levels, one quantized $(k)$ and the other continuous $(\sigma)$ \cite{Witten-cx}. (We will see in Section \ref{sec:CS} that quantization at level $k=1$ looks very much like quantization of an $SL(n,\R)$ theory, explaining the earlier observations about $SL(n,\R)$.)

The case $(k,p)=(0,1)$ was studied in \cite{DGG-index}. Here $L(0,1)$ should be understood as the geometry $S^2\times_{\sigma}S^1$, namely a fibration of $S^2$ over $S^1$ whose monodromy is a rotation of $S^2$ by an angle $1/\sigma$. The resulting partition function of $T_n[M]$ is a 3d index. Using field-theory computations from \cite{Kim-index, IY-index, KW-index}, it was argued that the 3d index of $T_n[M]$ is equivalent to $SL(n,\C)$ Chern-Simons theory at levels $(k,\sigma)=(0,\sigma)$; this was later proven in \cite{LY-S2, Yagi-S2}.

What about $k>1$? The analysis of \cite{CJ-S3} for $S^3_b$ was based on the fact that $S^3_b$ is a Hopf fibration of degree $k=1$ over $S^2$. Since $L(k,1)_b$ is a Hopf fibration of degree $k$, the analysis of \cite{CJ-S3} predicts that the $L(k,1)_b$ partition function of $T_n[M]$ should correspond to $SL(n,\C)$ Chern-Simons theory at level $k$ on $M$. Indeed, by quantizing a model Chern-Simons phase space in Section \ref{sec:CS}, and comparing to field-theory results on $L(k,1)_b$ partition functions \cite{BNY-Lens, IY-Lens, IMY-fact}, we will see explicitly that the $L(k,1)_b$ partition function corresponds to Chern-Simons theory at levels
\be  k =k\,,\qquad \sigma = k \frac{1-b^2}{1+b^2}\,. \label{levelsk} \ee

The general case of $L(k,p)_b$ is harder to identify with a standard Chern-Simons theory. Partition functions of 3d $\CN=2$ theories on $L(k,p)_b$ for $p\neq \pm 1$ have also not been fully described in the literature yet. In Section \ref{sec:kp} we will propose a state-integral model that computes $L(k,p)_b$ invariants of triangulated 3-manifolds $M$. The model is uniquely determined by requirements of holomorphic factorization and its annihilation by certain difference operators (see below). It would be nice to find a concrete interpretation of the resulting invariants in terms of complex Chern-Simons theory.%
\footnote{The problem seems related to one in \emph{compact} Chern-Simons theory. Compact Chern-Simons at level $k$ on a knot complement $M=S^3\bs K$ computes Jones polynomials of knots $J_K(q)$, with $q = \exp(2\pi i/k)$. Mathematically, these polynomials (and their definition via quantum groups) make sense for any primitive $k$-th root of unity $q=\exp(2 \pi i p/k)$, and it has not yet been understood how to interpret these other roots of unity physically.}

\subsection*{State-sum models}

An algorithm for computing some theories $T_2[M]$ was proposed in \cite{DGG}. The algorithm requires a sufficiently ``good''  topological ideal triangulation $\mb t= \{\Delta_i\}_{i=1}^N$ of an oriented 3-manifold $M$, and constructs $T_2[M]$ by gluing together simple canonical theories $T_\Delta$ associated to each tetrahedron,
\be T_2[M,\mb t]  = T_{\Delta_1}\otimes\cdots\otimes T_{\Delta_N}\big/\sim\,,\qquad T_{\Delta_i} \equiv (\text{free chiral multiplet})\,. \label{DGG-alg} \ee
Physically, the gluing involves gauging global symmetries and adding some superpotential interactions among theories $T_{\Delta_i}$. The output is an abelian 3d gauge theory $T_2[M,\mb t]$ that (conjecturally) flows to the desired superconformal theory $T_2[M]$ in the infrared.
 The algorithm was extended in \cite{DGG-Kdec} to compute theories $T_n[M]$ for $n>2$, and involves the very same $T_\Delta$ building blocks, but more refined triangulations.

By composing the ``DGG'' algorithm of \cite{DGG, DGG-Kdec} with evaluation of a squashed lens space partition function,
\be \CZ_b^{(k,p)}[M]_n := \CZ[ T_n[M],\, L(k,p)_b ]\,, \label{eval} \ee
one should obtain a state-integral model for the $L(k,p)_b$ invariants of $M$. That is,
\be \CZ_b^{(k,p)}[M,\mb t]_n  =  \CZ_b^{(k,p)}[\Delta_1]\times\cdots\times \CZ_b^{(k,p)}[\Delta_N]\big/\sim\, \label{inducedSIM} \ee
where each $\CZ_b^{(k,p)}[\Delta_i]$ is a canonical wavefunction associated to a tetrahedron, and ``$\sim$'' is a symplectic reduction operation, implemented by performing some integrals and constraining the arguments of the $\CZ_b^{(k,p)}[\Delta_i]$.
The original DGG algorithm of \cite{DGG} was \emph{engineered} so that the induced state-integral for $S^3_b=L(1,1)_b$ would be identical to a state integral already formulated for $SL(2,\C)$ Chern-Simons theory in \cite{hikami-2006, DGLZ, Dimofte-QRS} (following insights of \cite{gukov-2003}).%
\footnote{The state integral of \cite{hikami-2006, DGLZ,Dimofte-QRS}, defined only perturbatively there, was promoted to a convergent, non-perturbative invariant in \cite{KashAnd, AK-new}. The connection between the two invariants has not been fully clarified yet in the literature. We hope that with the analysis of convergence (of the former) in Section \ref{sec:SS}, the equivalence of the two invariants will become obvious.} %
In the case of $S^2\times_\sigma S^1 = L(0,1)_\sigma$, a new state-sum model for $SL(2,\C)$ Chern-Simons at level $k=0$ was obtained in \cite{DGG-index} and further studied in \cite{Gar-index, GHRS-index}. In this paper, we will use localization computations of $L(k,1)_b$ partition functions from \cite{BNY-Lens, IY-Lens, IMY-fact} to develop the full machinery of a state-integral model for $SL(n,\C)$ Chern-Simons theory at any levels $(k,\sigma)$.

\subsection*{An irreducible caveat}

The state-integral models induced by \eqref{inducedSIM} provide the first non-perturbative computations of wavefunctions in complex Chern-Simons theory at level $k>1$, with one important caveat. As emphasized recently in \cite{D-volume, CDGS}, the algorithm \eqref{DGG-alg} of \cite{DGG,DGG-Kdec}, based on ideal triangulations, often only produces a subsector of the full theory $T_n[M]$. Thus, the state-integral models induced by \eqref{inducedSIM} only contain a subsector of the full $SL(n,\C)$ Chern-Simons theory on $M$. This subsector contains a quantization of the \emph{irreducible} flat connections on $M$, rather than all flat connections. The subsector cannot be a full TQFT, allowing gluing along arbitrary boundaries, since such gluing would require all flat connections to be considered. Nevertheless, there is evidence that this subsector is a consistent truncation of complex Chern-Simons theory on a fixed 3-manifold $M$,%
\footnote{In particular, the analysis of Chern-Simons integration cycles in \cite[Sec. 5]{Wit-anal}, each cycle being labelled by a flat connection on $M$, suggests that irreducible flat connections can mix with reducible ones, but not the other way around.} %
and corresponds to a restricted TQFT for ideally triangulated 3-manifolds, allowing gluing only along faces of tetrahedra (not along ideal boundary, which is always nonempty), subject to certain regularity requirements. The restricted TQFT was rigorously constructed in \cite{GHRS-index,AK-new} for $k=0,1$. 
As a concrete example, our state-integral models allow us to construct partition functions for most knot complements $M=S^3\bs K$ by gluing ideal tetrahedra together one at a time, but they do \emph{not} provide a way to perform Dehn surgery on knots, to glue two knot complements together, or more generally to construct partition functions for closed 3-manifolds. One hopes that once an algorithmic definition for the full theory $T_n[M]$ becomes known, composing it with an evaluation functor \eqref{eval} will lead to true TQFT's.

\subsection*{Properties}

The $L(k,p)_b$ state-integral models have several wonderful, unifying properties. These properties can conveniently be illustrated by considering the partition function of a tetrahedron, which takes the form
\be \CZ^{(k,p)}_b[\Delta](\mu,m) := (qx^{-1};q)_\infty(\tilde q\tilde x^{-1};\tilde q)_\infty\,, \label{defZD} \ee
where
\be (z;q)_\infty = \begin{cases} \prod_{j=0}^\infty (1-q^jz) & |q|<1 \\
  \prod_{j=1}^\infty (1-q^{-j}z)^{-1} & |q|>1\,. \end{cases} \label{defqf} \ee
This partition function depends on a complex variable $\mu\in \C$ and an integer $m\in \Z/k\Z$, as well as the parameters $(k,p;b)$. The variables appearing on the RHS are%
\footnote{When $k=0$ these expressions become $q=\exp\frac{2\pi}{\sigma} = \tilde q^{-1}$ and $x=q^{m/2}e^\mu,\,\tilde x=q^{m/2}e^{-\mu}$, see \cite{DGG-index}.}
\footnote{There may appear to be an asymmetry between the definitions of $x$ and $\tilde x$, since $p$ appears in $x$ but $r$ does not appear in $\tilde x$. This is not so. Both $p$ and $r$ are units in $\Z/k\Z$. If we simply rescale $m\to r m$, the definition of $\tilde x$ acquires an $r$ and the definition of $x$ looses the $p$. Writing $x$ and $\tilde x$ as in \eqref{defqx} is simply a choice that must be made, and will have also appear in the definition of shift operators below.} %
\be \label{defqx}
\begin{array}{c} \displaystyle q = \exp \frac{2\pi i}{k}(b^2 + p)\,,\qquad \tilde q=\exp\frac{2\pi i}{k}(b^{-2}+r)\,;\\[.3cm] \displaystyle
   x = \exp \frac{2\pi i}{k} (-ib\mu - pm)\,,\qquad \tilde x=\exp\frac{2\pi i}{k}(-ib^{-1}\mu+m)\,,\end{array}
\ee
where $rp\equiv 1$ (mod $k$).  Notice that all dependence on $(k,p)$ is hidden in the definitions of $q,\tilde q,x,\tilde x$. The relation between $q=:e^{2\pi i\tau}$ and $\tilde q=:e^{2\pi i\tilde\tau}$ is a modular transformation (composed with a reflection),
\be \tilde \tau = \frac1k(b^{-2}+r) = - \varphi(\tau)\qquad\text{with}\; \varphi=\begin{pmatrix} r & s \\ -k & p \end{pmatrix} \subset SL(2,\Z)\,,\quad \tau = \frac1k(b^2+p)\,, \label{mod}\ee
so that $|\tilde q|>1$ whenever $|q|<1$ and vice-versa;
while the relation between $x$ and $\tilde x$ is a Jacobi transformation (together with a translation).

For $k=1$, the modular matrix is $\varphi=TST$. Since the $T$'s act trivially, we could equivalently take $\varphi=S$.
At $k=1$, the function \eqref{defZD} is a Barnes double-gamma function, also known as Faddeev's noncompact quantum dilogarithm \cite{faddeev-1994}. While the definition \eqref{defZD} only makes sense (a priori) for $b\in \C\bs(\R\cup i\R)$, it extends to a meromorphic function of $\mu\in\C$ and $b$ in the half-space $\Re(b)>0$. In particular, the two regimes $\Im(b)>0$ and $\Im(b)<0$ can be smoothly connected. We will show in Section \ref{sec:tet} that this behavior persists for all $k>1$.

The partition function $\CZ^{(k,p)}_b[M]_n$ of the $L(k,p)_b$ state-integral model for a triangulated manifold $M$ will be a sum (over $m$'s) and integral (over $\mu$'s) of a product of tetrahedron partition functions \eqref{defZD}. It will depend on pairs of vectors of un-summed/un-integrated variables $(\vec \mu,\vec m)=(\mu_1,...,\mu_d,m_1,...,m_d)\in \C^d \times (\Z/k\Z)^d$, which can again be re-grouped as $x_i = \exp\frac{2\pi i}{k}(-ib+pm)$ and $\tilde x_i=\exp \frac{2\pi i}{k}(-ib^{-1}+m)$. The number $d$ of such pairs depends on the topology of the boundary of $M$ and on the rank $n-1$ of the gauge group. For example, if $M=S^3\bs K$ is a knot complement there will be $d=n-1$ pairs of variables parametrizing eigenvalues of the $SL(n,\C)$ holonomy around one cycle of the boundary $\pd M\simeq T^2$.
We argue in this paper that
\begin{itemize}

\item[\bf 1.] The integrals appearing in the calculation of $\CZ_b^{(k,p)}[M]_n$ have well defined contours, and converge provided that the triangulation admits a positive angle structure, much as in \cite{Gar-index, GHRS-index}, \cite{KashAnd, AK-new} for $k=0,1$. Physically, this translates to a positivity requirement for $U(1)_R$ charges of operators in $T_n[M]$. Mathematically, the analysis of convergence is simplified by the introduction of functional spaces labelled by angle structures (Section \ref{sec:p}), in which functions $\CZ^{(k,p)}_b[M]_n$ are naturally valued.

\item[\bf 2.] The state-integrals are invariant under 2--3 moves that change the triangulation of $M$, so long as both initial and final triangulations admit a positive angle structure. This implies a ``local'' independence of triangulation. (It is not known whether any two positive ideal triangulations are related by a string of 2--3 moves passing only through positive ideal triangulations; so, strictly speaking, we do not prove global triangulation independence. There may be ways around this, as indicated by \cite{GHRS-index, AK-new}.)

Physically, the (local) 2--3 move follows from the mirror symmetry between 3d SQED and the XYZ model \cite{AHISS}. It translates to a relation of the form (see Sections \ref{sec:23}, \ref{sec:23-p} for details)
\begin{align} &  \frac{1}{k}\sum_{n=0}^{k-1}\int d\nu\, e^{\tfrac{2i\pi}{k}(-\mu_2\nu+ pm_2n)+\cdots} \CZ_b^{(k,p)}[\Delta](\nu,n)\,\CZ_b^{(k,p)}[\Delta](\mu_1-\nu,m_1-n) \\
 & =e^{(\cdots)} \CZ_b^{(k,p)}[\Delta](\mu_1,m_1)\,\CZ_b^{(k,p)}[\Delta](\mu_2,m_2)\, \CZ_b^{(k,p)}[\Delta](i(b+b^{-1})-\mu_1-\mu_2,r-1-m_1-m_2)\,,
\notag\end{align}
generalizing the 5-term relation for the quantum dilagarithm. For $p=1$, this relation was discussed in \cite{IY-Lens, IMY-fact}.

\item[\bf 3.] Each partition function $\CZ^{(k,p)}_b[M]_n$ should be annihilated by two sets of commuting difference operators, schematically $\{\CL_a(\x_i,\y_i;q^{\frac12})\}$ and $\{\CL_a(\tilde \x_i,\tilde \y_i;\tilde q^{\frac12})\}$, which depend on $M,n$, but not on the particular choice of $(k,p)$.
Here $\x_i,\y_i,\tilde \x_i,\tilde \y_i$ act via multiplication and shifts of the pairs of variables $(\mu_i,m_i)$,
\be \label{ops-intro}
 \begin{array}{ll} \x_i = x_i = e^{\frac{2\pi b}{k}\mu_i-\frac{2\pi i}{k}pm}\,,\qquad &\y_i = \exp\big(ib \tfrac{\pd}{\pd\mu_i}-\tfrac{\pd}{\pd m_i}\big) \\[.2cm]
\tilde\x_i = \tilde x_i = e^{\frac{2\pi b^{-1}}{k}\mu_i+\frac{2\pi i}{k}m}\,,\qquad &\tilde\y_i = \exp\big(ib^{-1} \tfrac{\pd}{\pd\mu_i}+r\tfrac{\pd}{\pd m_i}\big) \end{array}
\ee
(thus $\y_i$ shifts $\mu_i$ by $ib$ and $m_i$ by $-1$, etc.), forming a representation of two mutually commuting $q$- and $\tilde q$-Weyl algebras,
\be \y_i\x_j = q^{\delta_{ij}}\x_j\y_i\,,\qquad \tilde\y_i\tilde\x_j = \tilde q^{\delta_{ij}}\tilde\y_j\tilde\x_i\,;\qquad
\y_i\tilde \x_j = \tilde \x_j\y_i\,,\qquad \tilde\y_i\x_j=\x_j\tilde\y_i\,. \label{Weyl-intro} \ee
The $\CL_a$ are Laurent polynomials, and can be thought of as generating left ideals in $q$- and $\tilde q$-commutative rings.

For the tetrahedron partition function, it is easy to see that, \cf\ \cite{Dimofte-QRS}
\be (\y+\x^{-1}-1) \CZ^{(k,p)}_b[\Delta] \;=\; (\tilde\y+\tilde\x^{-1}-1)\CZ^{(k,p)}_b[\Delta] \;=\; 0\,.\ee
In general, the $\CL_a$ are a quantization of the algebraic variety of framed%
\footnote{We refer to \cite{DGG-Kdec, DGV-hybrid} for details on framing, which was inspired by the 2d constructions of \cite{FG-Teich}. Framing is extra data that resolves and localizes moduli spaces of flat connections, allowing one (for example) to obtain a nontrivial moduli space for a single ideal tetrahedron.} %
flat $SL(n,\C)$ connections that extend (irreducibly) from the boundary of $M$ into its interior \cite{gukov-2003}. For $SL(2,\C)$, this is (a component of) the classical A-polynomial of \cite{cooper-1994}.

Physically, the $\x_i,\y_i,\tilde x_i,\tilde y_i$ are electric and magnetic line operators localized at north and south poles of the lens space $L(k,p)_b$, \ie\ at $z=0$ or $w=0$ in \eqref{lens}, \cf\ \cite{GMNIII, DGG-index}. The relations $\CL_a\simeq 0$ are Ward identities.

Mathematically, the relations $\CL_a\simeq 0$ can be used to show (among other things) that partition functions $\CZ^{(k,p)}_b[M]_n(\vec \mu,\vec m)$ are meromorphic functions of $\vec\mu\in \C^d$ for each fixed $\vec m$, and to characterize their asymptotics.

\item[\bf 4.] Conjecturally, partition functions $\CZ^{(k,p)}_b[M]_n$ admit a factorization into holomorphic blocks
\be \CZ^{(k,p)}_b[M]_n  = \sum_{\alpha} B_\alpha(x,q^{\frac12})  B_\alpha(\tilde x,\tilde q^{\frac12})\,.  \label{blocks} \ee
where $\{\alpha\}$ is the set of irreducible flat $SL(n,\C)$ connections on $M$, assumed to be a finite set once appropriate boundary conditions are imposed. (Such a factorization is manifest for the tetrahedron partition function in \eqref{defZD}.) The $B_\alpha(x,q^{\frac12})$ are meromorphic functions of $x$ and $|q^{\frac12}|<1$ or $|q^{\frac12}|>1$ that solve the holomorphic difference equations $\CL_a(\x,\y;q^{\frac12})B_\alpha=0$ in the representation $\x B(x,q^{\frac12}) = xB(x,q^{\frac12})$ and $\y B(x,q^{\frac12}) = B(qx,q^{\frac12})$. Notably, the blocks depend on $M,n$, but \emph{not} on the choice of $(k,p)$ --- all dependence on $(k,p)$ enters indirectly via \eqref{defqx}.

Holomorphic blocks for $k=0,1$ were introduced in \cite{Pasquetti-fact, BDP-blocks}, and discussed for $L(k,1)_b$ partition functions in \cite{IMY-fact}. The general physical idea is that the lens space $L(k,p)_b$ can be glued together from two solid tori $D^2\times S^1$, with the $SL(2,\Z)$ element $\left(\begin{smallmatrix} r & s\\ -k & p \end{smallmatrix}\right)$ from \eqref{mod} specifying how the boundaries are to be identified. Each holomorphic block $B_\alpha$ is a partition function of $T_n[M]$ on a solid torus, with a fixed choice of vacuum $\alpha$ at the boundary $\pd(D^2\times S^1)$.%
\footnote{Since $T_n[M]$ is not a topological theory, it is not immediately obvious that only vacua $\alpha$ will contribute to the sum \eqref{blocks}. One needs, for example, to show that there is a to deform $L(k,p)_b$ into into $D^2\times S^1$ geometries connected by an infinitely long cylindrical region (thus projecting to ground states, or vacua), without modifying the partition function. Methods of \cite{CDFZ-geom} should be useful for entertaining a general analysis of this sort; for $k=1$, \ie\ for $S^3_b$, factorization was proven in \cite{AMRP-fact}.
} %

\item[\bf 5.] Finally, we note in passing that the asymptotics of $\CZ_b^{(k,p)}[M]_n$ as $b\to0$ seem to be related to asymptotics of colored Jones (or HOMFLY) polynomials taken around the $k$-th root of unity $q=e^{2\pi ip/k}$, as in \cite{QMF}.
This is an interesting twist on the standard $q\to 1$ asymptotics of Jones polynomials that enter the volume conjecture \cite{kashaev-1997, Mur-Mur, gukov-2003}, and which were already argued to match the $k=1$ state-integral model in \cite{DGLZ, DG-quantumNZ}. We will study asymptotics for general $(k,p)$ in a separate companion paper.

\end{itemize}

\subsection*{Organization}

We proceed in Section \ref{sec:CS} to discuss some basic features of quantization of complex Chern-Simons theory, with a real polarization. We focus on a model phase space of the form $\C^*\times \C^*$ (as appropriate for tetrahedra in state-integral models), and will find holomorphic and antiholomorphic copies of the operator algebra \eqref{ops-intro}.

In Sections \ref{sec:p} and \ref{sec:tet}, we make some necessary preparations for the definition of $L(k,1)_b$ state-integral models. In particular, we introduce the notion of functional spaces $\CH_\p$ labelled by convex symplectic polytopes $\p$, in which state integrals are naturally valued. These polytopes $\p$ govern the decay of functions, and ultimately guarantee that integrals are done along unique, convergent contours. They come to life in Section \ref{sec:tet} as the positive-angle polytopes for individual tetrahedra.

In Section \ref{sec:SS} we define the $L(k,1)_b$ state-integral model, computing level-$k$ Chern-Simons partition functions, and explain its local topological invariance under 2--3 moves. We give some examples of state integrals and holomorphic-block factorization in Sections \ref{sec:ex}--\ref{sec:blocks}. Then, in Section \ref{sec:kp}, we extend the state-integral to general $L(k,p)_b$.

\subsection*{Complementary results}

This paper has been coordinated to appear alongside three other works on non-perturbative aspects of Chern-Simons theory with complex gauge group. Geometric quantization of complex Chern-Simons theory at level $k$ is studied in \cite{AG-HW}. The quantization there uses a \emph{complex} polarization (in contrast to the real polarization invoked in the present paper). This causes Chern-Simons Hilbert spaces to depends on the complex structure of an underlying surface, with different choices of complex structure related by the projectively flat Hitchin-Witten connection.
Such a situation is familiar in the standard geometric quantization of compact Chern-Simons theory.
Mapping-class-group actions induced by the Hitchin-Witten connection are analyzed in \cite{A-MCG} and \cite{AK-complexCS}. The latter uses ideal triangulations to generate the mapping class group action, which should be very closely related to the approach of the present paper.

\section{Quantization of a model phase space}
\label{sec:CS}

In this section, we perform the level $(k,\sigma)$ quantization of a model phase space
\be \CP = \C^*\times \C^*=\{(x,y)\}\,,\qquad \Omega=\frac{dy}{y}\wedge\frac{dx}{x}\,, \label{PD}\ee
with holomorphic symplectic form $\Omega$. (To actually quantize the space we will need to choose a real symplectic form, and $k$ and $\sigma$ determine this choice.) Quantization produces an algebra of operators $\x,\y,\tilde\x,\tilde\y$ acting on a functional space $\CH^{(k)}$. We will see that $\CH^{(k)}$ naturally contains functions $\CZ(\mu,m)$ depending on a continuous variable $\mu$ and an integer $m\in \Z/k\Z$, matching the structure of $L(k,1)_b$ partition function of the tetrahedron theory.  More generally, the $L(k,1)_b$ partition functions of theories $T_n[M]$ on ideally triangulated $M$ belong (roughly) to products $(\CH^{(k)})^{\otimes d}$. The present analysis mirrors \cite[Sec. 6]{DGG-index}, which studied the case $k=0$.

In what sense is $\CP$ a model phase space? It appears most directly as an open subvariety of the space of $(P)SL(2,\C)$ Chern-Simons theory on the boundary of an ideal tetrahedron \cite{Dimofte-QRS, DGG-Kdec}. To be precise,
\be \begin{array}{l@{\!}l} \CP \subset \{\,&\text{framed flat $PGL(2,\C)$ connections on a 4-punctured sphere}\\
   &\text{with unipotent holonomy at the punctures}\}\,,
\end{array}
\ee
where $x,y$ are $\C^*$-valued cluster coordinates on the moduli space \cite{Fock-Teich, FG-Teich} and the holomorphic symplectic form $\Omega$ agrees with that of Atiyah and Bott \cite{AtiyahBott-YM}, $\Omega \sim \int_{\pd\Delta} {\rm Tr}[\delta\CA\wedge\delta \CA]$. More generally, the moduli space of framed flat $PGL(n,\C)$ connections on the boundary of any of 3-manifold with an ideal triangulation has open patches of the form $(\CP)^d$.  An example at the opposite end of the spectrum from a single tetrahedron is a knot complement, with smooth torus boundary; the space of flat $SL(n,\C)$ connections on a torus has the form $(\CP)^{n-1}/S_n$. (Upon quantization, the quotient by the Weyl group $S_n$ simply acts as a projection to invariant wavefunctions.)

\subsection{Diagonalizing $\Re(t\Omega)$}

Chern-Simons theory at levels $(k,\sigma)$ quantizes the model phase space $\CP$ with respect to the \emph{real} symplectic form \cite{Witten-cx}
\be \omega_{k,\sigma} := \frac{1}{4\pi} \big( t\, \Omega + \tilde t\, \ol \Omega \big)\,;\qquad
 t:= k+\sigma\,,\quad \tilde t:=k-\sigma\,.  \label{wks} \ee
Setting $x = \exp X$, $y=\exp Y$ and expanding, we get
\be \omega_{k,\sigma} = \frac k{2\pi} (d\,\Re\, Y\wedge d\,\Re\, X-d\,\Im\, Y\wedge d\,\Im\,X) +\frac{i\sigma}{2\pi} (d\,\Re\, Y\wedge d\,\Im\, X + d\,\Im\, Y\wedge d\,\Re\, X)\,.\ee
One term here has a nontrivial period, $\frac{1}{2\pi}\int_{S^1\times S^1} d\,\Im\, Y\wedge d\,\Im\, X = 2\pi$. Thus, in order to get an integral class $\frac1{2\pi}\omega_{k,\sigma}\in H^2(\CP[\pd \Delta],\Z)$, as necessary for pre-quantization, we must require $\boxed{k\in \Z}$. We will assume $k\geq 0$, noting that changing the sign of $k$ is equivalent to reversal of orientation.
We also observe that $\omega_{k,\sigma}$ is manifestly real when $\boxed{\sigma \in i\R}$\,; for such values, Chern-Simons theory becomes unitary, with respect to a standard Hermitian structure. A second unitary branch, with respect to a non-standard Hermitian structure, arises when $\boxed{\sigma\in \R}$ \cite{Witten-cx}.

It is extraordinarily convenient to reparameterize%
\footnote{Such a reparameterization, for $k=1$, arose naturally when compactifying the 6d $(2,0)$ on $S^3_b$ to obtain complex Chern-Simons in \cite{CJ-S3}. It is also reminiscent of the dependence of the canonical parameter on the twisting parameter in 4d Langlands-twisted Yang-Mills theory \cite{Kapustin-Witten}, which can be used to engineer complex Chern-Simons theory \cite{Wfiveknots}, \cf\ \cite[Sec. 6]{BDP-blocks}.} %
the level $\sigma$ as
\be \sigma = k\, \frac{1-b^2}{1+b^2}\,, \label{skb} \ee 
where $b$ is a complex number with $\Re(b)>0$.
Note that $\sigma$ is imaginary when $|b|=1$ and real when $b^2\in \R$. One advantage of this parametrization is that the holomorphic and anti-holomorphic inverse-couplings of Chern-Simons theory become
\be  \frac{4\pi i}{t} = \frac{4\pi i}{k+\sigma} = \frac{2\pi i}{k}(1+b^2)\,,\qquad  \frac{4\pi i}{\tilde t} = \frac{4\pi i}{k-\sigma} =\frac{2\pi i}{k}(1+b^{-2})\,. \label{invt} \ee
Another advantage is that upon choosing a square-root $b=\sqrt{b^2}$ and reparameterizing $x$ and $y$ as
\be \label{defxmu} \begin{array}{c} \ds x = \exp\frac{2\pi i}{k}(-ib\mu-m)\,,\quad \bar x= \exp\frac{2\pi i}{k}(-ib^{-1}\mu+m)\,, \\[.3cm]
\ds y=\exp\frac{2\pi i}{k}(-ib\nu-n)\,,\quad \bar y= \exp\frac{2\pi i}{k}(-ib^{-1}\nu+n)\,,  \end{array}\ee
the symplectic form diagonalizes:
\be \omega_{k,\sigma} =  \frac{2\pi}{k} (d\nu\wedge d\mu- dn\wedge dm)\,. \label{wdiag} \ee
Here $(m,n)$ are real and periodic ($m\sim m+k$, $n\sim n+k$), and for the standard Hermitian structure $(\mu,\nu)$ are real. When we analytically continue $\mu,\nu$ (and $b$) away from real (unit-circle) values, we will define $\tilde x,\tilde y$ by the RHS of \eqref{defxmu}, using `tildes' instead of `bars' since the variables are no longer related by standard complex conjugation.

\subsection{Operator algebra}
\label{sec:ops}

It is easy to quantize $(\CP,\omega_{k,\sigma})$ in the $(\mu,m;\nu,n)$ parametrization. The coordinates get promoted to operators $\bmu,\m;\bnu, \n$ with canonical commutation relations following from \eqref{wdiag},
\be [\bnu, \bmu] = -\frac{k}{2\pi i}\,,\qquad [\n,\m] = \frac{k}{2\pi i}\,;\qquad [\bnu,\m]=[\bmu,\n]=0\,. \label{comm-mu} \ee
Alternatively, setting
\be \label{defxmuhat} \begin{array}{c} \ds \x = \exp\frac{2\pi i}{k}(-ib\bmu-\m)\,,\quad \tilde \x= \exp\frac{2\pi i}{k}(-ib^{-1}\bmu+\m)\,, \\[.3cm]
\ds \y=\exp\frac{2\pi i}{k}(-ib\bnu-\n)\,,\quad \tilde \y= \exp\frac{2\pi i}{k}(-ib^{-1}\bnu+\n)\,,  \end{array}\ee
and
\be \label{defqq}  q = \exp\frac{4\pi i}{t}=\exp\frac{2\pi i}{k}(1+b^2)\,,\qquad \tilde q = \exp\frac{4\pi i}{\tilde t}=\exp\frac{2\pi i}{k}(1+b^{-2})\,, \ee
we find $q$- and $\tilde q$-Weyl algebras
\be \label{xycomm} \y\x = q\x\y\,,\quad \tilde\y\tilde\x=\tilde q\tilde\x\tilde\y\,;\qquad \y\tilde\x=\tilde\x\y\,,\quad \tilde\y\x=\x\tilde\y\,.\ee

Since $n$ and $m$ are both periodic and canonically conjugate, the operators $\m$, $\n$ must be constraint to have quantized spectra (in any representation). Indeed, standard arguments show that $\m$ and $\n$ take integer values
\be \m,\n \in \Z/k\Z\,. \label{quantmn} \ee
In contrast, the non-periodic variables $\mu,\nu$ are unconstrained. In the standard Hermitian structure, all operators $\bmu,\m;\bnu,\n$ are Hermitian, and the spectra of $\mu,\nu$ should be real.

We observe that when $k=1$, the operators $\m,\n \equiv 1$ are completely fixed. Alternatively, the arguments $e^{2\pi i m/k}$, $e^{2\pi i n/k}$ of $x$ and $y$ trivialize. In this case, the quantization of $\CP$ looks identical to the quantization of $\CP_\R = \R\times\R=\{(\mu,\nu)\}$ parametrized by $\mu,\nu$, with symplectic form $2\pi\,d\mu\wedge d\nu$, which would correspond not to $SL(2,\C)$ but to $SL(2,\R)$ Chern-Simons theory.

\subsection{Naive representation}
\label{sec:CS-rep}

Given these quantization conditions, one expects the quantization of $\CP$ at levels $(k,\sigma)$ to produce a Hilbert space of the form
\be \CH^{(k,\sigma)} \approx  L^2(\R)\otimes_\C V_k\,,  \label{L2} \ee 
where $V_k\simeq \C^k$ is a $k$-dimensional vector space. For example, elements of $\CH^{(k,\sigma)}$ could be represented as functions $f(\mu,m)$ of a continuous variable $\mu$ and a quantized variable $m\in \Z/k\Z$. Operators $\bmu,\m;\bnu,\n$ act on these functions as
\bse \label{op-action}
\be  \begin{array}{ll} \bmu f(\mu,m) = \mu f(\mu,m)\,,\qquad &e^{\frac{2\pi i}{k}\m} f(\mu,m) = e^{\frac{2\pi i}{k}m}f(\mu,m)\,, \\[.2cm]
 \bnu f(\mu,m) = -\frac{k}{2\pi i}\pd_\mu f(\mu,m)\,,\qquad &e^{\frac{2\pi i}{k}\n} f(\mu,m) = f(\mu,m+1)\,.
\end{array}
\ee
Equivalently,
\be \begin{array}{ll} \x f(\mu,m) = xf(\mu,m)\,,\qquad &\tilde\x f(\mu,m) = \tilde xf(\mu,m)\,, \\[.2cm]
 \y f(\mu,m) = f(\mu+ib,m-1)\,, \quad & \tilde\y f(\mu,m) = f(\mu+ib^{-1},m+1)\,; \end{array} \ee
or, reparameterizing functions directly in terms of variables $x$ and $\tilde x$,
\be \begin{array}{ll} \x f(x,\tilde x) = xf(x,\tilde x)\,,\qquad &\tilde\x f(x,\tilde x) = \tilde xf(x,\tilde x)\,, \\[.2cm]
 \y f(x,\tilde x) = f(qx,\tilde x)\,, & \tilde\y f(x,\tilde x) = f(x,\tilde q\tilde x)\,. \end{array} \ee
\ese

\subsection{Comparison to partition functions of $T_n[M]$}
\label{sec:CS-comp}

The theories $T_n[M]$ associated to 3-manifolds by the DGG construction \cite{DGG, DGG-Kdec} always have a UV description as abelian gauge theories (with nontrivial Chern-Simons terms and superpotential couplings). Their partition function of $T_n[M]$ on a lens space $L(k,1)_b$ \cite{BNY-Lens} depends on some $d$ pairs of variables $(\mu_i,m_i)$. The $\mu_i$ are real masses associated to each $U(1)$ factor in the maximal torus of the flavor symmetry group of $T_n[M]$, while the $m_i$ label background holonomies that can be turned on for each $U(1)$ in the maximal torus. Since the fundamental group of $L(k,1)_b$ is $\Z/k\Z$, the Wilson lines must take values $m_i\in \Z/k\Z$. We find that the lens-space partition functions of $T_n[M]$ (roughly) take values in
\be  \CZ_b^{(k,1)}[M]_n   \;\in\; (\CH^{(k,\sigma)})^{\otimes d}\,, \label{ZinH} \ee
exactly as desired in order for these partition functions to correspond to wavefunctions in complex Chern-Simons theory at levels $(k,\sigma)$.

From the perspective of $T_n[M]$ on a lens space, it is most natural to have $b^2\in \R_+$, as in \eqref{lens}. This corresponds to the non-standard Hermitian structure in Chern-Simons theory. Once partition functions are computed, they can be analytically continued to other values of~$b$.

The operator algebra of Section \ref{sec:ops} also has a natural field-theory interpretation, along the lines described in \cite{DGG,DGG-index} (following \cite{GMNIII, AGGTV, DGOT}) for $k=0$ and $k=1$. Namely, $\x$ and $\tilde\x$ are half-BPS flavor Wilson lines, for a given $U(1)$ in the flavor symmetry group of $T_n[M]$, localized on circles at the north or south poles of the lens space $L(k,1)_b$. In terms of \eqref{lens}, these circles are defined by $z=0$ and $w=0$, respectively. Similarly, $\y$ and $\tilde \y$ are magnetic or ``disorder'' operators for flavor-symmetry gauge fields, localized on the circles $z=0$ and $w=0$, respectively. They act on partition functions precisely as in \eqref{op-action}.%
\footnote{In some cases, one can also understand the line operators $\x,\y,\tilde\x,\tilde\y$ as dimensional reductions of the surface operators described in \cite{RY-Lens}.}

\section{Functional spaces labelled by angle polytopes}
\label{sec:p}

In Section \ref{sec:CS}, we found that quantization of Chern-Simons theory at levels $(k,\sigma)$ seems to associate a functional space $\CH^{(k,\sigma)}\simeq L^2(\R)\otimes V_k$ to the classical phase space $\C^*\times \C^*$, and that $L(k,1)_b$ partition functions of 3-manifold theories $T_n[M]$ depend on the right variables to belong to this space, or to products thereof.

In practice, one quickly realizes that $L(k,1)_b$ partition functions of $T_n[M]$ are \emph{not} always square-integrable. Thus they do not exactly belong to $L^2(\R)\otimes V_k$. Moreover, the partition functions themselves are expressed as noncompact integrals (of products of tetrahedron partition functions) and \emph{a priori} the integration contours are not always well defined.  We fix these problems by slightly modifying the representation of the operator algebra from Section \ref{sec:CS-rep}: we introduce a family of functional spaces $\CH_\p$ labelled by symplectic polytopes $\p$, in which partition functions will naturally be valued, and which render all integrals well defined.

In the present section we describe the formal properties of the spaces $\CH_\p$, focusing in particular on the action of affine symplectic transformations. Such transformations intertwine different representations of the operator algebra from Section \ref{sec:CS}. Physically, they descend from Witten's symplectic action on 3d $\CN=2$ theories $T_n[M]$ \cite{Witten-sl2}.
Later (Section \ref{sec:SS}) we identify $\p$ with the angle polytope of an ideal triangulation, and with a certain positivity structure on dimensions of operators in theories $T_n[M]$.

Throughout this section we fix an integer $k\geq 1$ and a complex number $b$ with $\Re(b) > 0$. We also set (here and in the rest of the paper)
\be \boxed{Q:=b+b^{-1}\,,\qquad \mathfrak Q :=\Re(Q)>0}\,.\ee

\subsection{Definitions}
\label{sec:defP}

Let us endow $\R^{2N}\simeq T^*(\R^N)$ with the canonical symplectic structure $\omega$ and choose Darboux coordinates $(\vec\alpha,\vec\beta) = (\alpha_1,...,\alpha_N,\beta_1,...,\beta_N)$ such that
\be  \omega = \sum_i d\beta_i\wedge d\alpha_i\,. \ee
The symplectic space $\R^{2N}$ can be thought of the universal cover of the ``angle-space'' of the phase space $\CP^N = (\C^*\times\C^*)^N$ from \eqref{PD}. Indeed, using $(\mu_i,m_i)$ and $(\nu_i,n_i)$ to parametrize $\CP^N$ as in \eqref{defxmu}, we may identify
\be \alpha_i\, \leftrightarrow \,\Im(\mu_i)\,,\qquad \beta_i\, \leftrightarrow \,\Im(\nu_i)\,. \label{anglemu} \ee

We define an \emph{open convex symplectic polytope} (or simply \emph{polytope}) $\p$ of dimension $2N$ to be an open subset of $(\R^{2N},\omega)$ cut out by a finite set of strict linear inequalities. Let $\pi(\p)$ denote the projection of $\p$ to the base of $T^*(\R^N)$, with coordinates $\vec\alpha$, and $\text{strip}(\p) \subset \C^N$ be the strip
\be \text{strip}(\p) := \{\,\vec\mu\in \C^N\,|\, \Im(\vec\mu) \in \pi(\p)\,\}\,. \ee
We can then define the functional space
\be  \begin{array}{ll}\CH_\p := &
\big\{\,\text{holomorphic functions $f:\, \text{strip}(\p) \to \C$ s.t.} \\[.1cm]
&\;\;\text{$\forall$ $(\vec\alpha,\vec\beta)\in\p$, the function $e^{-\frac{2\pi}{k}\vec\mu\cdot\vec\beta}f(\vec\mu+i\vec\alpha) \in \CS(\R^N)$ is Schwarz class}\,\big\}\,.
\end{array} \label{Hp1}
\ee
This means that $e^{-\frac{2\pi}{k}\vec\mu\cdot\vec\beta}f(\vec\mu+i\vec\alpha)$ decays exponentially as $|\vec\mu|\to\infty$, $\vec\mu\in \R^N$.%
\footnote{In order to generalize and extend some of the arguments of the present paper, it may be useful to extend from $\CS(\R^N)$ to $L^2(\R^N)$, or to corresponding spaces of distributions. We leave this to future work.} %
The condition on exponential decay can be rephrased more intuitively by using $N$-tuples of operators $\vec\bmu,\vec\bnu$ acting as in (\ref{op-action}a), that is, $\bmu_i=\mu_i\cdot$ and $\bnu_i = -\frac{k}{2\pi i}\pd_{\mu_i}$. The space $\CH_\p$ contains precisely those functions $f$ such that
\be    \boxed{\exp\Big[\frac{2\pi}{k}(\vec\alpha\cdot\vec\bnu-\vec\beta\cdot\vec\bmu)\Big]\cdot  f   \; \in \CS(\R^N)\qquad \forall\;(\vec\alpha,\vec\beta)\in \p\,.} \label{Hp2} \ee
In this form, the condition is manifestly invariant under Fourier transform. Indeed, one should think of \eqref{Hp2} as simultaneously bounding the decay of $f$ and its Fourier transform, with bounds given by the polytope $\p$.

For quantization at level $k$, it is also useful to introduce the twisted space
\be \CH^{(k)}_\p := \CH_\p \otimes_\C (V_k)^{\otimes N} \simeq (\CH_\p)^{\oplus k^N} \,, \ee
where $V_k \simeq \C^k$ is a $k$-dimensional vector space.

\subsection{The action of $ISp(2N,\Z)$}
\label{sec:p-Weil}

The affine symplectic group $ISp(2N,\Z)$ acts naturally both on symplectic polytopes $\p\subset\R^{2N}$ and  on the spaces $\CH_\p$. Indeed, \emph{every $g\in ISp(2N,\Z)$ induces an isomorphism
\be \boxed{ g:\; \CH_\p  \overset{\sim}{\longrightarrow} \CH_{g(\p)} \qquad\text{and}\qquad  \CH_\p^{(k)}  \overset{\sim}{\longrightarrow} \CH_{g(\p)}^{(k)}} \,, \label{giso} \ee
uniquely defined up to multiplication by a constant} (to be discussed momentarily). The isomorphisms come essentially from the Weil representation \cite{Shale-rep, Weil-rep}, and constitute an action of $ISp(2N,\Z)$ on the spaces $\{\CH_\p\}$, \ie\ $(g_1g_2)\CH_p = g_1\cdot (g_2\cdot \CH_p)\simeq \CH_{g_1g_2(\p)}$.

We can define \eqref{giso} by using a set of generators of the symplectic group. For $Sp(2N,\Z)$, thought of as the group of matrices $g$ that preserve the symplectic form
$gJ_{2N}g^T = J_{2N}$ with $J_{2N} = \left(\begin{smallmatrix} \mb 0 & -\mb I \\ \mb I & \mb 0  \end{smallmatrix}\right)$, we can choose generators of the form
\be \label{sympgens}
\begin{array}{c@{\quad}c@{\quad}c}
\text{S-type} & S_i:=\begin{pmatrix} \mb 0 & \text{diag}(0,...,\overset{i}{-1},...,0) \\
 \text{diag}(0,...,\overset{i}{1},...,0) & \mb 0 \end{pmatrix}\,, & i=1,...,N\,; \\[.6cm]
\text{T-type} & T(\mb B):=\begin{pmatrix} \mb I & \mb 0 \\ \mb B & \mb I \end{pmatrix}\,, & \mb B = \mb B^T \in M(N,\Z)\,; \\[.6cm]
\text{U-type} & U(\mb A):= \begin{pmatrix} \mb A & \mb 0 \\ \mb 0 & \mb A^{-1,T} \end{pmatrix}\,, & \mb A \in GL(N,\Z)\,.
\end{array}
\ee
Here all generators are written in terms of $N\times N$ blocks.

We take $Sp(2N,\Z)$ to act on $\R^{2N}$ in the defining representation, multiplying coordinates (viewed as column vectors) on the left. It is convenient to think of these as passive transformations, so that $g:\begin{pmatrix}\vec\alpha\\\vec\beta\end{pmatrix}\mapsto g^{-1}\begin{pmatrix}\vec\alpha'\\\vec\beta'\end{pmatrix}$, or $\begin{pmatrix}\vec\alpha'\\\vec\beta'\end{pmatrix}= g\begin{pmatrix}\vec\alpha\\\vec\beta\end{pmatrix}$.
The affine extension $ISp(2N,\Z) \simeq Sp(2N,\Z)\ltimes\Z^{2N}$ is generated by the above matrices together with affine shifts, which act passively on $\R^{2N}$ as
\be  \sigma_\alpha(\vec t\,):\; \begin{pmatrix}\vec\alpha\\\vec\beta\end{pmatrix} \mapsto \begin{pmatrix} \vec\alpha'-\frac{\mathfrak Q}{2}\vec t\, \\ \vec\beta' \end{pmatrix}\,,\qquad  \sigma_\beta(\vec t\,):\; \begin{pmatrix}\vec\alpha\\\vec\beta\end{pmatrix} \mapsto \begin{pmatrix} \vec\alpha' \\ \vec\beta'-\frac{\mathfrak Q}{2}\vec t\, \end{pmatrix}\,,\qquad \vec t\in \Z^N\,. \ee
(In this paper, we will only consider affine shifts by multiples of $\mathfrak Q/2$. This quantization condition will actually be essential when studying general Lens spaces $L(k,p)$ in Section~\ref{sec:kp}.)
The action on $\R^{2N}$ is faithful, and can be used to define the affine symplectic group.
Altogether, we see that for $g\in ISp(2N,\Z)$ the transformed polytope $g(\p)$ is defined by
\be (\vec\alpha',\vec\beta')\in g(\p) \quad\Leftrightarrow \quad g^{-1}(\vec\alpha',\vec\beta')\in \p\,. \ee

The corresponding Weil-like action on functions $f(\vec \mu,\vec m)\in \CH_\p^{(k)}$ is defined by
\be
(S_i\cdot f)(\vec\mu,\vec m)  \ds := \frac{1}{k}\sum_{n\in \Z/k\Z} \int d\nu\;  e^{\tfrac{2\pi i}{k}(-\mu_i\nu+m_in)}\,   f(\mu_1,...,\overset{(i)}{\nu},...,\mu_N;m_1,...,\overset{(i)}{n},...,m_N)\,; \notag \ee
\vspace{-.4cm}
\be\begin{array}{rl}
  (T(\mb B)\cdot f)(\vec\mu,\vec m) &:=  (-1)^{\vec m\cdot\mb B\cdot\vec m}e^{\tfrac{i\pi}{k}(-\vec\mu\cdot\mb B\cdot\vec\mu+\vec m\cdot\mb B\cdot\vec m)} f(\vec\mu,\vec m)\,; \\[.2cm]
  (U(\mb A)\cdot f)(\vec\mu,\vec m)  &:= (\det\mb A)^{-\frac12} f(\mb A^{-1}\vec\mu,\mb A^{-1}\vec m)\,;\\[.2cm]
  (\sigma_\alpha(\vec t\,)\cdot f)(\vec\mu,\vec m)  &:= f(\vec \mu - \frac{iQ}{2}\vec t,\vec m)\,;\\[.2cm]
  (\sigma_\beta(\vec t\,)\cdot f)(\vec\mu,\vec m) &:=  e^{\tfrac{\pi Q}{k} \vec\mu \cdot \vec t}\, f(\vec\mu,\vec m)\,;
\end{array} \label{repf}
\ee
where the integration contour in the action of the $S_i$ (essentially a Fourier transform) is chosen to lie on any line $\nu\in \R+i\alpha_i$ such that the hyperplane $\{\alpha=\alpha_i\}$ intersects $\p$.

From the definition of $\CH_\p^{(k)}$, it follows that when $f\in \CH_\p^{(k)}$ the integral in $S_i\cdot f$ converges absolutely, with a unique result, so long as the hyperplane $\{\alpha=\alpha_i,\beta=-\Im(\nu)\}$ intersects $\p$. Indeed, it follows from standard analysis%
\footnote{For example, this is an application of the Paley-Wiener theorem. Note that the finite sum over $n$ is irrelevant here.} %
that $S_i\cdot f \in \CH_{S_i(\p)}^{(k)}$ whenever $f\in \CH_\p^{(k)}$. Formally, this result is a consequence of the fact that the operator condition \eqref{Hp2} involves a symplectic pairing that is invariant under any symplectic transformation,
\be \vec\alpha\cdot\vec\bnu -\vec\beta\cdot\vec\bmu = \langle (\vec\alpha,\vec\beta),(\vec\bmu,\vec\bnu)\rangle = \langle g(\vec\alpha,\vec\beta), g(\vec\bmu,\vec\bnu)\rangle  \qquad \forall\; g\in Sp(2N,\Z)\,, \ee
together with invariance of the Schwarz space under Fourier transform. To see that $S_i:\CH^{(k)}_\p\to \CH^{(k)}_{S_i(\p)}$ is an isomorphism, one has simply to observe that $S_i$ is invertible, with $S_i\cdot(S_i\cdot f(\vec\mu,\vec m)) = f(-\vec \mu,-\vec m)$.

We have argued that S-type transformations induce the isomorphism claimed on the RHS of \eqref{giso}. (The corresponding isomorphism on the LHS follows by setting $k=1$.) The remaining generators of $ISp(2N,\Z)$ also induce the desired isomorphisms, in a fairly trivial way. For $g\in Sp(2N,\Z)$ this again follows from the symplectically invariant operator condition \eqref{Hp2}. For affine shifts, it is clear that the polytope $\p$ just gets translated.

Once integration contours are defined, one can verify that relations of the affine symplectic group hold in the representation \eqref{repf}, by repeating arguments for the standard Weil representation. To be precise, the relations hold up to multiplicative (projective) factors. These factors have two origins. First, within $Sp(2N,\Z)$ itself, the S-type and U-type transformations come with ambiguous factors of $\pm 1$. (Indeed, the Weil representation is an honest representation of the metaplectic group, and is only a projective representation of the symplectic group.) Second, when acting on functions, the affine shifts $\sigma_\alpha$, $\sigma_\beta$ don't quite commute properly with each other or with T-type elements. With the restriction that affine shifts only come in integer multiples of $iQ/2$, the overall projective ambiguity is of the form
\be  (-1)^{a_1}\,e^{a_2\tfrac{i\pi}{k}\tfrac{Q^2}{4}}\,,\qquad a_1,a_2\in \Z\,. \label{amb1} \ee

We emphasize that when $k\geq 2$ the action in \eqref{repf} is necessarily that of $ISp(2N,\Z)$ rather than $ISp(2N,\R)$. That is, although affine shifts might take arbitrary real values, the symplectic transformations must be integral. Otherwise, the exponential prefactors involved in $S_i\cdot f$ and $T(\mb B)\cdot f$ would not be well-defined functions of $\vec m\in (\Z/k\Z)^N$, and the argument $\mb A^{-1}\vec m$ in $U(\mb A)\cdot f$ might not be defined at all.

\subsection{A bit of physics: symplectic actions on theories}
\label{sec:H-phys}

From the point of view of complex Chern-Simons theory, the symplectic isomorphisms of \eqref{giso} relate different representations of the same Chern-Simons ``Hilbert'' space. They are an essential ingredient in any consistent quantization. Had the Hilbert space been $(L^2(\R)\otimes_\C V_k)^{\otimes N}$ as in \eqref{L2}, there would have been no need to keep track of polytopes $\p$, since the standard Weil representation preserves $L^2(\R)$ and automatically provides the requisite isomorphisms. In our case, since wavefunctions are not standard elements in $L^2(\R)$, we had to do a bit more work.

From the point of view of 3d superconformal theories $T_n[M]$, the symplectic action of \eqref{giso} is a manifestation of an affine symplectic action on \emph{theories}. This action was first described in \cite{Witten-sl2}, and exploited in \cite{DGG}. It can be understood as arising from electric-magnetic duality in a 4d ``bulk'' abelian gauge theory coupled to the flavor currents of a 3d theory $T_n[M]$ on its boundary. Indeed, the 4d duality group is $Sp(2N,\Z)$, and can be extended to $ISp(2N,\Z)$ in the presence of a boundary. The fact that symplectic transformations provide \emph{isomorphisms} of spaces as in \eqref{giso} follows from the fact that electric-magnetic duality is an equivalence of 4d theories and their boundary conditions.

The symplectic action on theories $T_n[M]$ induces an action on $L(k,p)_b$ partition functions.
In the case of $L(k,1)_b$ it can be read off from the rules of \cite{IMY-fact} (and led us to the prescription in \eqref{repf}).
For example, T-type transformations arise from adding abelian Chern-Simons terms to $T_n[M]$, and these Chern-Simons produce the quadratic-exponential prefactor in \eqref{repf}. It was in \cite{IMY-fact} that several crucial signs coming from these Chern-Simons terms were systematically analyzed. The fact that the symplectic action only exists for $ISp(2N,\Z)$ rather than $ISp(2N,\R)$ in general follows from the fact that abelian Chern-Simons terms in $T_n[M]$ have integer levels (or from integrality of a 4d electric-magnetic duality group).

\subsection{Intertwining the operators}

The affine symplectic transformations $g:\CH_\p^{(k)} \overset{\sim}\longrightarrow \CH_{g(\p)}^{(k)}$ intertwine a symplectic action in the operator algebra generated by $\x_i,\y_i,\tilde\x_i,\tilde\y_i$ as in Section \ref{sec:CS-rep} (\cf\ \cite{Dimofte-QRS} for $k=1$). In other words, if $\CO$ is any element of the operator algebra, acting on a wavefunction $f \in \CH_\p^{(k)}$, then
\be g\cdot [\CO\,f] = (g\cdot \CO)\, (g\cdot f)\,. \label{ginter}  \ee
Clearly the intertwining action is just conjugation, $g\cdot \CO = g\CO g^{-1}$.

As a technical point, we note that operators $\x,\y,\tilde\x,\tilde \y$ do not preserve symplectic polytopes $\p$, but shift them slightly in units of $b$ or $b^{-1}$. If $\CO$ is a sum of monomials in generators of the operator algebra and $f\in \CH_\p^{(k)}$, then $\CO\,f \in \CH_{\p'}^{(k)}$, where $\p'$ is an intersection of copies of $\p$ shifted by the monomials in $\CO$. The intertwining identity \eqref{ginter} is meant to hold in $\CH_{g(\p')}^{(k)}$.

Working out the action on formal \emph{logarithms} of generators of the operator algebra, we find that it is just the defining representation of $ISp(2N,\Z)$, implemented via passive transformations. Thus, in terms of column vectors,
\bse \label{glogop}
\be g: \begin{pmatrix} \log\vec\x \\ \log\vec\y \end{pmatrix} \mapsto g^{-1}\begin{pmatrix} \log\vec\x' \\ \log\vec\y' \end{pmatrix}\,, \qquad 
g: \begin{pmatrix} \log\vec{\tilde\x} \\ \log\vec{\tilde\y} \end{pmatrix} \mapsto g^{-1}\begin{pmatrix} \log\vec{\tilde\x}' \\ \log\vec{\tilde\y}' \end{pmatrix}\,,\qquad
g\in Sp(2N,\Z)\,, \ee
while affine shifts are slightly rescaled,
\be \sigma_\alpha(\vec t\,) : \begin{array}{c} \begin{pmatrix} \log\vec\x \\ \log\vec\y \end{pmatrix} \mapsto \begin{pmatrix} \log\vec\x' - \frac{i\pi bQ}{k}\vec t\, \\ \log\vec\y' \end{pmatrix} \\[.5cm]
\begin{pmatrix} \log\vec{\tilde\x} \\ \log\vec{\tilde\y} \end{pmatrix}  \mapsto
  \begin{pmatrix} \log\vec{\tilde\x}'-\frac{ i\pi b^{-1}Q}{k}\vec t\, \\ \log\vec{\tilde\y}' \end{pmatrix} \end{array}\,,\qquad
  \sigma_\beta(\vec t\, ): \begin{array}{c} \begin{pmatrix} \log\vec\x \\ \log\vec\y \end{pmatrix} \mapsto \begin{pmatrix} \log\vec\x' \\ \log\vec\y'  - \frac{i\pi b Q}{k}\vec t\, \end{pmatrix} \\[.5cm]
\begin{pmatrix} \log\vec{\tilde\x} \\ \log\vec{\tilde\y} \end{pmatrix}  \mapsto
  \begin{pmatrix} \log\vec{\tilde\x}' \\ \log\vec{\tilde\y}' -\frac{i\pi b^{-1} Q}{k}\vec t\, \end{pmatrix} \end{array}\,.
\ee
\ese
The transformations of the operators themselves result from exponentiating \eqref{glogop}. For example (leaving out the primes on the RHS),
\be \label{gop}
\begin{array}{rl} S_i:(\x_i,\y_i,\tilde\x_i,\tilde\y_i)&\mapsto (\y_i,\x_i^{-1},\tilde\y_i,\tilde\x_i^{-1})\,,\\[.2cm]
T(\mb B): (\x_i,\y_i,\tilde\x_i,\tilde \y_i)&\mapsto (\x_i,\, (-q^{-\frac12})^{\mb B_{ii}}(\x^{-\mb B})_i\y_i,\, \tilde\x_i,(-\tilde q^{-\frac12})^{\mb B_{ii}}(\tilde\x^{-\mb B})_i\y_i)\,, \\[.2cm]
U(\mb A):  (\x_i,\y_i,\tilde\x_i,\tilde \y_i)&\mapsto ((\x^{\mb A^{-1}})_i,\, (\y^{\mb A^T})_i,\,
(\tilde\x^{\mb A^{-1}})_i,\, (\tilde\y^{\mb A^T})_i)\,, \\[.2cm]
\sigma_\alpha(\tfrac Q2 e_i) : (\x_i,\y_i,\tilde\x_i,\tilde\y_i) &\mapsto (q^{-\frac12}\x_i,\,\y_i,\,\tilde q^{-\frac12}\tilde \x_i,\,\tilde\y_i)\,,\\[.2cm]
\sigma_\beta(\tfrac Q2 e_i) : (\x_i,\y_i,\tilde\x_i,\tilde\y_i) &\mapsto (\x_i,\,q^{-\frac12}\y_i,\,\tilde \x_i,\, \tilde q^{-\frac12}\tilde\y_i)\,,
\end{array}
\ee
where on the RHS we've omitted primes and used notation $(\x^{\mb A})_i:=\prod_j \x_j^{\mb A_{ij}}$ (etc.); and the shifts are by vectors $\vec t = e_i$ with entry $1$ in the $i$-th place and zero elsewhere. We have also introduced the square roots $q^{\frac12}:=\exp \frac{i\pi}{k}(1+b^2)$ and $\tilde q^{\frac12}:=\exp\frac{i\pi}{k}(1+b^{-2})$.%
\footnote{To compare this to the operator algebra and transformations in \cite{Dimofte-QRS,DGG,DGG-index}, one must change signs $(q^{\frac12},\tilde q^{\frac12})\to (-q^{\frac12},-\tilde q^{\frac12})$. The present sign convention is somewhat more convenient for general $k\geq 2$.}

\subsection{Products and symplectic quotients}
\label{sec:Hred}

To prepare for the definition of the state-integral models, we describe two more elementary operations on the spaces $\CH_\p^{(k)}$.

The first is an embedding
\be \label{pprod}
\begin{array}{ccc} \CH_\p^{(k)} \times \CH_{\p'}^{(k)} & \longrightarrow & 
  \CH_{\p\times \p'}^{(k)} \\[.2cm]
  f(\vec\mu;\vec m)\,,\; f'(\vec\mu';\vec m') & \mapsto & \tilde f(\vec\mu,\vec\mu';\vec m,\vec m') =  f(\vec\mu;\vec m) f'(\vec\mu';\vec m')
 \end{array} \ee
corresponding to a multiplication of wavefunctions. If $\p\subset \R^{2N}$ and $\p'\subset \R^{2N'}$ (so that $f$ and $f'$ depend on $N$ and $N'$ pairs of variables, respectively), then the decay of the product $ff'$ is governed by the product polytope $\p\times \p'\subset \R^{2(N+N')}$.

The second operation is symplectic reduction. It suffices to consider a simple, special case. Given a wavefunction $f\in \CH_\p^{(k)}$, let $\text{red}_if$ be obtained from $f$ by setting the $i$-th pair of variables to zero,
\be\begin{array}{l} (\text{red}_i f)(\mu_1,...,\mu_{N-1};m_1,...,m_{N-1}) \\[.2cm] \hspace{1in}:= f(\mu_1,...,\mu_{i-1},0,\mu_i,...,\mu_{N-1};m_1,...,m_{i-1},0,m_i,...,m_{N-1})\,. \end{array} \ee
Then $e^{-\frac{2\pi}{k}\vec\mu\cdot\vec\beta'}(\text{red}_if)(\vec\mu+i\vec\alpha',\vec m)$ decays exponentially as long as $(\alpha_1',...,\alpha_{i-1}',0,\alpha_i',...,\alpha_{N-1}';$ $\beta_1',...,\beta_{i-1}',\beta^*,\beta_i',...,\beta_{N-1}')\in \p$ for any $\beta^*$, \ie\ as long as
\be (\vec\alpha',\vec\beta') \in \text{red}_i\p := \text{proj}_{\beta_i}(\p|_{\alpha_i=0})\,, \ee
where $\text{red}_i\p \subset \R^{2N-2}$ is obtained by projecting $\p\subset \R^{2N}$ along the $\beta_i$ direction and intersecting with the hyperplane $\{\alpha_i=0\}$. Put differently, $\text{red}_i\p$ is the image of $\p$ under symplectic reduction in $\R^{2N}$, with respect to the moment map $\alpha_i$. Therefore,
\be \label{pred} \text{red}_i:\, \CH_\p^{(k)} \to \CH_{\text{red}_i\p}^{(k)}\,. \ee

\section{Tetrahedron}
\label{sec:tet}

The fundamental building block of state-integral models is the partition function of a tetrahedron, \emph{a.k.a.} the lens-space partition function of the tetrahedron theory $T_\Delta$. In this section we discuss several of its more important properties. In particular, we see that the partition function is naturally an element of $\CH_\p^{(k)}$, where $\p$ is the positive-angle polytope associated to the boundary of a tetrahedron.

\subsection{Analytic properties}
\label{sec:tet-anal}

The $L(k,1)_b$ partition function of the basic tetrahedron theory $T_\Delta$ can be defined as \cite{IMY-fact}
\be \label{Dk1}
\CZ_b^{(k,1)}[\Delta](\mu,m) = (qx^{-1};q)_\infty (\tilde q\tilde x^{-1};\tilde q)_\infty =\left\{\begin{array}{cl}
 \ds\prod_{j=0}^\infty \frac{1-q^{j+1}x^{-1}}{1-\tilde q^{-j}\tilde x^{-1}} & |q|<1 \\[.2cm]
 \ds\prod_{j=0}^\infty \frac{1-\tilde q^{j+1}\tilde x^{-1}}{1-q^{-j}x^{-1}} & |q|>1 \end{array}\right.\,,
 \ee
where $x,\tilde x,q,\tilde q$ are as in \eqref{defqx} or \eqref{defxmu}--\eqref{defqq}. The products on the RHS make sense for any $b$ such that $\Re(b)>0$ and $\Im(b)\neq 0$; the products then define a meromorphic function of $\mu\in \C$ for each $m\in \Z/k\Z$. The dependence on $b$ is analytic in each regime $\Im(b)>0$ and $\Im(b)<0$ (corresponding to $|q|<1$ and $|q|>1$), and it is actually possible to match smoothly across the half-line $b\in \R_{> 0}$. This is a well-known property for $k=1$, where $\CZ_b^{(1,1)}[\Delta]$ is the Barnes double-gamma function (rediscovered as Faddeev's noncompact quantum dilogarithm), and admits an alternative integral representation%
\footnote{The function in \eqref{ZFad} is related to a function $s_b$ that sometimes occurs in the literature as $\CZ^{(1,1)}_b[\Delta](\mu,0) = e^{\tfrac{i\pi}{2}\big(\tfrac{iQ}{2}-\mu\big)^2+\tfrac{i\pi}{24}(b^2+b^{-2})} s_b(\tfrac{iQ}{2}-\mu)$.}
\be \CZ^{(1,1)}_b[\Delta](\mu,0)  = \exp\bigg(\frac14 \int_{\R+i\epsilon}\frac{dz}{z} \frac{e^{2\pi i\mu z+\pi Q z}}{\sinh(\pi bz)\sinh(\pi b^{-1}z)} \bigg)  \label{ZFad} \ee
that converges in a neighborhood of the half-line $b\in \R_{> 0}$, and for $0<\Im(\mu)<\Re(b+b^{-1})$ (\cf\ \cite{DGLZ} and references therein). For general $k$, we use the relation
\be \label{prodk}
 \CZ_b^{(k,1)}[\Delta](\mu,m) = \prod_{\gamma,\delta\in\Gamma(k,1;m)} \CZ_b^{(1,1)}[\Delta]\big(\tfrac1k(\mu+ib\gamma+ib^{-1}\delta),0\big)\,,\ee
where the product is over integers $0\leq \gamma,\delta < k$ such that $\gamma-\delta\equiv m$ (mod $k$). Then many properties of the $k=1$ partition function, such as continuation across $b\in\R_{>0}$, automatically lift to $\CZ_b^{(k,1)}[\Delta][\mu,m]$.

The lattice product \eqref{prodk} actually matches the initial definitions of $L(k,1)_b$ partitions in the physics literature \cite{BNY-Lens}. Indeed, $L(k,1)_b  = S^3_b/\Z_k$ is an orbifold, and partition functions on $L(k,1)_b$ can be obtained from those on $L(1,1)_b=S^3_b$ by applying a suitable projection. This projection leads to the product of \eqref{prodk}.

The poles and zeroes of $\CZ_b^{(k,1)}[\Delta](\mu,m)$, as a function of $\mu$, lie on the torsor
\be \label{poles}
 \mu\in \big\{ ib\alpha + ib^{-1}\beta\,\big|\;\alpha,\beta\in\Z,\;\; \alpha-\beta=-m \;(\text{mod $k$})\big\} \qquad\text{with}\quad \begin{array}{r@{\quad}l} \text{zeroes:} & \alpha,\beta \geq 1 \\[,2cm]
\text{poles:} & \alpha,\beta \leq 0 \end{array}\,.
\ee
This follows by direct computation from \eqref{Dk1} or by using \eqref{prodk}.

The asymptotic behavior of the partition function as $|\Re(\mu)|\to \infty$ (with fixed $\Im(\mu)$) is 
\be  \label{mu-ass}
\CZ^{(k,1)}_b[\Delta](\mu,m) = \begin{cases} O(1)  & \Re(\mu) \to +\infty \\
\exp\big[ \frac{i\pi}{k}(\mu-\frac i2Q)^2 +O(1) \big]&  \Re(\mu)\to-\infty\,, \end{cases}
\ee
with $Q=b+b^{-1}$ as usual.
This follows from \eqref{prodk} and similar asymptotic behavior of the $k=1$ partition function. Physically, the two limits in \eqref{mu-ass} descend from the fact that when the real mass of the chiral in the theory $T_\Delta$ is taken to $+\infty$ or $-\infty$, the theory is effectively described by a single Chern-Simons term, at level zero or $-1$, respectively.

We also have an exact identity
\be \CZ_b^{(k,1)}[\Delta](\tfrac{iQ}{2}+\mu,m)\;  \CZ_b^{(k,1)}[\Delta](\tfrac{iQ}{2}-\mu,-m) =  (-1)^me^{\tfrac{i\pi}{k}(\mu^2-m^2)+\tfrac{2\pi i}{24 k}(b^2+b^{-2})-\tfrac{2\pi i}{12}(k-k^{-1})}\,, \label{ZZCS} \ee
which follows because the product on the LHS is a ratio of Jacobi theta functions whose arguments related by the modular transformation $T\varphi T$, with $\varphi=\left(\begin{smallmatrix}1 & 0 \\ -k & 1\end{smallmatrix}\right)$. Alternatively, it may be deduced from the standard inversion identity for $\CZ_b^{(1,1)}[\Delta]$. It expresses the fact that two chirals with opposite flavor charges (and appropriate R-charges) can be coupled by a superpotential and integrated out, leaving behind a pure Chern-Simons term at level $+1$. (Remember that such supersymmetric Chern-Simons terms also appeared in the T-type transformations of \eqref{repf}.)

Finally, it is easy to see from \eqref{Dk1} that the tetrahedron partition function at any $k$ is annihilated by difference operators
\be \label{diffD}
  (\y+\x^{-1}-1)\,\CZ_b^{(k,1)}[\Delta](\mu,m) =  (\tilde\y+\tilde\x^{-1}-1)\,\CZ_b^{(k,1)}[\Delta](\mu,m)=0\,, \ee
acting as in \eqref{op-action}.
These are Ward identities for the theory $T_\Delta$ on $L(k,1)_b$. Alternatively, from the point of view of complex Chern-Simons theory, they are quantizations of the classical Lagrangian
\be \{\CL(x,y) = y+x^{-1}-1=0\} \,\subset \CP\,, \ee
with $\CP =\C^*\times\C^*$ the model phase space of Section \ref{sec:CS}. As noted in Section \ref{sec:CS}, $\CP$ coincides with the phase space of $SL(2,\C)$ Chern-Simons theory on an ideal tetrahedron. The condition $y+x^{-1}-1=0$ characterizes the framed flat connections on the boundary of the tetrahedron that extend to the interior \cite{Dimofte-QRS, DGG-Kdec}. The difference relations \eqref{diffD} confirm that our identification of $\CZ^{(k,1)}_b[\Delta]$ with the Chern-Simons wavefunction of a tetrahedron is reasonable.

\subsection{Angle polytope}
\label{sec:tet-p}

Using the analytic properties above, we can identify the functional space $\CH_{\p[\Delta]}^{(k)}$, labelled by $\p[\Delta]\subset \R^2$, that the tetrahedron partition function belongs to. Recall the definition \eqref{Hp1} of $\CH_\p$. We take $(\alpha,\beta)$ as Darboux coordinates on $\R^2$.

The characterization of poles in \eqref{poles} shows that, for all $m\in \Z/k\Z$, $\CZ^{(k,1)}_b[\Delta](\mu,m)$ is holomorphic in the (semi-infinite) strip $\Im(\mu)>0$. Thus the first inequality cutting out $\p[\Delta]$ is $\alpha>0$. We then use \eqref{mu-ass} to calculate that when $\mu$ is real
\be \Big|e^{-\frac{2\pi}{k}\beta\mu}\CZ^{(k,1)}_b[\Delta](\mu+i\alpha,m)\Big| \sim \begin{cases}
\exp{-\frac{2\pi}{k}\beta\mu} & \mu\to \infty \\
\exp{-\frac{2\pi}{k}\mu(\alpha+\beta-\mathfrak{Q}/2)} & \mu\to-\infty \end{cases}\,,
\ee
with 
\be \boxed{ \mathfrak Q := \Re(Q) = \Re(b+b^{-1})}\,.\ee
Therefore, $\CZ_b^{(k,1)}[\Delta] \in \CH_{\p[\Delta]}^{(k)}$ with
\be \p[\Delta] = \{\, \alpha>0,\, \beta> 0,\, \alpha+\beta < \mathfrak Q/2\, \} \subset \R^2\,. \label{pD} \ee

\begin{figure}[htb]
\centering
\includegraphics[width=4in]{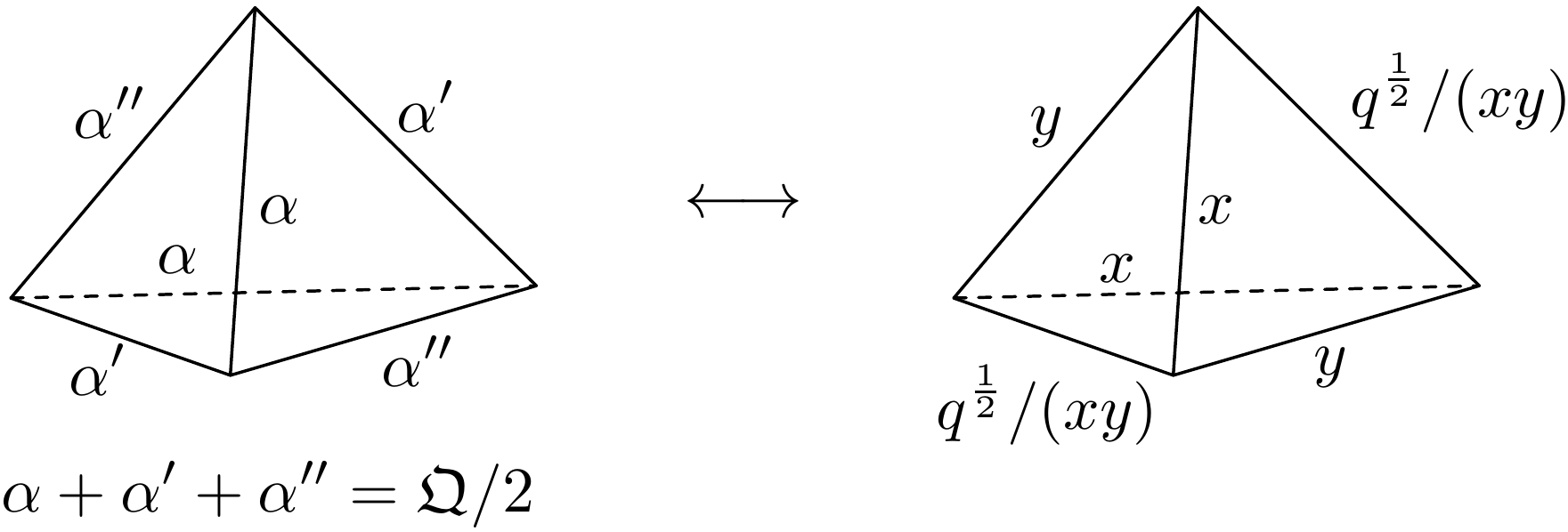}
\caption{Rescaled angle structure on an ideal tetrahedron: $\alpha,\alpha',\alpha''>0$, subject to $\alpha+\alpha'+\alpha''=\mathfrak Q/2$. The angles $\alpha,\alpha''$ are identified \eqref{anglemu} as the imaginary parts of $\mu,\nu$, where $x=\exp \frac{2\pi i}{k}(-ib\mu-m)$ and $y=\exp\frac{2\pi i}k(-ib\nu-n)$ parametrize the tetrahedron phase space.}
\label{fig:tet-ang}
\end{figure}

The polytope $\p[\Delta]$ could be written more symmetrically as $\{\alpha,\alpha',\alpha''>0\}\cap\{\alpha+\alpha'+\alpha''=\mathfrak Q/2\}$, with (say) $\beta=\alpha''$. We see that $\p[\Delta]$ is equivalent to the set of positive ``angles'' labeling the edges of an ideal tetrahedron, such that opposite edges have equal angles, and the sum of angles around any vertex is equal to $\mathfrak Q/2$ (Figure \ref{fig:tet-ang}). In the context of hyperbolic geometry, where $\Delta$ is an ideal \emph{hyperbolic} tetrahedron,%
\footnote{Note that in general we build manifolds using topological ideal tetrahedra (supporting flat connections), not hyperbolic ideal tetrahedra.} %
this is usually called the set of positive angle structures. (More precisely, it is the rescaled quantities $(2\pi b\alpha,2\pi b\alpha',2\pi b\alpha'')$ that correspond to classical angles; for real $b$ their sum around any vertex is $2\pi b\mathfrak Q/2 = \pi(1+b^2)$, which is a $b$-deformation of the classical value $\pi$.) There is a natural symplectic structure on the set of angles on a tetrahedron, first described by Neumann and Zagier \cite{NZ}, which coincides with the symplectic structure on the ambient space $\R^2$ containing $\p[\Delta]$. It agrees with the symplectic structure on the actual phase space $\CP=\C^*\times\C^*$ associated to the boundary of the tetrahedron, under the identification \eqref{anglemu}.

\subsection{Rotation by a third}

The coordinates $(x,y)$ of the classical phase space $\CP$ are themselves associated to edges of the tetrahedron, as in Figure \ref{fig:tet-ang}. 
(Classically, $x$ and $y$ are formed by computing certain cross-ratios of the framing data of a flat $SL(2,\C)$ connection.)
The precise association is a \emph{choice}. There are two other choices, related by a cyclic permutation $\rho$, which preserves the orientation of the tetrahedron and classically sends
\be (x,y) \overset{\rho}\mapsto (y,\,q^{\frac12}/(xy)) \overset{\rho}\mapsto (q^{\frac12}/(xy),\,x) \overset{\rho}\mapsto (x,y)\,. \ee
This $\rho$ is just the $ISp(2,\Z)$ transformation $\rho = \sigma_\alpha(1)S_1T(1)$, where $S_1=\left(\begin{smallmatrix} 0 & -1 \\ 1 & 0 \end{smallmatrix}\right)$ and $T(1)=\left(\begin{smallmatrix} 1 & 0 \\ 1 & 1 \end{smallmatrix}\right)$.

Quantum mechanically, $\rho$ acts on wavefunctions and on the operator algebra as described in Section \ref{sec:p}. In order to preserve the cyclic symmetry of the tetrahedron, its action must \emph{preserve} the tetrahedron wavefunction (\cf\ \cite[Sec. 6]{Dimofte-QRS}). As a preliminary check, note that $\rho$ does preserve the angle polytope $\rho\cdot \p[\Delta] = \p[\Delta]$ (as it should, since $\rho$ is just a rotation of the tetrahedron), so
\be \rho:\, \CH_{\p[\delta]}^{(k)} \to \CH_{\p[\delta]}^{(k)}\,.\ee
We then compute:
\begin{align} 
\big(\rho\cdot \CZ^{(k,1)}_b[\Delta]\big)(\mu,m) &= \frac{1}{k}\sum_{n\in\Z/k\Z} \int d\nu\, (-1)^n e^{\frac{i\pi}{k}[-\nu^2-2(\mu-iQ/2)\nu +n^2+2mn]}\,\CZ^{(k,1)}_b[\Delta](\nu,n) \notag \\
&= C_3^{-1}\, \CZ^{(k,1)}_b[\Delta](\mu,m)\,, \label{ST}
\end{align}
where
\be C_3 := \exp\Big[ \tfrac{2\pi i}{24k}Q^2+\tfrac{2\pi ik}{24}\Big]\,. \label{C3} \ee
The integration contour on the first line of \eqref{ST} is dictated by the rules of Section \ref{sec:p} to be along $\nu\in \R+i\alpha$ for any $0\leq \alpha \leq \mathfrak Q/2$, and converges so long as $\alpha<\mathfrak Q/2-\Im(\mu)$. The crucial equality in the second line follows by computing residues, as in Appendix \ref{app:23}. Thus, the tetrahedron partition function \emph{is} preserved, up to a prefactor $C_3$.

We also observe that $\rho$ preserves the quantum operators entering the equations \eqref{diffD}, up to a left-hand factor, \eg,
\be \rho\cdot(\y+\x^{-1}-1) = -q\,\x^{-1}\y^{-1}  + \y^{-1}-1 = -\y^{-1}(\y+\x^{-1}-1)\,.\ee
(The quantum version of the transformation $\rho(y)=q^{\frac12}/(xy)$ is $\rho(\y) = q^{\frac12}(-q^{\frac12})\x^{-1}\y^{-1} = -q\,\x^{-1}\y^{-1}$, as per \eqref{gop}.) The preservation of the difference equation, together with the intertwining property \eqref{ginter}, provides another route to proving the identity \eqref{ST}.

Physically, the identity \eqref{ST} reflects the 3d duality (mirror symmetry) between the theory $T_\Delta$ of a free chiral and the theory $\rho\cdot T_\Delta$ of a $U(1)$ gauge field coupled to a chiral, \emph{a.k.a.} a free vortex \cite{DGG}. Since the duality holds for 3d theories, it must hold for any $L(k,p)_b$ partition function, and we've just verified that it does.

\section{The state-integral model}
\label{sec:SS}

We have described various ingredients that we need to build a state-integral model for complex Chern-Simons theory at general level $k$: ``Hilbert'' spaces and operator algebras, a representation of the affine symplectic group on them, the partition function $\CZ[\Delta]$ of a tetrahedron, and the difference operators that annihilate $\CZ[\Delta]$. By the 3d-3d correspondence, each of these ingredients descends from 3d $\CN=2$ theories on the lens space $L(k,1)_b$, or from operators and transformations in such theories. The DGG prescriptions \cite{DGG, DGG-Kdec} for constructing a theory $T_n[M]$ then translate to a state-integral model for complex Chern-Simons theory, which we now describe.

Alternatively, we observe that the various ``ingredients'' in preceding sections are formally identical to those used to construct a $k=1$ state-integral model in \cite{Dimofte-QRS}. Thus, we may simply repeat the rules of \cite{Dimofte-QRS} to build a level-$k$ state-integral model. Our new analysis of angle polytopes ensures that integration contours, formerly ambiguous, are now well defined.

We define the state integral in Section \ref{sec:SS-wf}, given the gluing data and positivity of ideal triangulations that is reviewed in Sections \ref{sec:SS-geom}--\ref{sec:SS-angle}. Various properties of the state-integral model are then discussed, including invariance under 2--3 moves, (conjectured) full independence of triangulation, difference equations, and analytic continuation to a meromorphic function on the full complex space of $\vec\mu$ variables.

\subsection{Topological data}
\label{sec:SS-geom}

Let $M$ be an oriented 3-manifold with nonempty \emph{framed} boundary, in the sense of \cite{DGG-Kdec, DGV-hybrid}. Namely, the boundary is split $\pd M = (\pd M)_{\rm big}\cup_{S^1_a} (\pd M)_{\rm small}$ into \emph{big} pieces (having any genus, at least one hole, and strictly negative Euler characteristic); and \emph{small} pieces (which are discs, annuli, or tori); with the two types of pieces glued together along holes $\{S^1_a\}$. Let $\mb t$ be a topological ideal triangulation of the framed 3-manifold $M$ with $\hat N$ tetrahedra $\{\Delta_i\}_{i=1}^{\hat N}$. This is equivalent to a tiling of $M$ by truncated tetrahedra, such that the truncated vertices of tetrahedra tile $(\pd M)_{\rm small}$ and the big hexagonal faces of truncated tetrahedra tile $(\pd M)_{\rm big}$. The orientation of $M$ induces an orientation on each tetrahedron $\Delta_i$.

For example, one could take $M$ to be a hyperbolic knot complement with an ideal hyperbolic triangulation, whose cusp (at the knot) has been regularized by a horocycle (thereby truncating all tetrahedra). The boundary is a small torus, $\pd M = (\pd M)_{\rm small} \simeq T^2$. On the other extreme, a single truncated tetrahedron is a framed 3-manifold $M$, whose big boundary is a 4-holed sphere and whose small boundary consists of four discs.

The basic piece of data associated to such a triangulation is a matrix of gluing equations for a framed flat $PGL(n,\C)$ structure. For $n=2$ these coincide with Thurston's gluing equations for ideal hyperbolic triangulations \cite{thurston-1980}, and for $n>2$ they were described in \cite{BFG-sl3, GGZ-sln, DGG-Kdec}. The classical gluing equations take the form
\be \label{NZ0}
\begin{pmatrix} \vec u \\ \vec C-2\pi i \\\hline \vec v \end{pmatrix} = \left(\begin{array}{c@{\;}|@{\;}c} \mb A & \mb B \\\hline \hat{ \mb C} & \hat{\mb D} \end{array}\right) \begin{pmatrix} \vec Z \\ \vec Z'' \end{pmatrix} + i\pi\begin{pmatrix} \vec t_u \\\hline \vec t_v \end{pmatrix} 
\ee
where $(\vec Z,\vec Z'')$ are $N:={n+1\choose 3}\cdot \hat N$ pairs of logarithmic shape parameters associated to edges%
\footnote{For $n>2$ one must replace ``edges of tetrahedra'' with ``vertices of octahedra'' in an $n$-decomposition of the tetrahedra.} %
of the tetrahedra; $(\vec u,\vec v)$ are parameters associated to the boundary of $M$, such as logarithmic holonomies around meridian and longitude cycles of a knot complement; $\vec C$ are parameters associated to internal edges of the triangulation; and the affine transformation \eqref{NZ0} relates the former to the latter. 

\begin{wrapfigure}{r}{2in}
\centering
\includegraphics[width=1.5in]{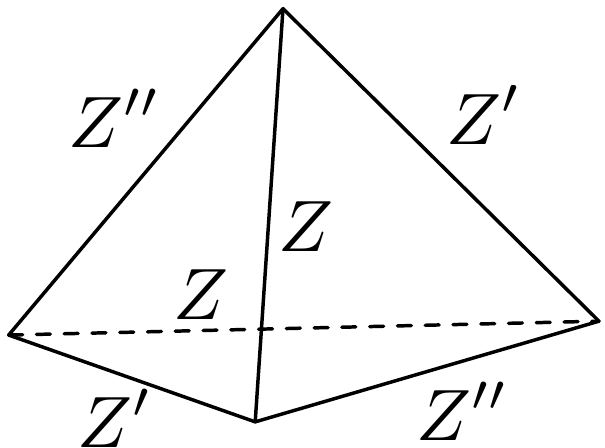}
\caption{Assignment of classical edge parameters to an ideal tetrahedron}
\label{fig:tet-Z}
\end{wrapfigure}

The construction of the gluing equations \eqref{NZ0} given a triangulation $\mb t$ of a general framed 3-manifold (at any $n\geq 2$) is discussed in \cite{DGG-Kdec, DGV-hybrid}, to which we refer the reader for  details. In the case $n=2$, the basic idea is to first label the edges each tetrahedron $\Delta_i$ with parameters $Z_i,Z_i',Z_i''$, equal on opposite edges, running counterclockwise around each vertex, and summing to $i\pi$, as in Figure \ref{fig:tet-Z}. Then the parameter $C_j$ associated to an internal edge $I_j$ of $\mb t$ simply equals the sum of edge-parameters $(Z_i,Z_i',Z_i'')$ on edges of tetrahedra identified with $I_j$. The boundary parameters $\vec u,\vec v$ fall into two classes. For every external edge $E_k$ of the triangulation (lying on the big boundary) there is a boundary parameter equal to the sum of edge-parameters on edges identified with it. For each torus or annulus component $a$ of the small boundary, there is a pair $(u_a,v_a)$ parameterizing logarithms of the eigenvalues of a flat $PGL(2,\C)$ connection around A and B cycles; then $u_a$, $v_a$ are equal (roughly) to sums and differences of the $(Z_i,Z_i',Z_i'')$ along paths representing the A and B cycles.
After using $Z_i+Z_i'+Z_i''=i\pi$ to eliminate all the $Z_i'$,
the relations between boundary variables and edge-parameters can be written in the form \eqref{NZ0},
where $\vec t_u,\vec t_v$ and the matrix blocks $\mb A,\mb B,\hat{\mb C},\hat{\mb D}$ contain \emph{integers}.

Suppose that the number of linearly independent pairs of boundary variables appearing in \eqref{NZ0} is $d$.
A fundamental result of Neumann and Zagier \cite{NZ} was that, in the case of hyperbolic knot complements and $n=2$, the matrix $\left(\begin{smallmatrix} \mb A & \mb B \\  \hat{\mb C} & \hat{\mb D} \end{smallmatrix}\right)$ in \eqref{NZ0} has rank $N+d$ and can be completed (after deleting some linearly dependent rows, and adding $N-d$ new rows at the bottom) to a full \emph{symplectic} matrix of rank $2N$, of the form
\be \begin{pmatrix} \mb A & \mb B \\ \mb C & \mb D \end{pmatrix} \in Sp(2N,\Q)\,. \label{NZ1} \ee
The result has been extended to all manifolds with small torus boundaries \cite{neumann-combinatorics}, to general framed 3-manifolds \cite{DV-NZ}, and to $n>2$ \cite{Guilloux-PGL, GZ-gluing}. For knot complements in homology spheres and $n=2$, it follows from \cite[Thm 4.2]{neumann-combinatorics} that the completion \eqref{NZ1} can be arranged to take values in $Sp(2N,\Z)$. \emph{We will assume below that this is always the case.} For general rational-valued matrices, some extra choices would need to be made in order to define a state sum --- essentially because the Weil-like action of Section \ref{sec:p} is ill defined for $Sp(2N,\Q)$.

We recall that the symplectic property of the gluing equations (the ability to complete \eqref{NZ0} to a symplectic matrix \eqref{NZ1}) has a very natural, geometric meaning. The moduli space of flat connections on $\pd M$ is parameterized by the $\vec u,\vec v$, and the (holomorphic, non-degenerate) Atiyah-Bott Poisson structure is linear in the $\vec u,\vec v$ variables. It can be extended trivially to a Poisson structure on the space parameterized by $\vec u,\vec v, \vec C$ by setting all the $C_j$'s to be central elements. Then the symplectic property of \eqref{NZ0} is equivalent to the fact that this extended Poisson structure on the boundary of $M$ is compatible with the product Poisson structure on boundaries of the individual tetrahedra, under which $\{Z_i'',Z_j\} = \delta_{ij}$ (and so on cyclically). Yet another way to say this is that the moduli space of flat connections on $\pd M$, locally of the form $\CP^d$, is a symplectic reduction of the moduli space of flat connections on individual tetrahedra $\sqcup_i\pd\Delta_i$, locally of the form $\CP^N$, where the internal-edge functions $\vec C$ are used as moment maps.

The symplectic completion \eqref{NZ1}, together with a trivial completion of the shift vector $(\vec t_u; \vec t_v)\to (\vec t_u;\vec t_v,0,...,0) \in \Z^{2N}$ (just filling in zeroes) defines an affine symplectic transformation
\be \label{defg}
 g = g(M,\mb t) := \sigma_\alpha(\vec t_u)\sigma_\beta(\vec t_v,0,...,0) \circ  \begin{pmatrix} \mb A & \mb B \\ \mb C & \mb D \end{pmatrix} \in ISp(2N,\Z)\,. \ee
This is the transformation we use to construct the state integral.

\subsection{Angle structures}
\label{sec:SS-angle}

Closely related to the gluing equations is an angle structure. Recall from Section \ref{sec:tet-p} that the positive-angle polytope for a tetrahedron $\Delta$ is given by $\p[\Delta]=\{\alpha>0,\alpha'>0,\beta:=\alpha''>0,\alpha+\alpha'+\alpha''=\mathfrak Q/2\} \subset \R^2$, where $\alpha,\alpha',\alpha''$ are angles labeling its edges. 

Assume for the moment that $n=2$.
If $M$ is a framed 3-manifold triangulated by $N$ tetrahedra $\Delta_i$, then we associate angles $\alpha_i,\alpha_i',\alpha_i''=:\beta_i$ to the edges of each $\Delta_i$, equal on opposite edges and summing to $\mathfrak Q/2$. We place these angles long the same edges that were labelled $(Z_i,Z_i',Z_i'')$ in Section \ref{sec:SS-geom}. Indeed, after a rescaling, we may identify $\alpha_i\sim\Im(Z_i),\beta_i\sim\Im(Z_i'')$.
The angle polytope for the disjoint collection of tetrahedra is just
\begin{align} \p[\sqcup_i\Delta_i] &= \{\alpha_i>0,\alpha_i'>0,\alpha_i''>0\quad\forall\;i\} = \{\alpha_i>0,\beta_i>0,\alpha_i+\beta_i< \mathfrak Q/2\quad\forall\; i\}  \notag \\
 &= \prod_i \p[\Delta_i] \;\subset \R^{2N}\,.\end{align}
The space of positive angle structures for the whole triangulation $\mb t$ is, by definition, the intersection of $\p[\sqcup_i\Delta_i]$ with the hyperplanes enforcing the condition that the sum of angles around every internal edge $I_j$ in the triangulation is equal to $\mathfrak Q$.
Thus,
\be \label{posang} \text{positive angles for $\mb t$}:\qquad \p[\sqcup_i\Delta_i]\cap \{\text{sum around $j$-th edge}=\mathfrak Q\;\;\forall\,j\} \,.\ee
If the space \eqref{posang} is nonempty, we say that $\mb t$ admits a positive angle structure.

Notice that by applying the affine symplectic transformation $g$ from \eqref{defg} to $\p[\sqcup_i\Delta_i]$ we obtain an isomorphic polytope in a copy of $\R^{2N}$, whose coordinates are naturally identified with (rescaled) imaginary parts $(\vec\alpha';\vec\beta') \sim (\Im(\vec u),\Im(\vec C)-\mathfrak Q;\, \Im(\vec v),*)$. The hyperplanes of \eqref{posang} are just coordinate hyperplanes $\{\alpha_i'=0\}_{i=d+1}^N$. By intersecting with these hyperplanes and then projecting along the dual coordinates $\beta_i$, $i=d+1,...,N$, we obtain the symplectically reduced polytope
\be \p[M,\mb t] := \text{red}_{d+1,...,N} \big( g\cdot \p[\sqcup_i\Delta_i] \big)\,. \label{posangr} \ee
The triangulation admits a positive angle structure if and only if $\p[M,\mb t]$ is nonempty.

For $n>2$, the same definitions apply, replacing ``edges of tetrahedra'' with ``vertices of octahedra,''
namely the octahedra in an $n$-decomposition.  It is the vertices that are labelled by ``angles.'' The internal-edge constraints become internal-vertex constraints, but still have the same algebraic structure. Symplectic reduction leads to a polytope $\p[M,\mb t]_n$, and we continue to say that $\mb t$ admits a positive angle structure (for given $n$) if $\p[M,\mb t]_n$ is nonempty. A triangulation $\mb t$ admits a positive angle structure for all $n\geq 2$ if it admits one for $n=2$, since in the $n$-decomposition of each tetrahedron all octahedra can be given the equal (positive) angles, as in \cite[Section 7.2.1]{DGG-Kdec}.

\subsection{The wavefunction}
\label{sec:SS-wf}

Suppose we are given a framed 3-manifold $M$ with a triangulation $\mb t$ that admits a positive angle structure, for some $n\geq 2$. Let $g=g(M,\mb t)$ be the symplectic completion of the gluing transformation \eqref{defg}. Then the level-$k$ state-integral partition function of $(M,\mb t)$ is
\be \label{defZ}  \boxed{\CZ^{(k,1)}_b[M,\mb t]_n := \text{red}_{d+1,...,N}\Big[ g\cdot \prod_{i=1}^N \CZ^{(k,1)}_b[\Delta_i] \Big] \;\subset \;\CH^{(k)}_{\p[M,\mb t]_n}}\,. \ee

Let us unravel this definition a bit. We start with partition functions $\CZ_b^{(k,1)}(\zeta_i,s_i)$ for each tetrahedron (with $s_i\in \Z/k\Z$), and take their product as in \eqref{pprod} to obtain a product wavefunction in $\CH^{(k)}_{\p[\sqcup_i\Delta_i]}$. We apply to this the affine symplectic transformation $g$, in the Weil-like representation of Section \ref{sec:p-Weil}, producing a wavefunction $g\cdot\prod_i \CZ_b^{(k,1)}[\Delta_i]\subset\CH^{(k)}_{g\cdot\p[\sqcup_i\Delta_i]}$. This transformed wavefunction depends on $N$ new pairs of coordinates $(\vec\mu,\vec\gamma;\,\vec m,\vec c)\in \C^N\times (\Z/k\Z)^N$, corresponding to the classical combinations of shapes $(\vec u,\vec C-2\pi i)$.
Then we apply symplectic reduction \eqref{pred}, which sets all the coordinates for internal edges to zero, $\gamma_j = c_j = 0$, thereby enforcing the gluing. This last operation is well defined precisely due to the existence of a positive angle structure. Indeed, since the intersection \eqref{posang} is nonempty (or, equivalently, \eqref{posangr} is nonempty), the function $g\cdot\prod_i \CZ_b^{(k,1)}[\Delta_i]$ is holomorphic in a neighborhood of $\{\gamma_j= 0$ $\forall\,j\}$ (for any $\vec c$). Typically, $g\cdot\prod_i \CZ_b^{(k,1)}[\Delta_i]$ is given by some integral that converges absolutely when $\gamma_j= 0$. The result of this integral is guaranteed to be an element of $\CH^{(k)}_{\p[M,\mb t]_n}$.

Given a fixed triangulation $\mb t$, it is easy to check that the state integral \eqref{defZ} is independent of various choices that enter the definition of the gluing transformation \eqref{defg}. These choices are delineated explicitly in (\eg) \cite[Sec. 5]{Dimofte-QRS} or \cite[Sec. 3]{DG-quantumNZ}. In particular:
\begin{itemize}
\item Due to invariance of the tetrahedron wavefunction under $\rho$ in \eqref{ST}, the state integral cannot depend on how tetrahedra are labelled with shapes $Z_i,Z_i',Z_i''$ or angles $\alpha_i,\alpha_i',\alpha_i''$, as long as the cyclic ordering of the labels agrees with the induced orientation.

\item The rows of the gluing transformation \eqref{defg} corresponding to boundary variables $(\vec u,\vec v)$ and to the symplectic completion are ambiguous --- some the internal-edge rows $(C_j-2\pi i)$ may be added to them. The ambiguity changes the state integral $g\cdot\prod_i \CZ_b^{(k,1)}[\Delta_i]$ by factors that drop out after setting to zero the corresponding internal-edge variables $(\gamma_j,c_j)$.

\item The rows corresponding to internal edges may be also be modified by $GL(N-d,\Z)$ transformations. (For example, different subsets of the internal edges may be chosen to provide independent gluing constraints.) This just re-parametrizes the pairs $(\gamma_j,c_j)$, and the choice again drops out in the reduction step.

\end{itemize}
There are a few multiplicative (projective) ambiguities that enter the state integral. Since all the affine shifts in the transformation \eqref{posangr} are by integer multiples of $iQ/2$, the Weil-like representation comes with the inherent ambiguity \eqref{amb1}. Moreover, the tetrahedron wavefunction is only invariant under $\rho$ up to the factor $C_3$ from \eqref{C3}. Altogether, for a fixed triangulation, the state integral \eqref{defZ} is well defined up to a constant prefactor of the form
\be  (-1)^{a_1} e^{a_2\frac{i\pi}{k}(b+b^{-1})^2} C_3^{a_3}\,,\qquad a_1,a_2,a_3\in \Z\,. \label{amb} \ee

The state-integral wavefunction \eqref{defZ} \emph{does} depend on which linearly independent boundary parameters are chosen to parametrize the space of framed flat connections on $\pd M$, and which appear as $u$'s vs $v$'s. For example, for a knot complement, this is a choice of basis for A and B cycles on the torus boundary. Different choices are related by $ISp(2d,\Z)$ transformations, which act in the usual Weil-like representation.

\subsection{Invariance under 2-3 moves}
\label{sec:23}

\begin{figure}[htb]
\centering
\includegraphics[width=5.5in]{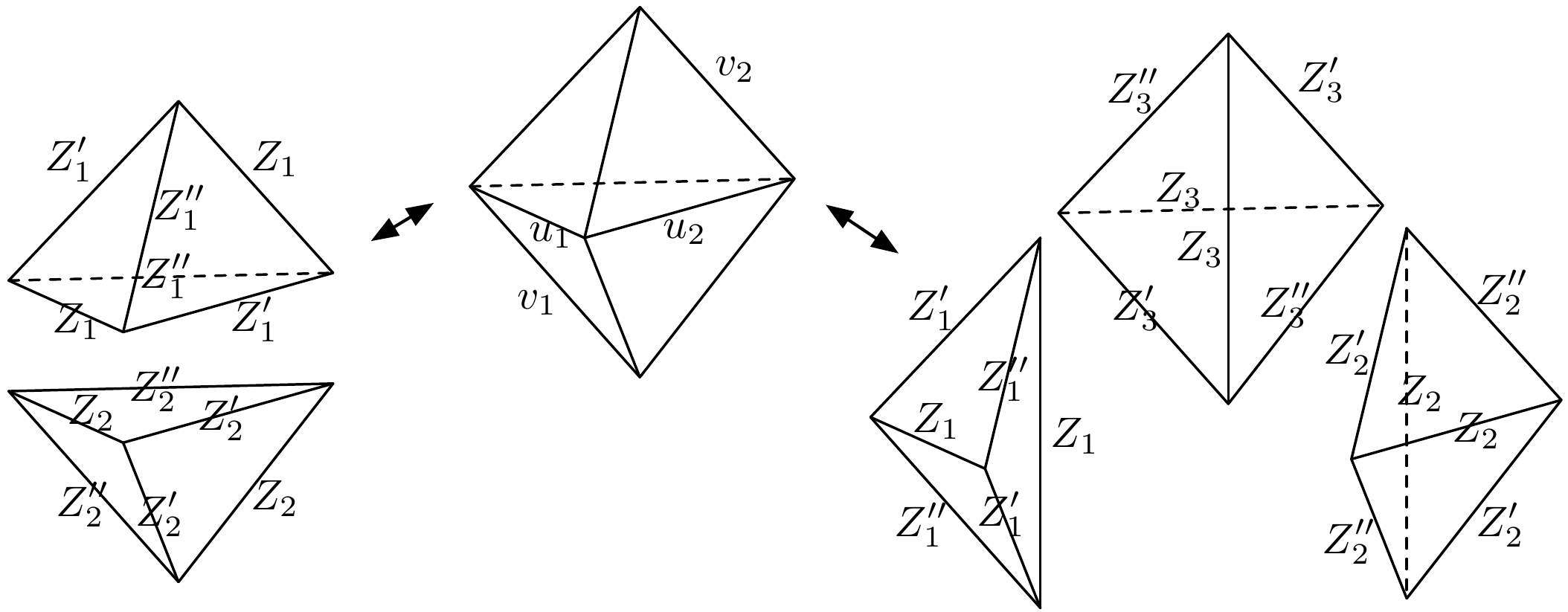}
\caption{Labeling of external edges of the bipyramid, and labeling of tetrahedra that compose it.}
\label{fig:23}
\end{figure}

We conjecture that \eqref{defZ} is also independent of the choice of triangulation $\mb t$, so long as the triangulation admits a positive angle structure. We can prove a local version of this statement. Namely, if $\mb t_2$ and $\mb t_3$ are two triangulations of $M$ related by a 2--3 Pachner move (as in Figure~\ref{fig:23}), and \emph{both} $\mb t_2$ and $\mb t_3$ admit positive angle structures, then $\CZ^{(k,1)}_b[M,\mb t_2]_n = \CZ^{(k,1)}_b[M,\mb t_2]_n$, modulo prefactors of the form \eqref{amb}. This follows from an equality of state integrals for the bipyramid at the core of the 2--3 move (which in turn descends, via 3d-3d correspondence, from the fundamental mirror symmetry between 3d $\CN=2$ SQED and the XYZ model). Unfortunately, it is not known whether any two triangulations of a 3-manifold $M$ that admit a positive angle structure are related by a chain of 2--3 moves admitting positive angle structures at every step.%
\footnote{In the opposite direction, starting from an arbitrary triangulation of $M$, it is always possible to perform a finite number of 2--3 moves and obtain a new triangulation that does \emph{not} admit a positive angle structure, \eg\ a triangulation in which two adjacent faces of a single tetrahedron are identified, trapping an internal edge. This is the type of situation that must be avoided.} %
Thus, global triangulation independence is only conjectured for now.%
\footnote{If it were possible to relax angle structures to be non-negative rather than strictly positive, it would follow that \eqref{defZ} is an invariant of hyperbolic cusped manifolds \cite{GHRS-index}. Alternatively, methods of \cite{AK-new} for $k=1$ might be extended to $k>1$ to prove triangulation independence.}

We'll consider the bipyramid for $n=2$. (When $n>2$, the Pachner move decomposes into a sequence of elementary 2--3 moves on octahedra \cite{DGG-Kdec}, which are formally equivalent to the $n=2$ identity discussed here.)
In order to calculate the state-integral partition function of the bipyramid, we choose independent, canonically conjugate parameters $(u_1,u_2,v_1,v_2)$ on its edges, as shown. The partition function will depend on pairs $(\mu_1,m_1)$, $(\mu_2,m_2)$, with $m_1,m_2\in \Z/k\Z$. Decomposing the bipyramid into two tetrahedra produces, labelled as shown on the left, gives an $ISp(4,\Z)$ transformation
\be  \begin{pmatrix} u_1 \\ u_2\\\hline v_1\\ v_2 \end{pmatrix} = \left( \begin{array}{cc@{\;}|@{\;}cc}
 1 &1 & 0 & 0 \\ -1 & 0 & -1 & 1 \\\hline 0 & 0 & 0 & 1 \\ 1 & 0 & 0 & 0 \end{array}\right)
  \begin{pmatrix}  Z_1 \\   Z_2 \\\hline  Z_1'' \\  Z_2'' \end{pmatrix}
   + i\pi \begin{pmatrix} 0  \\ 1 \\\hline 0 \\ 0 \end{pmatrix}
   \;=: g_2 \begin{pmatrix}  Z_1 \\   Z_2 \\\hline  Z_1'' \\  Z_2'' \end{pmatrix} \,,
\ee
which can be factored (say) as $g_2 = \sigma_\alpha(0,1)\,S_2\,T\left(\begin{smallmatrix} 0 & 0 \\ 0 & 1\end{smallmatrix}\right)\,U\left(\begin{smallmatrix} 1 & 1 \\ 1 & 0\end{smallmatrix}\right)$, leading to the state integral
\begin{align} \label{bip2}
\CZ^{(k,1)}_b[\text{bip}_2](\vec\mu,\vec m) &= \frac{1}{k}\sum_{n=0}^{k-1}\int d\nu\,(-1)^ne^{\frac{i\pi}{k}\big[-(2\mu_2-iQ)\nu-\nu^2+2m_2n+n^2\big]} \\
 &\hspace{1in} \times \CZ^{(k,1)}_b[\Delta](\nu,n)\;\CZ^{(k,1)}_b[\Delta](\mu_1-\nu,m_1-n)\,.
\notag \end{align}
The integral is to be done on along line with $0<\Im(\nu)<\Im(\mu_1)$, and converges absolutely so long as $\Im(\mu_1)+\Im(\mu_2)<\mathfrak Q/2$ and $\Im(\mu_2)>-\mathfrak Q/2$.
On the other hand, decomposing into three tetrahedra gives an $ISp(6,\Z)$ transformation
\be \small
\begin{pmatrix} u_1 \\ u_2 \\ C - 2\pi i \\\hline v_1\\ v_2 \\  * \end{pmatrix} = \left( \begin{array}{ccc@{\;}|@{\;}ccc}
 1 &0&0&0&0&0 \\ 0&1&0&0&0&0 \\ 1&1&1&0&0&0 \\\hline 0&0&-1&1&0&-1 \\ 0&0&-1&0&1&-1 \\ 0&0&1&0&0&1 \end{array}\right)
  \begin{pmatrix} \zeta_1 \\  \zeta_2  \\\zeta_3\\\hline \zeta_1'' \\ \zeta_2''\\\zeta_3'' \end{pmatrix}
   + i\pi \begin{pmatrix} 0 \\0  \\ -2 \\\hline 1 \\ 1 \\ 0 \end{pmatrix}
    \;=: g_3 \begin{pmatrix} \zeta_1 \\  \zeta_2  \\\zeta_3\\\hline \zeta_1'' \\ \zeta_2''\\\zeta_3'' \end{pmatrix} \,,
\ee
where $C=Z_1+Z_2+Z_3$ is now a parameter for the central edge. The state integral becomes
\begin{align} \CZ^{(k,1)}_b[\text{bip}_3](\vec\mu,\vec m) &= (-1)^{m_1+m_2}e^{\frac{i\pi}{k}\big[ -(i\frac Q2-\mu_1-\mu_2)^2+(m_1+m_2)^2\big]} \label{bip3} \\
 &\hspace{-.2in} \times \CZ^{(k,1)}_b[\Delta](\mu_1,m_1)\;\CZ^{(k,1)}_b[\Delta](\mu_2,m_2)\; \CZ^{(k,1)}_b[\Delta](iQ-\mu_1-\mu_2,-m_1-m_2)\,. \notag
\end{align}

The equality of \eqref{bip2} and \eqref{bip3} was already verified numerically by \cite{IMY-fact}. We prove in Appendix \ref{app:23} that for all $k$
\be \boxed{\CZ^{(k,1)}_b[\text{bip}_2](\vec\mu,\vec m)=  C_3\, e^{\frac{i\pi}{k}(\frac{iQ}{2})^2}\, \CZ^{(k,1)}_b[\text{bip}_3](\vec\mu,\vec m)}\,. \label{23} \ee

The equality \eqref{23}, however, has a hidden subtlety: as state-integral wavefunctions,  $\CZ^{(k,1)}_b[\text{bip}_2](\vec\mu,\vec m)$ and $\CZ^{(k,1)}_b[\text{bip}_3](\vec\mu,\vec m)$ take values in different functional spaces! The angle polytope $\p_2 := g_2\cdot (\p[\Delta]^2)$ (coming from gluing two tetrahedra) is the full positive-angle polytope that one would expect for the bipyramid --- meaning that all angles on exterior edges are positive, subject to the constraint that the sum around any $f$-valent vertex is $(f-2)\mathfrak Q/2$. The angle polytope $\p_3 := {\text{red}_3}\big[ g_3\cdot (\p[\Delta]^3)\big]$ coming from three tetrahedra, however, is \emph{strictly smaller},
\be \p_3 \subset \p_2\,,\qquad  \CH_{\p_2}^{(k)} \subset \CH_{\p_3}^{(k)}\,. \ee
This is easy to see by considering the angle $\alpha_1 \sim \Im(Z_1)$. In the bipyramid geometry, this angle can take any value $0<\alpha_1 < \mathfrak Q$, and all these values can actually be realized with a two-tetrahedron gluing (since $\alpha_1$ is a sum of two tetrahedron angles). 
But in the three-tetrahedron gluing, $\alpha_1$ comes from a single tetrahedron, so it is impossible realize values greater than $\mathfrak Q/2$.

In terms of the actual wavefunctions, the problem is that the product of three tetrahedron wavefunctions in \eqref{bip3} is technically only assured to be holomorphic and decay within the strip where $\Im(\mu_1),\, \Im(\mu_2)$, and $\mathfrak Q-\Im(\mu_1)-\Im(\mu_2)$ take values between $0$ and $\mathfrak Q/2$.  In reality, the strip can be extended to values between $0$ and $\mathfrak Q$, due to cancellations between the three factors. This must of course be the case, since once the equivalence \eqref{23} is known for wavefunctions in $\CH_{\p_3}^{(k)}$, analytic continuation and the fact that the left-hand side belongs to $\CH_{\p_2}^{(k)}$ implies that the right-hand side belongs to $\CH_{\p_2}^{(k)}$ as well.

Given an arbitrary framed 3-manifold $M$ with two triangulations $\mb t$, $\mb t'$ related by a 2--3 move, we can apply the identity \eqref{23} to equate the wavefunctions $\CZ^{(k,1)}_b[M,\mb t]_n$ and $\CZ^{(k,1)}_b[M,\mb t']_n$, so long as both $\mb t$ and $\mb t'$ admit positive angle structures. In this case, we can choose the positive structures to coincide (the positive structure for the triangulation $\mb t'$, on the `3' side of the move, uniquely determines a positive structure on $\mb t$). In this regime, all integrals defining $\CZ^{(k,1)}_b[M,\mb t]_n$ and $\CZ^{(k,1)}_b[M,\mb t']_n$ converge absolutely, and the identity \eqref{23} still holds despite being nested inside other integrals.

\subsection{Triangulation invariance and a maximal polytope $\p[M]$}
\label{sec:max}

We have conjectured that the state-integral partition functions $\CZ_b^{(k,1)}[M,\mb t]_n$ should not depend on the choice of (positive) triangulation $\mb t$. These partition functions, however, are all in slightly different functional spaces $\CH_{\p[M,\mb t]_n}^{(k)}$.
A better statement of the conjecture is that given a framed 3-manifold $M$ and integers $n\geq 2,\,k\geq 1$ there exists a \emph{maximal} polytope $\p[M]_n$ and a wavefunction $\CZ_b^{(k,1)}[M]_n \in \CH_{\p[M]_n}^{(k)}$, depending only on $M,\,n,\,k$, such that for any ideal triangulation $\mb t$,
\be  \p[M,\mb t]_n \subset \p[M]_n\,, \qquad\text{and}\qquad \CZ_b^{(k,1)}[M,\mb t]_n = \CZ_b^{(k,1)}[M]_n \quad\text{on $\CH_{\p[M,\mb t]_n}^{(k)} \supset \CH_{\p[M]_n}^{(k)}$}\,. \ee
In other words, any $\CZ_b^{(k,1)}[M,\mb t]_n$ can be analytically continued to $\CZ_b^{(k,1)}[M]_n$. The maximal polytope $\p[M]_n$ is the ``true'' angle polytope describing the decay properties of any $\CZ_b^{(k,1)}[M,\mb t]_n$ after analytic continuation.

We can guess the form of $\p[M]$ in some special cases.
If $M$ is an ideal polyhedron, then $\p[M]\subset \R^{2d}$ with $2d=2(\#\,\text{vertices})-6=(\#\,\text{edges})-(\#\,\text{vertices})$. The coordinates on $\R^{2d}$ are angles associated to the edges of the big-boundary triangulation, subject to the constraint that around an $f$-valent vertex the sum of angles equals $(f-2)\mathfrak Q/2$. (Alternatively, the sum of (angles $-$ $\mathfrak Q/2$) around any vertex equals $-\mathfrak Q$.) The symplectic structure on $\R^{2d}$ is the standard one from Teichm\"uller/cluster theory: the Poisson bracket of two edges equals the number of faces they share, counted with orientation. We expect that the polytope $\p[M]$ is simply cut out by the constraints that the angles on all big-boundary edges are positive. We already saw that this was true for a single tetrahedron, and for the bipyramid of Section~\ref{sec:23}.

\subsection{Meromorphic continuation via difference equations}

Given $M$, $n$, and a positive triangulation $\mb t$, the state integral $\CZ_b^{(k,1)}[M]$ is annihilated by two sets of mutually commuting polynomial difference operators $\{\CL_a(\vec\x,\vec\y;q^{\frac12})\}_a$ and $\{\CL_a(\vec{\tilde\x},\vec{\tilde\y};\tilde q^{\frac12})\}_a$, which are a classical quantization of (an open subset of) the classical space
\be \{\CL_a(x,y;q^{\frac12}=1) = 0\}  \approx \{\text{framed flat connections that extend from $\pd M$ to $M$}\}\,.\ee
(This space should be a $K_2$ Lagrangian subvariety of the space of framed flat connections on the boundary $\pd M$.%
\footnote{See \cite{DGG-Kdec, DV-NZ} for recent proofs of this in various contexts. Initial ideas about quantizing such a Lagrangian go back to \cite{gukov-2003, garoufalidis-2004}. The $K_2$ property appeared in \cite{Champ-hypA, Dunfield-Mahler, GS-quant}.} %
For example, when $M$ is a knot complement and $n=2$, there is a single pair of operators that quantizes the classical A-polynomial of $M$. As long as the operators $\CL_a$ are finite, nonzero polynomials, they can be used to extend the state integral $\CZ_b^{(k,1)}[M](\vec\mu,\vec m)$ to a \emph{meromorphic} function of $\vec\mu\in \C^d$ for all fixed $\vec m\in(\Z/k\Z)^d$.

Let us explain briefly how the operators $\CL_a$ are derived, following the formal arguments of \cite{Dimofte-QRS}.%
\footnote{This derivation also falls under the general theory of $q$-holonomic functions, \cf\ \cite{Zeil-AB}.} %
We have already seen that, for any $k\geq 1$, the tetrahedron partition function is annihilated by operators $\y+\x^{-1}-1$, $\tilde\y+\tilde\x^{-1}-1$. Thus, in constructing the state integral, the product partition function $\prod_i \CZ_b^{(k,1)}[\Delta_i]$ is annihilated by $\{\y_i+\x_i^{-1}-1\}_{i=1}^N$ and $\{\tilde\y_i+\tilde\x_i^{-1}-1\}_{i=1}^N$. In turn, the $ISp(2N,\Z)$-transformed partition function $g\cdot \prod_i \CZ_b^{(k,1)}[\Delta_i]$ is annihilated by the $g$-transformed version of these operators, using the action \eqref{gop}. Let  $\CI_0$ denote the left ideal generated by the $g$-transformed version of $\{\y_i+\x_i^{-1}-1\}_{i=1}^N$, in the $q$-commutative ring  $\Z(\x,\y,q^{\frac12})$.
Finally, in order to obtain operators that annihilate the symplectic reduction $\text{red}_{i=d+1,...,N}\big[g\cdot \prod_i \CZ_b^{(k,1)}[\Delta_i]\big]$, one must eliminate from $\CI_0$ all variables $\y_i$ and then set $\x_i$ for $i=d+1,...,N$; call the elimination ideal $\CI$. Then the operators $\CL_a(\vec\x,\vec\y;q^{\frac12})$ are the generators of $\CI$. Similarly, the $\CL_a(\tilde{\vec\x},\tilde{\vec\y};q^{\frac12})$ generate an ideal $\wt \CI$ that descends from $\{\tilde\y_i+\tilde\x_i^{-1}-1\}_{i=1}^N$.

The formal $g$-transformation and symplectic reduction steps here require a bit of care. The annihilation of $g\cdot \prod_i \CZ_b^{(k,1)}[\Delta_i]$ by $g$-transformed operators follows from manipulations under the integrals that define the $g$-action on wavefunctions. For operators in $\CI_0$ this involves deformations of the integration contour by finite multiples of $ib$. Fortunately, the convex, linear polytope $g\cdot \p[\sqcup_i\Delta_i]$, which determines how far integration contours can be shifted while maintaining a convergent integral, has volume $>C\, \mathfrak Q^d = C\,\Re(b+b^{-1})$ (for some positive $C$). As $b\to 0$, the polytope scales as $1/b$, and its volume diverges. Thus, it is possible to shift the contour as much as needed, so long as $b$ is sufficiently small; and $g$-transformed operators will annihilate $g\cdot \prod_i \CZ_b^{(k,1)}[\Delta_i]$.

Similarly, so long as $b$ is sufficiently small, the elimination ideal $\CI$ will annihilate the final state integral. Here there is an additional subtlety: working in the non-commutative ring $\Z(\x,\y,q^{\frac12})$, it is not totally obvious that the elimination ideal exists or is nonzero. We hope this will be clarified in future work.%
\footnote{Concrete examples of $\CI$ were given in (\eg) \cite{Dimofte-QRS, DGV-hybrid}. For knot complements and $n=2$, it was conjectured in \cite{Dimofte-QRS} that the generator of $\CI$ coincides with the inhomogeneous part of the recursion relation of \cite{gukov-2003, garoufalidis-2004} for the colored Jones.} %
 \emph{Assuming} that $\CI$ exists, and has a finite basis, it will annihilate $\CZ_b^{(k,1)}[M,\mb t]_n$.

One a set of (Laurent-polynomial) $q$-difference operators for $\CZ_b^{(k,1)}[M,\mb t]_n \in\CH_{\p[M,\mb t]_n}^{(k)}$ are obtained, they can be used to extend the state integral outside of the holomorphic strip \eqref{Hp1} specified by the polytope $\p[M,\mb t]_n$. Namely, $\CZ_b^{(k,1)}[M,\mb t]_n(\vec\mu,\vec m)$ gets extended to a meromorphic function on all of $\C^d$, for each fixed $\vec m$. It is not an entire function because the recursion relation coming from $\CL_a(\vec\x,\vec\y;q^{\frac12})$ may introduce poles. Using symmetry under $b\to b^{-1}$ and $m\to -m$, we find that this meromorphic function is also annihilated by the dual operator $\CL_a(\vec{\tilde\x},\vec{\tilde\y};\tilde q^{\frac12})$.

As a final remark, we emphasize (as claimed in the introduction) that the difference operators $\CL_a(\vec\x,\vec\y;q^{\frac12})$, $\CL_a(\vec{\tilde\x},\vec{\tilde\y};\tilde q^{\frac12})$ have no explicit dependence on $k$! The choice of $k\geq 1$ enters only the definition of $q,\tilde q$ and the action of $\x,\y,\tilde\x,\tilde\y$ in the basic tetrahedron operators. Thereafter, the process of obtaining the $\CL_a$'s is fully $k$-independent.

\subsection{Physical meaning of the angle polytope: R-charges of operators}

We have seen that the existence of a non-empty angle polytope $\p[M,\mb t]_n$ guarantees that the state integral $\CZ_b^{(k,1)}[M,\mb t]$ is well defined. In the context of the 3d-3d correspondence, we propose that the existence of a nonempty polytope is necessary in order for the UV gauge theory $T_n[M,\mb t]$ defined by a particular triangulation $\mb t$ to flow to a good superconformal theory in the IR --- presumably $T_n[M]$, or a subsector thereof.

Some motivation for this proposal comes simply from the interpretation of the state-integral $\CZ_b^{(k,1)}[M,\mb t]_n$ as a lens-space partition function of $T_n[M,\mb t]$.
Let us assume%
\footnote{The existence of a finite partition function would follow from $T_n[M]$ having enough global symmetry such that associated real mass deformations of the SCFT make it fully massive. We are assuming this is the case here. It is true (\eg) for $M$ a hyperbolic knot complement.} %
that there does exist a (finite) lens-space partition function $\CZ_b^{(k,1)}[M]_n$ for the superconformal IR theory $T_n[M]$. Since lens-space partition functions are invariant under IR flow, we must have $\CZ_b^{(k,1)}[M,\mb t]_n=\CZ_b^{(k,1)}[M]_n$ so long as $T_n[M,\mb t]$ flows to $T_n[M]$. But if the polytope $\p[M,\mb t]_n$ is empty and the state integral $\CZ_b^{(k,1)}[M,\mb t]_n$ cannot be made sense of, it indicates that $T_n[M,\mb t]$ may not actually flow to $T_n[M]$.

It is also instructive to analyze the $S^2\times_q S^1$ index of $T_n[M,\mb t]$, \emph{a.k.a.} the level-zero Chern-Simons partition function $\CZ^{(0,1)}[M,\mb t]_n(\vec m,\vec e;\, q^{\frac12})$. Recall \cite{DGG-index} that the index is a formal Laurent series in $q^{\frac12}$ for all $d$-tuples of integer magnetic and electric charges $(\vec m,\vec e)$. Mathematically, the meaning of the angle polytope at $k=0$ is simply that the shifted expression
\be   q^{\frac12(\vec\alpha\cdot\vec e-\vec\beta\cdot\vec m)} \CZ^{(0,1)}[M,\mb t]_n(\vec m,\vec e;\, q^{\frac12})  \label{shiftI} \ee
is well defined and only contains positive powers of $q$ for all $(\tfrac{\mathfrak Q}{2}\vec\alpha,\tfrac{\mathfrak Q}{2}\vec\beta)  \in \p[M,\mb t]_n$, \cf\ \cite{Gar-index}. Note that the prefactor in \eqref{shiftI} is simply comes from the  operator of \eqref{Hp2} acting in the representation that's appropriate for the index.

Physically, the prefactor of \eqref{shiftI} represents a redefinition of the $U(1)_R$ R-charge assignment in the UV gauge theory $T_n[M,\mb t]$. The R-charge is shifted by $\vec\alpha$ units of $U(1)^d$ flavor charge, and $\vec\beta$ units of dual magnetic charge.
(The magnetic shift is implemented by a mixed R-flavor background Chern-Simons term in the Lagrangian of $T_n[M,\mb t]$.)
In a supersymmetric (but not superconformal) theory, we are at liberty to choose the R-charge assignment as we like.
The index \eqref{shiftI} then calculates the trace $\Tr (-1)^R q^{\tfrac{R}{2}+j_3}$ in the space of BPS states on $S^2$, with electric charge $\vec e$ and magnetic flux $\vec m$.

Now, if $T_n[M,\mb t]$ flows to a superconformal fixed point $T_n[M]$ in the IR, then the same expression \eqref{shiftI} should equal the superconformal index of $T_n[M]$ with a distinguished choice $(\vec\alpha_{SCFT},\vec\beta_{SCFT})$ of R-charges. This is the choice that enters the superconformal algebra, obtained (for example) by $\CZ$-extremization \cite{Jafferis-Zmin}. Moreover, the superconformal algebra implies a BPS-like bound $\frac{R}{2}\pm j_3\geq 0$ guaranteeing that for the distinguished $(\vec\alpha_{SCFT},\vec\beta_{SCFT})$, the index \eqref{shiftI} contains only non-negative powers of $q$. Therefore, if $T_n[M,\mb t]$ flows to a superconformal fixed point, the polytope $\p[M]_n$ --- the maximal polytope discussed in Section \ref{sec:max}, which contains $\p[M,\mb t]$ --- must be nonempty, as it contains $(\vec\alpha_{SCFT},\vec\beta_{SCFT})$.
(Strictly speaking, since the BPS bound is a semi-strict inequality, this argument only shows that a ``non-negative'' analogue of $\p[M]_n$, defined using semi-strict inequalities, is nonempty. As long as $T_n[M]$ is an interacting SCFT, however, we expect that $(\vec\alpha_{SCFT},\vec\beta_{SCFT})$ lies in the interior of $\p[M]_n$.)

\section{Examples}
\label{sec:ex}

We illustrate the level-$k$ state integral with two quick examples.

\subsection{Figure-eight $\mb{4_1}$}

The figure-eight knot complement $M=S^3\bs\mb{4_1}$ is glued together from two tetrahedra, as in Thurston's original triangulation \cite{thurston-1980}. The gluing matrix for $n=2$ can be computed efficiently using \texttt{SnapPy} \cite{snappy}, and with one particular choice of tetrahedron labeling reads
\be \small  \begin{pmatrix} u\\C-2\pi i\\\hline v\\ * \end{pmatrix} = \left(\begin{array}{cc@{\;}|@{\;}cc}
   0 & -1 & -1 & -1 \\
 -1 & -1 & -2 & -2 \\\hline
 1 & 0 & 2 & 0 \\
 0 & 1 & 0 & 1 \\ \end{array}\right) \begin{pmatrix}  Z_1 \\  Z_2 \\\hline  Z_1''\\  Z_2'' \end{pmatrix} + i\pi \begin{pmatrix}1\\2\\\hline -1 \\ 0 \end{pmatrix}\,,
\ee
with the decomposition
$g=\sigma_\alpha(1,2)\,\sigma_\beta(-1,0)\,
U\left(\begin{smallmatrix} 1&0\\2&1\end{smallmatrix}\right)\,
T\left(\begin{smallmatrix} -2&-1\\-1&0\end{smallmatrix}\right)\,
S_1\,
T\left(\begin{smallmatrix} 1&0\\0&0\end{smallmatrix}\right)\,
U\left(\begin{smallmatrix} 0&1\\-1&1\end{smallmatrix}\right)$\,.
The resulting level-$k$ state integral is
\begin{align} \label{Z41}
 \CZ_b^{(k,1)}[\mb{4_1},\mb t](\mu,m) &= \frac{1}{k}\sum_{s\in \Z/k\Z} \int d\sigma\, (-1)^{s+m}e^{\frac{i\pi}{k}\big[-\mu^2-(\sigma-\frac{iQ}{2})^2+m^2+s^2\big]} \\
&\hspace{.5in} \times \CZ^{(k,1)}_b[\Delta](\sigma+\mu,s+m)\,\CZ^{(k,1)}_b[\Delta](\sigma-\mu,s-m)\,, \notag
\end{align}
with the integral done along anywhere inside the strip $|\Im(\mu)|< \Im(\sigma)<\mathfrak Q/2$. The result is a function $\CZ_b^{(k,1)}[\mb{4_1},\mb t]\in \CH_{\p[\mb{4_1},\mb t]}^{(k)}$, where
\be \label{p41} \p[\mb{4_1},\mb t]  = \{ |\beta| < \mathfrak Q/2-2|\alpha| \} \subset \R^2\,. \ee
Analyzing the actual decay of the function \eqref{Z41} shows that $\p[\mb{4_1},\mb t]$ is already maximal, in the sense of Section \ref{sec:max}, \ie\ $\p[\mb{4_1},\mb t]=\p[\mb{4_1}]$. The polytope $\p[\mb{4_1}]$ is a diamond, with the same shape as the Newton polygon of the A-polynomial $A_{\mb{4_1}} = \ell-(m^4-m^2-2-m^{-2}-m^{-4})+\ell^{-1}$.

For $k=1$, the state integral \eqref{Z41} first appeared, without a prescribed contour, in \cite{hikami-2006}, and has been seen in many places since. A Fourier transform of \eqref{Z41} (which is just the wavefunction in a different representation, related by $S\in Sp(2,\Z)$) appeared together with a contour in \cite{KashAnd}.

The difference operator that annihilates \eqref{Z41} for any $k$ is the $\hat A$-polynomial in Eqn (1.8) of \cite{Dimofte-QRS}, with the substitution $(\hat\ell,\hat m^2,q^{\frac12})\to (\y,\x,-q^{\frac12})$:
\be \CL_{\mb{4_1}}(\x,\y;q^{\frac12}) = (q^{\frac12}\x-q^{-\frac12}\x^{-1})\y^{-1}-(\x-\x^{-1})(\x^{-2}-\x^{-1}-q-q^{-1}-\x+\x^2)+(q^{-\frac12}\x-q^{\frac12}\x^{-1})\y\,. \ee
It is clearly a quantization of $A_{\mb 4_1}$ (setting $q^{\frac12}=-1$ gives $\CL_{\mb 4_1}(m^2,-\ell;1)=(m^2-m^{-2})A_{\mb 4_1}$) and coincides with the homogeneous part of the recursion relation for the colored Jones polynomial of the figure-eight knot \cite{garoufalidis-2004}. For sufficiently small $b$, the relation $\CL_{\mb{4_1}}(\x,\y;q^{\frac12})\CZ_b^{(k,1)}[M](\mu,m)=0$ can be used to analytically continue $\CZ_b^{(k,1)}[M]$ to a meromorphic function of $\mu\in \C$. The analytically continued function also satisfies the dual relation $\CL_{\mb{4_1}}(\tilde\x,\tilde\y;\tilde q^{\frac12})\CZ_b^{(k,1)}[M](\mu,m)=0$.

\begin{figure}[htb]
\centering
\includegraphics[width=2.3in]{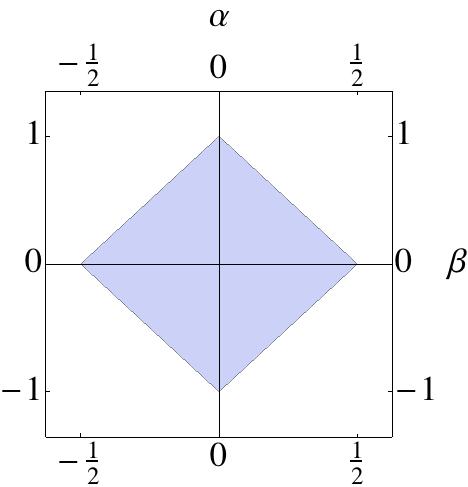}\hspace{.5in} \includegraphics[width=2.3in]{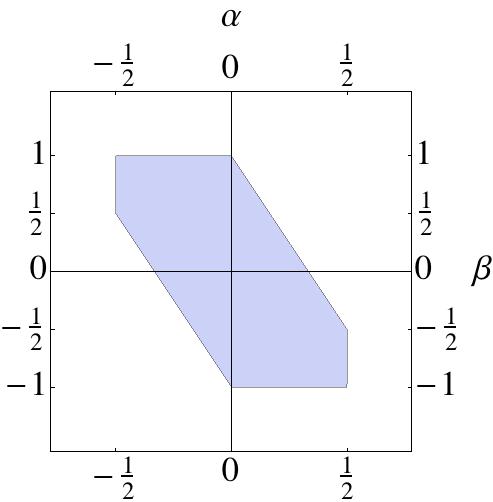}
\caption{Angle polytopes $\p[\mb{4_1},\mb t]=\p[\mb{4_1}]$ (left) and $\p[\mb{5_2},\mb t]$ (right), with axes  in units of $\mathfrak Q/2$.}
\label{fig:p4152}
\end{figure}

\subsection{$\mb{5_2}$ knot}

The partition function of the $\mb{5_2}$ knot complement $M=S^3\bs\mb{5_2}$, at $n=2$, is very similar to that of the figure-eight complement. Now $M$ can be triangulated by three tetrahedra. The gluing matrix is
$g=\sigma_\alpha(-1,-1,0)\,
U\left(\begin{smallmatrix} 1&0&0\\1&0&-1\\0&-1&2\end{smallmatrix}\right)\,
T\left(\begin{smallmatrix}2&-2&1\\-2&1&0\\1&0&0\end{smallmatrix}\right)\,S_1\,
T\left(\begin{smallmatrix} 1&-1&0\\-1&0&0\\0&0&0\end{smallmatrix}\right)\,
U\left(\begin{smallmatrix}-1&0&0\\-1&1&0\\-1&0&1\end{smallmatrix}\right)$
(given a convenient choice of labeling), leading to the state-integral model
\begin{align} \label{Z52}
 \CZ_b^{(k,1)}[\mb{5_2},\mb t](\mu,m) &= \frac{1}{k}\sum_{s\in \Z/k\Z} \int d\sigma\, (-1)^{s+m}e^{\frac{i\pi}{k}\big[\mu^2-(\sigma-\frac{iQ}{2})^2-m^2+s^2\big]} \\
&\hspace{.5in} \times \CZ^{(k,1)}_b[\Delta](\sigma,s)\, \CZ^{(k,1)}_b[\Delta](\sigma+\mu,s+m)\,\CZ^{(k,1)}_b[\Delta](\sigma-\mu,s-m)\,, \notag
\end{align}
with integration contour anywhere within the strip $|\Im(\mu)|<\sigma<\mathfrak Q/2$.
Now $\CZ_b^{(k,1)}[\mb{5_2},\mb t]\in \CH_{\p[\mb{5_2},\mb t]}^{(k)}$, where
 the angle polytope resulting from the gluing is 
\be \p[\mb{5_2},\mb t]  =  \{|\alpha|<\tfrac{\mathfrak Q}4,\,|\beta|<\tfrac{\mathfrak Q}{2},\, |\beta+3\alpha|<\tfrac{\mathfrak Q}{2}\} \subset \R^2\,.\ee
It is a hexagon, depicted in Figure \ref{fig:p4152}.

The difference operator that annihilate \eqref{Z52} appears in Eqn. (6.40) of \cite{BDP-blocks} (with the substitution $q^{\frac12}\to -q^{\frac12}$), and is a quantization of the A-polynomial $A_{\mb{5_2}}(\ell,m^2) = m^{14}\ell^3+m^4(1-m^2+2m^6+2m^8-m^{10})\ell^2-(1-2m^2-2m^4+m^8-m^{10})\ell+1$. Once again, the angle polytope has roughly the same shape as the Newton polygon of the A-polynomial.

\section{Holomorphic blocks}
\label{sec:blocks}

The level-$k$ state-integral partition functions $\CZ_b^{(k,p)}[M]$ defined in \eqref{defZ} should admit a decomposition as a (finite) sum of products of \emph{holomorphic blocks},
\be  \label{blocksk} \CZ_b^{(k,p)}[M](\vec \mu,\vec m) = \sum_\alpha B_\alpha[M,n](\vec x,q^{\frac12}) B_\alpha[M,n](\vec{\tilde x},\tilde q^{\frac12})\,, \ee
where the blocks $B_\alpha$ depend on $k,b$ only implicitly through the usual variables $\vec x,\vec{\tilde x},q,\tilde q$. Holomorphic blocks were introduced in \cite{BDP-blocks}, following \cite{Pasquetti-fact}, to unify $L(0,1)_b$ and $L(1,1)_b$ partition functions. A formula like \eqref{blocksk} was conjectured to hold for all $k$ in \cite{IMY-fact}.

While decomposition into holomorphic blocks is still a conjecture mathematically, it is well motivated physically on both sides of the 3d-3d correspondence. In complex Chern-Simons theory at level $k$, \eqref{blocksk} represents a holomorphic-antiholomorphic factorization of the partition function. Indeed, we saw in Section \ref{sec:CS} that $q$ and $\tilde q$ are the exponentiated holomorphic and antiholomorphic inverse-couplings of Chern-Simons theory, while $x$ and $\tilde x$ are natural holomorphic and antiholomorphic $\C^*$ variables on a Chern-Simons phase space. The finite sum is over flat $SL(n,\C)$ connections $\alpha$ on $M$, which are the classical solutions of Chern-Simons theory --- critical points of the complex Chern-Simons path integral.

On the $T_n[M]$ side of the 3d-3d correspondence, the decomposition \eqref{blocksk} represents a factorization of the supersymmetric partition function on $L(k,1)_b$. Topologically, the lens space $L(k,1)_b$ is a union of two solid tori $(D^2\times S^1)\cup_{\varphi}(D^2\times S^1)$, whose boundaries are identified using the element
\be  \varphi = \begin{pmatrix} 1 & 0 \\ -k & 1 \end{pmatrix}  \;\in SL(2,\Z) \ee
in the mapping class group of the torus $T^2\simeq \pd(D^2\times S^1)$. More precisely, the identification is a composition of $\varphi$ and an orientation reversal. One then expects the partition function of $T_n[M]$ on $L(k,1)_b$ to be to be an inner product of wavefunctions on solid tori,
\be \CZ_b^{(k,1)}[M]_n  = \big\langle \CZ_{D^2\times S^1}(T_n[M])\,\big|\,\CZ_{D^2\times S^1}(T_n[M])\big\rangle\,. \label{LDD} \ee
A priori, the Hilbert space in which this inner product is taken is enormous: it is the space of all physical states of $T_n[M]$ on a torus. However, due to supersymmetry, it appears%
\footnote{For $k=0$, the argument is an extension of the topological/anti-topological fusion construction of \cite{CV-tt*}, \cf\ \cite{CGV}; for $k=1$, factorization was derived in \cite{AMRP-fact}. For $k>1$, a careful physical argument is still required, and methods of \cite{CDFZ-geom} should prove useful in making it.} %
that $\big\langle \CZ_{D^2\times S^1}(T_n[M])\,\big|$ actually takes values in a finite-dimensional subspace of the full Hilbert space, namely the subspace $\CH_{\rm BPS}$ of supersymmetric ground states. By the 3d-3d correspondence, a basis for $\CH_{\rm BPS}$ is indexed by flat $SL(n,\C)$ connections $\alpha$ on $M$; so again we recover a factorization \eqref{blocksk}.

Note that in the decomposition \eqref{LDD}, the two wavefunctions $\big\langle \CZ_{D^2\times S^1}(T_n[M])\,\big|$ and $|\,\CZ_{D^2\times S^1}(T_n[M])\big\rangle$ depend holomorphically on parameters $q$ and $\tilde q$, respectively.  These are modular parameters of the respective boundary tori, related by $\varphi\in SL(2,\Z)$ (together with a reflection) as in \eqref{mod}. This coincides perfectly with the relation between holomorphic and antiholomorphic inverse-couplings in complex Chern-Simons theory at level $k$. Notice that $|q|<1$ if any only if $|\tilde q|>1$.

The holomorphic blocks $B_\alpha$ were conjectured in \cite{BDP-blocks} to have several wonderful properties. In particular:
\begin{itemize}
\item $B_\alpha(\vec x,q^{\frac12})$ is a meromorphic function of $\vec x$ and $q^{\frac12}$ for $|q|<1$ or $|q|>1$.
\item The functions $B_\alpha(\vec x,q)$ have a natural boundary at $|q^{\frac12}|=1$, but their values for $|q|<1$ and $|q|>1$ are ``naturally'' related. Concretely, in examples, this meant that $B_\alpha(\vec x,q^{\frac12})$ could be expressed as a single $q,x$-series that converged both for $|q|<1$ and $|q|>1$. (Physically, $B_\alpha(\vec x,q)$ is an index in the infinite-dimensional space of states of $T_n[M]$ on a disc; taking $|q|<1$ or $|q|>1$ corresponds to regrouping these infinite states in slightly different ways so as to make the index converge.)
\item For any $k\geq 0$, each product $B_\alpha(\vec x,q^{\frac12}) B_\alpha(\vec{\tilde x},\tilde q^{\frac12})$ is a function of $b$ that can be continued smoothly across the half-line $b\in \R_{>0}$ (with $q^{\frac12}=e^{i\pi(b^2+1)/k}$, $\tilde q^{\frac12}=e^{i \pi (b^{-2}+1)/k}$ as usual).
\item The set of functions $\{B_\alpha(\vec x,q^{\frac12})\}_\alpha$ form a basis of solutions to the $q$-difference equations $\CL_a(\vec\x,\vec\y;q^{\frac12})B_\alpha(\vec x,q^{\frac12})=0$.
\end{itemize}

\subsection{Knot complements}

We have verified via residue calculations (as in Appendix \ref{app:23}) that the holomorphic blocks found in \cite{BDP-blocks} for the figure-eight and $\mb{5_2}$ knots reproduce the state integrals $\CZ_b^{(k,1)}[M]_2$. For the figure-eight knot, the precise factorization of \eqref{Z41} is
\begin{align} \label{B41}
&\CZ_b^{(k,1)}[\mb{4_1},\mb t](\mu,m) \\ &\qquad= C_3^{-1}e^{\tfrac{2\pi i}{k}(-\mu^2+m^2)}\Big[
 e^{\tfrac{\pi}{k}Q\mu}J(x,x^2;q)J(\tilde x,\tilde x^2;\tilde q)+e^{-\tfrac{\pi}{k}Q\mu}J(x^{-1},x^{-2};q)J(\tilde x^{-1},\tilde x^{-2};\tilde q)\Big]\,, \notag
\end{align}
where
\be J(x,y;q) := (qy;q)_\infty \sum_{n=0}^\infty \frac{x^n}{(q^{-1})_n(qy;q)_n}\,,\qquad |q|<1\quad\text{or}\quad |q|>1 \ee
is the $q$-Bessel function (see \cite[Sec. 5.2.1]{BDP-blocks} and properties and references therein). Note that the prefactors depending on $\mu,m$ can also be holomorphically factorized (trivially) into Jacobi theta-functions using \eqref{ZZCS}. The two blocks correspond to the two irreducible flat $SL(2,\C)$ connections on the figure-eight knot complement, with fixed meridian eigenvalues on the boundary; these are solutions to $A_{\mb{4_1}}(\ell,m^2)=0$ at fixed $m^2$.

For the $\mb{5_2}$ knot, the factorization of \eqref{Z52} is
\begin{align} \label{B52}
&\CZ_b^{(k,1)}[\mb{5_2},\mb t](\mu,m) = \Big[
 (-1)^me^{\frac{i\pi}{k}(\mu^2-m^2)} \CG(x,x^{-1},1;q)\,\CG(\tilde x,\tilde x^{-1},1;\tilde q) \\ &\qquad+
 e^{-\frac{\pi}{k}Q\mu+\frac{i\pi}{4k}Q^2}\CG(x,x^2,x;q) \,\CG(\tilde x,\tilde x^2,\tilde x;\tilde q) +
 e^{\frac{\pi}{k}Q\mu+\frac{i\pi}{4k}Q^2}\CG(x^{-1},x^{-2},x^{-1};q) \,\CG(\tilde x^{-1},\tilde x^{-2},\tilde x^{-1};\tilde q) \Big] \notag
\end{align}
where
\be \CG(x,y,z;q) := (qx;q)_\infty (qy;q)_\infty\sum_{n=0}^\infty \frac{z^n}{(q^{-1})_n(qx;q)_n(qy;q)_n}\,,\qquad |q|<1\quad\text{or}\quad |q|>1\,. \ee
Now the three blocks correspond to the three irreducible flat $SL(2,\C)$ connections on the $\mb{5_2}$ knot complement.

\subsection{$G(q)$ and $g(q)$ series when $\mu=m=0$}

The decompositions into blocks for the examples above make sense for generic values of $\mu$. As $\mu$ varies, the ``natural'' basis of blocks may undergo wall crossing (or a Stokes phenomenon) across real codimension-one walls, as discussed in \cite{BDP-blocks}. At complex codimension-one loci in the $\mu$-plane, the decomposition into blocks can break down completely. This happens, for example for $\mu=m=0$. Then (for example) the two terms in \eqref{B41} each diverge, while their sum remains finite.

At such degenerate loci, a different factorization is appropriate. For $k=1$ it was studied in \cite{GK-Gg} in a class of examples that includes the figure-eight-knot state integral. It was shown there that, for $k=1$,
\be \CZ_b^{(1,1)}[\mb 4_1,\mb t](0,0) =  \frac{1}{2}e^{-\tfrac{i\pi}{6}(b^2+b^{-2})}\big[b^{-1}g(q)G(\tilde q^{-1})+bg(\tilde q^{-1})G(q)\big]\,,   \label{Gg} \ee
with series
\be g(q) := \sum_{n=0}^\infty (-1)^n \frac{q^{\frac12n(n+1)}}{(q)_n^2}\,,\qquad
  G(q):= \sum_{n=0}^\infty \Big( 1+2n-4\sum_{s=0}^\infty \frac{q^{s(n+1)}}{1-q^s}\Big)(-1)^n \frac{q^{\frac12n(n+1)}}{(q)_n^2}\,. \ee
that converge as long as the arguments are inside the unit circle.
We expect that a similar formula holds for all $k$. Indeed, a bit of numerical experimentation suggests that
\be \CZ_b^{(k,1)}[\mb 4_1,\mb t](0,0) =  \frac{i}{2}e^{-\tfrac{i\pi}{6k}(Q^2+k^2)}\big[b^{-1}g(q)G(\tilde q^{-1})+b\,g(\tilde q^{-1})G(q)\big]\,.   \label{Gg-k} \ee

\section{Extension to $L(k,p)_b$}
\label{sec:kp}

The $L(k,p)_b$ partition functions of 3d $\CN=2$ gauge theories have not yet been systematically computed for values other than $p\equiv \pm 1$ (mod $k$). Nevertheless, we know many properties that such partition functions should satisfy; for example:
\begin{itemize}
\item They must admit an $ISp(2N,\Z)$ action (since this descends from an action on 3d theories).
\item They should admit holomorphic factorization (since $L(k,p)_b = (D^2\times S^1)\cup_\varphi(D^2\times S^1)$ for any $p$, with an appropriate $\varphi\in SL(2,\Z)$).
\item They should be acted on by two mutually commuting algebras of difference operators, coming from line operators localized at $|z|=0$ or $|w|=0$ in the lens-space geometries.
\item They should then be annihilated by the canonical difference operators $\CL_a(\vec\x,\vec\y;q^{\frac12})$ and $\CL(\vec{\tilde \x},\vec{\tilde\y};\tilde q^{\frac12})$ (since these come from Ward identities that only detect a neighborhood of $|z|=0$ or $|w|=0$ in the lens-space geometries).
\end{itemize}
It turns out that these properties are enough to construct partition functions for the tetrahedron theory, an $ISp(2N,\Z)$ action, and thus a consistent way to glue tetrahedron partition functions into an $L(k,p)_b$ partition function for general $T_n[M]$. (Physically, this is equivalent to producing a partition function for any abelian 3d $\CN=2$ gauge theory.) We describe these various ingredients and the resulting state-integral model $\CZ_b^{(k,p)}[M,\mb]_n$ in this section. It takes values in the same functional spaces $\CH_{\p[M,\mb t]}^{(k)}$ as for $p=1$.

Evidently, the $(k,p)$ state-integral models correspond to some twisted version of complex Chern-Simons theory at level $k$. It would be interesting to work out the full physical meaning of this.

\subsection{Holomorphic variables}

Let us start with the ansatz that an $L(k,p)_b$ partition function admits a decomposition into holomorphic blocks \eqref{blocksk}. By setting $q=e^{2\pi i\tau}$ and $\tilde q=e^{2\pi i\tilde \tau}$, where $\tau,\tilde \tau$ are the modular parameters of the boundaries of two copies of $D^2\times S^1$ that form $L(k,p)_b$, it follows that we must have
\be \label{phikp}
\tilde\tau = -\varphi\cdot\tau\,,\qquad  \varphi = \begin{pmatrix} r & s \\ -k & p \end{pmatrix}\in SL(2,\Z)\,, \ee
where $rp\equiv 1$ (mod $k$). This is satisfied if
\be q = \exp\frac{2\pi i}{k}(p+b^2)\,,\qquad \tilde q=\exp\frac{2\pi i}{k}(r+b^{-2})\,. \label{defqp} \ee
These variables depend only on the conjugacy class of $p,r$ modulo $k$, and reduce to our previous expressions when $p,r\equiv 1$.

When defining an affine symplectic action on partition functions, we saw that it was actually necessary to choose square roots of $q$ and $\tilde q$. There is a canonical choice when $p=r\equiv \pm 1$. Otherwise the cleanest way to define these square roots and the full symplectic action is to lift $p,r$ to integers modulo $2k$ such that
\be \boxed{pr\equiv 1\;\;(\text{mod}\,2k)}\,,  \label{kp2} \ee
or equivalently require $\varphi\in \Gamma^0(2)$ in \eqref{phikp}. We will henceforth assume that $p,r$ satisfy \eqref{kp2}. The corresponding roots are
\be q^{\frac12} := \exp\frac{i\pi}{k}(p+b^2)\,,\qquad \tilde q^{\frac12}:=\exp\frac{i\pi}{k}(r+b^{-2})\,.\ee

We must also encode continuous and discrete variables $(\mu\in \C$, $m\in \Z/k\Z)$ in holomorphic/anti-holomorphic parameters $x,\tilde x$, and modify the operator actions (\ref{op-action}a-b) in such a way that the canonical relations \eqref{xycomm} and (\ref{op-action}c) are preserved. To this end, we define
\be x = \exp \frac{2\pi i}{k}(-ib\mu - pm)\,,\qquad \tilde x= \exp\frac{2\pi i}{k}(-ib^{-1}\mu+m)\,,  \label{defxp} \ee
and operators
\bse \label{op-action-p}
\be  \begin{array}{ll} \bmu f(\mu,m) = \mu f(\mu,m)\,,\qquad &e^{\frac{2\pi i}{k}\m} f(\mu,m) = e^{\frac{2\pi i}{k}m}f(\mu,m)\,, \\[.2cm]
 \bnu f(\mu,m) = -\frac{k}{2\pi i}\pd_\mu f(\mu,m)\,,\qquad &e^{\frac{2\pi i}{k}\n} f(\mu,m) = f(\mu,m+r)\,,
\end{array}
\ee
or equivalently, with $\x=e^{\frac{2\pi i}{k}(-ib\bm\mu-p\mb m)}$, $\tilde\x=e^{\frac{2\pi i}{k}(-ib^{-1}\bm\mu+\mb m)}$, $\y=e^{\frac{2\pi i}{k}(-ib\bm\nu-p\mb n)}$, $\tilde\y=e^{\frac{2\pi i}{k}(-ib^{-1}\bm\nu+\mb n)}$,
\be \begin{array}{ll} \x f(\mu,m) = xf(\mu,m)\,,\qquad &\tilde\x f(\mu,m) = \tilde xf(\mu,m)\,, \\[.2cm]
 \y f(\mu,m) = f(\mu+ib,m-1)\,, \quad & \tilde\y f(\mu,m) = f(\mu+ib^{-1},m+r)\,. \end{array} \ee
Then, by virtue of $pr\equiv 1$ (mod $k$),
\be \begin{array}{ll} \x f(x,\tilde x) = xf(x,\tilde x)\,,\qquad &\tilde\x f(x,\tilde x) = \tilde xf(x,\tilde x)\,, \\[.2cm]
 \y f(x,\tilde x) = f(qx,\tilde x)\,, & \tilde\y f(x,\tilde x) = f(x,\tilde q\tilde x)\,, \end{array} \ee
\ese
and the canonical commutation relations  $\y\x = q\x\y\,,\; \tilde\y\tilde\x=\tilde q\tilde\x\tilde\y\,,\; \y\tilde\x=\tilde\x\y\,,\; \tilde\y\x=\x\tilde\y$ hold. Note that since $r$ is a unit in $\Z/k\Z$, we could send $m\mapsto rm$ to obtain an equivalent parameterization of wavefunctions $f(\mu,m)$. After this reparameterization, the roles of $p$ and $r$ above (which may look a little asymmetric) are exchanged.

In the case of multiple variables $\vec\mu\in \C^N$, $\vec m\in \C^N$, we set $x_i = \exp\frac{2\pi i}{k}(-ib\mu_i-pm_i)$, $\tilde x_i = \exp\frac{2\pi i}{k}(-ib^{-1} \mu_i+m_i)$, and take tensor products of the operator algebra, as usual.

\subsection{Tetrahedron partition function}

In order for the tetrahedron partition function to be compatible with holomorphic block decomposition, it must take the canonical form
\be  \CZ_b^{(k,p)}[\Delta](\mu,m) := (qx^{-1};q)_\infty (\tilde q\tilde x^{-1};\tilde q)_\infty\,,\qquad |q|<1\;\text{or}\;|q|>1\,, \label{tet-p} \ee
with $x,\tilde x,q,\tilde q$ depending on $k,p,r,b$ as above. This partition function is automatically annihilated by $\y+\x^{-1}-1$ and $\tilde\y+\tilde\x^{-1}-1$.

Beautifully, \eqref{tet-p} is a skewed lattice product
\be \CZ_b^{(k,1)}[\Delta](\mu,m) = \prod_{\gamma,\delta\in \Gamma(k,p;m)} \CZ_b^{(1,1)}[\Delta]\big(\frac1k(\mu+ib\gamma+ib^{-1}\delta),0\big)\,,  \label{prod-p} \ee
where $\Gamma(k,p;m):=\{\gamma,\delta\in\Z\,\big|\, 0\leq \gamma,\delta < k,\; p\gamma-\delta\equiv pm\;(\text{mod}\,k)\,\}$.  Equivalently, multiplying the constraint $\gamma-r\delta\equiv m$ by $r$, we can write it as  $\gamma-r\delta\equiv m$ (mod $k$). The formula \eqref{prod-p} together with known properties of Faddeev's noncompact quantum dilogarithm \eqref{ZFad} imply that 1) $\CZ_b^{(k,p)}[\Delta](\mu,m)$ extends to a meromorphic function of $\mu\in \C$ for all $b$ with $\Re(b)>0$;   2) the zeroes and poles of $\CZ_b^{(k,p)}[\Delta](\mu,m)$ lie on the torsor
\be \label{poles-p}
 \mu\in \big\{ ib\alpha + ib^{-1}\beta\,\big|\;\alpha,\beta\in\Z,\;\; -p\alpha+\beta=pm \;(\text{mod $k$})\big\} \qquad\text{with}\quad \begin{array}{r@{\quad}l} \text{zeroes:} & \alpha,\beta \geq 1 \\[,2cm]
\text{poles:} & \alpha,\beta \leq 0 \end{array}\,;
\ee
and 3) the large-$\mu$ behavior of $\CZ_b^{(k,p)}[\Delta](\mu,m)$ is identical to that of $\CZ_b^{(k,1)}[\Delta](\mu,m)$ in \eqref{mu-ass}. Therefore,
\be \CZ_b^{(k,p)}[\Delta]\in \CH_{\p[\Delta]}^{(k)}\,, \label{HD-p}  \ee
with the \emph{same} angle polytope $\p[\Delta]$ as for $p=1$.

We also have the inversion identity
\begin{align}  \label{ZZCS-p} &\CZ_b^{(k,p)}[\Delta](\tfrac{iQ}{2}+\mu,\tfrac{r-1}{2}+m)\;  \CZ_b^{(k,p)}[\Delta](\tfrac{iQ}{2}-\mu,\tfrac{r-1}{2}-m)  \\ &\hspace{2in} =  (-1)^{m}e^{\frac{i\pi}{k}(\mu^2-pm^2)+\frac{2\pi i}{24 k}(b^2+b^{-2})+\frac{i\pi}{6k}\kappa(k,r)}\,,  \notag\end{align}
coming from the fact that the LHS is a ratio of Jacobi theta functions related by the modular transformation $T\varphi T$. Alternatively, this follows from inversion for $\CZ_b^{(1,1)}$ and the product \eqref{prod-p}. The integer $\kappa$ is given by $\kappa(k,r)= 3(k-1)^2-\frac{12}{k}\sum_{j=0}^{k-1}j\big(rj+\frac{r-1}{2}\;\text{mod\,k}\big) $.  (Working modulo $k$, there is a much simpler formula $\kappa(k,r)\equiv \frac12(1-k^2)(p+r)$. Also, modulo $3k$, $\kappa(k,r)\equiv \frac32 k(k-1)-6k S(-p,k)$, where $S(-p,k)$ is a Dedekind sum as in~\eqref{Ded}.)

\subsection{Affine symplectic action}

The $ISp(2N,\Z)$ action on spaces $\CH_\p^{(k)}$ that we found for $p=1$ can also be modified slightly to preserve the crucial intertwining property in the operator algebra \eqref{gop}, given the modified operators \eqref{op-action-p}.  Indeed, simply requiring \eqref{gop} to hold uniquely determines the action of generators of $ISp(2N,\Z)$ up to normalization, and requiring the group relations of $ISp(2N,\Z)$ to hold (almost) fixes the normalizations.

To replace \eqref{repf}, we define
\be
(S_i\cdot f)(\vec\mu,\vec m)  \ds := \frac{1}{k}\sum_{n\in \Z/k\Z} \int d\nu\;  e^{\tfrac{2\pi i}{k}(-\mu_i\nu+p\,m_in)}\,   f(\mu_1,...,\overset{(i)}{\nu},...,\mu_N;m_1,...,\overset{(i)}{n},...,m_N)\,; \notag \ee
\vspace{-.4cm}
\be\begin{array}{rl}
  (T(\mb B)\cdot f)(\vec\mu,\vec m) &:=  (-1)^{p\,\vec m\cdot\mb B\cdot\vec m}e^{\tfrac{i\pi}{k}(-\vec\mu\cdot\mb B\cdot\vec\mu+p\,\vec m\cdot\mb B\cdot\vec m)} f(\vec\mu,\vec m)\,; \\[.2cm]
  (U(\mb A)\cdot f)(\vec\mu,\vec m)  &:= (\det\mb A)^{-\frac12} f(\mb A^{-1}\vec\mu,\mb A^{-1}\vec m)\,;\\[.2cm]
  (\sigma_\alpha(\vec t\,)\cdot f)(\vec\mu,\vec m)  &:= f(\vec \mu - i\tfrac Q2\vec t,\vec m+\tfrac{1-r}{2}\vec t)\qquad (\vec t\in \Z^N)\,;\\[.2cm]
  (\sigma_\beta(\vec t\,)\cdot f)(\vec\mu,\vec m) &:=  e^{\tfrac{2\pi i}{k}\big[- \tfrac{iQ}2 \vec t \cdot \vec \mu+\tfrac{1-p}{2}\vec t \cdot \vec m\big]}\, f(\vec\mu,\vec m) \qquad (\vec t\in \Z^N)\,.
\end{array} \label{repf-p}
\ee
It is essential in defining these transformations to use integers $p,r$ such that $pr\equiv 1$ modulo $2k$ (as in \eqref{kp2}) rather than simply modulo $k$. For example, having $p,r$ odd and $s=\frac1k(1-pr)$ even ensures that the kernel of the basic T-type transformation $T(1)= (-1)^{pm}e^{\tfrac{i\pi}k(-\mu^2+pm^2)}\times$\; is an honest function of $m\in \Z/k\Z$ and commutes as it should with affine shifts. Moreover, since the affine shifts act on $m$ as well as $\mu$ when $p,r\neq 1$, they must be quantized as indicated.
The modified action acquires a projective ambiguity of the form
\be (-1)^{a_1} e^{a_2 \tfrac{i\pi Q^2}{4k}} e^{a_2' \tfrac{i\pi}{k}}\,,\qquad a_1,a_2,a_2'\in \Z\,, \label{amb-p} \ee
which becomes \eqref{amb1} when $p,r=1$ (the $2k$-th root of unity parametrized by $a_2'$ is proportional to $(p-1)(r-1)$). 

In addition to the symplectic action on wavefunctions, one may also extend the product and symplectic reduction operations of Section \ref{sec:Hred} to general $k,p$. They are completely unmodified.

\subsection{State sum, rotations, and 2--3 moves}
\label{sec:23-p}

Given the $L(k,p)_b$ tetrahedron partition function and $ISp(2N,\Z)$ action proposed above, satisfying essentially the same properties as for $p=1$,
we can immediately define a state integral
\be \label{defZ-p}  \boxed{\CZ^{(k,p)}_b[M,\mb t]_n = \text{red}_{d+1,...,N}\Big[ g\cdot \prod_{i=1}^N \CZ^{(k,1)}_b[\Delta_i] \Big] \;\subset \;\CH^{(k,p)}_{\p[M,\mb t]_n}}\,. \ee
It has the same basic properties as \eqref{defZ}. In particular, it is well defined as long as $\p[M,\mb t]$ is nonempty. By construction, it is annihilated by the same difference operators $\CL_a(\vec\x,\vec\y;q^{\frac12})$ and $\CL_a(\vec{\tilde\x},\vec{\tilde\y};\tilde q^{\frac12})$ as for $p=1$, which can then be used to analytically continue to the whole space $\vec\mu\in \C^d$.

Invariance under re-labeling of tetrahedra and under local 2--3 moves follows from generalizations of the basic identities \eqref{ST} and \eqref{23}. We prove in Appendix \ref{app:23} that
\begin{align} 
\big(\rho\cdot \CZ^{(k,p)}_b[\Delta]\big)(\mu,m) &= \frac{1}{k}\sum_{n=0}^{k-1} \int d\nu\, (-1)^n e^{\tfrac{i\pi}{k}\big[-\nu^2-(2\mu-iQ)\nu +p\,n^2+p\,(2m+1-r)n\big]}\,\CZ^{(k,1)}_b[\Delta](\nu,n) \notag \\
&= e^{-\tfrac{i\pi}{12k}(Q^2-2)-\tfrac{i\pi}{4}+i\pi S(-p,k)} \CZ^{(k,p)}_b[\Delta](\mu,m)\, \label{ST-p}
\end{align}
and
\be \CZ^{(k,p)}_b[\text{bip}_2](\vec\mu,\vec m)=  C_3(k,p,r)\, \CZ^{(k,p)}_b[\text{bip}_3](\vec\mu,\vec m)\,, \label{23-p} \ee
where
\begin{align} \label{bip2-p}
\CZ^{(k,p)}_b[\text{bip}_2](\vec\mu,\vec m) &= \frac{1}{k}\sum_{n=0}^{k-1}\int d\nu\,(-1)^ne^{\tfrac{i\pi}{k}\big[-(2\mu_2-iQ)\nu-\nu^2+p\,(2m_2+1-r)n+p\,n^2\big]} \\
 &\hspace{1in} \times \CZ^{(k,p)}_b[\Delta](\nu,n)\;\CZ^{(k,p)}_b[\Delta](\mu_1-\nu,m_1-n)\,,
\notag \end{align}
\begin{align} \CZ^{(k,p)}_b[\text{bip}_3](\vec\mu,\vec m) &= (-1)^{m_1+m_2+\frac{r-1}{2}}e^{\tfrac{i\pi}{k}\big[ -(i\tfrac Q2-\mu_1-\mu_2)^2+p\,(\tfrac{r-1}{2}-m_1-m_2)^2\big]} \label{bip3-p} \\
 &\hspace{-.2in} \times \CZ^{(k,p)}_b[\Delta](\mu_1,m_1)\;\CZ^{(k,p)}_b[\Delta](\mu_2,m_2)\; \CZ^{(k,p)}_b[\Delta](iQ-\mu_1-\mu_2,r-1-m_1-m_2)\,; \notag
\end{align}
and
\be C_3(k,p,r) = \exp\Big[ -\tfrac{i\pi}{6k}(Q^2-2)-\tfrac{i\pi}{4}+i\pi S(-p,k)-\tfrac{i\pi}{6k}\kappa(k,r)\Big]\,\approx  \exp\tfrac{-i\pi}{6 k}\Big[Q^2+p+r-2\Big]\,, \label{defC3} \ee
where $S(-p,k)$ is a Dedekind sum as in \eqref{Ded}, $\kappa$ was defined below \eqref{ZZCS-p}, and the  approximation in \eqref{defC3} holds up to an 8-th root of unity.
It follows that the total projective ambiguity of the state integral, as defined here, is at worst of the form
\be   e^{\tfrac{2\pi i}{24\,k}a_1+ \tfrac{2\pi i}{4k}Q^2a_2}\,,\qquad a_i\in \Z\,.
\ee

\subsection{Holomorphic blocks}

A direct residue calculation also shows that some partition functions of the proposed $L(k,p)_b$ state-integral model have the desired canonical holomorphic-block decompositions. For example, in the case of the figure-eight knot complement, we have 
\begin{align}  &\CZ^{(k,p)}_b[\mb{4_1},\mb t](\mu,m) = e^{-\tfrac{i\pi}{12k}(Q^2+p+r-2)} e^{\tfrac{2\pi i}{k}(-\mu^2+pm^2)} \\
 &\quad \times\Big[ e^{\tfrac{i\pi}{k}(-iQ\mu+(1-p)m)} J(x,x^2;q)\,J(\tilde x,\tilde x^2;\tilde q) 
  + e^{-\tfrac{i\pi}{k}(-iQ\mu+(1-p)m)}J(x^{-1},x^{-2};q)\,J(\tilde x^{-1},\tilde x^{-2};\tilde q)\,\Big]\,,\notag
\end{align}
up to a twenty-fourth root of unity.

\subsection*{Acknowledgements}

The author would like to thank Jorgen Andersen, Stavros Garoufalidis, Daniel Jafferis, Rinat Kashaev, Sara Pasquetti, Roland van der Veen, and Don Zagier for pleasant and insightful discussions related to the ideas in this paper. The author gratefully acknowledges support from the Simons Center for Geometry and Physics, Stony Brook University and the Mathematisches Forschunginstitut Oberwolfach at which some of the research for this paper was performed. The author's work is supported by DOE grant DE-SC0009988, and also in part by ERC Starting Grant no. 335739 ``Quantum fields and knot homologies,'' funded by the European Research Council under the European Union's Seventh Framework Programme.

\appendix

\section{The $L(k,p)$ pentagon identity}
\label{app:23}

Here we give an elementary proof of the most general 2--3 move \eqref{23-p}. Specializing $p=1$ recovers the 2--3 move in Chern-Simons theory at level $k$ \eqref{23}. Sending $\mu_1\to\infty$ (or simply repeating the calculation with a slightly simpler integral) produces the identities \eqref{ST}, \eqref{ST-p} that ensure the tetrahedron wavefunction is invariant under cyclic rotations. The basic method (essentially summation of residues) that we use to analyze the 2--3 move here can also be used to derive the holomorphic block expressions in the text.

We start with the integral
\begin{align} \label{bip2-p-A}
\CZ^{(k,p)}_b[\text{bip}_2](\vec\mu,\vec m) &= \frac{1}{k}\sum_{n=0}^{k-1}\int d\nu\,(-1)^ne^{\tfrac{i\pi}{k}\big[-(2\mu_2-iQ)\nu-\nu^2+p\,(2m_2+1-r)n+p\,n^2\big]} \\
 &\hspace{1in} \times \CZ^{(k,p)}_b[\Delta](\nu,n)\;\CZ^{(k,p)}_b[\Delta](\mu_1-\nu,m_1-n)\,.
\notag \end{align}
Let us assume that $\Im(b)>0$ (and as always $\Re(b)>0$) and write the tetrahedron partition functions as
\be \CZ^{(k,p)}_b[\Delta](\nu,n) = \frac{(qy^{-1};q)_\infty}{(\tilde y^{-1};\tilde q^{-1})_\infty}\,,\qquad 
 \CZ^{(k,p)}_b[\Delta](\mu_1-\nu,m_1-n) = \frac{(qyx_1^{-1};q)_\infty}{(\tilde y\tilde x_1^{-1};\tilde q^{-1})_\infty}\,,
 \ee
where (as usual) $y=e^{\frac{2\pi i}{k}(-ib\nu-pn)}$, $\tilde y=e^{\frac{2\pi i}{k}(-ib^{-1}\nu+n)}$, $x_i=e^{\frac{2\pi i}k(-ib\mu_i-pm_i)}$,  $\tilde x_i=e^{\frac{2\pi i}{k}(-ib^{-1}\mu_i+m_i)}$; and $q=e^{\frac{2\pi i}{k}(p+b^2)}$, $\tilde q=e^{\frac{2\pi i}{k}(r+b^{-2})}$. The function $\CZ^{(k,p)}_b[\Delta](\nu,n)$ at fixed $n$ has poles in the lower half-plane at $\nu = -ibs-ib^{-1}t$ for integers $s,t\geq 0$ and $s-rt\equiv n$ (mod $k$), which translates to
\be \text{poles}:\qquad y=q^{-s}\,; \tilde y=\tilde q^{-t}\,,\qquad s,t\geq 0,\quad s-rt\equiv n \;(\text{mod $k$})\,.\ee
The function $\CZ^{(k,p)}_b[\Delta](\mu_1-\nu,m_1-n)$  has poles in the upper half-plane. Assuming that $\Im(\mu_2)>\mathfrak Q$ and $\Im(\mu_1)>\mathfrak Q-\Im(\mu_2)$ we can close the integration contour in the lower half-plane, enclosing all the poles of $\CZ^{(k,p)}_b[\Delta](\nu,n)$ and none of the poles of $\CZ^{(k,p)}_b[\Delta](\mu_1-\nu,m_1-n)$. (We can then analytically continue to other values of $\mu_1,\mu_2$, as well as $b$.)

The residue of $\CZ^{(k,p)}_b[\Delta](\nu,n)$ at the $(s,t)$-th pole is
\be  \frac{kb}{2\pi} \frac{(q)_\infty}{(\tilde q^{-1})_\infty} \frac{1}{(q)_s(\tilde q)_t} = \frac{k}{2\pi} e^{-\frac{i\pi}{12k}(b^2+b^{-2})+\frac{i\pi}{4}+i\pi S(-p,k)} \frac{1}{(q)_s(\tilde q)_t}\,, \label{Ded} \ee
with $(q)_a = (q;q)_a:=\prod_{i=1}^a(1-q^a)$ (more generally $(x;q)_a:=\prod_{i=0}^{a-1}(1-q^ix)$); $(q)_\infty = q^{-\frac1{24}}\eta(q)$; and the Dedekind sum $S(-p,k) := - \sum_{i=1}^{k-1} \frac{i}{k}\big(\frac{ip}{k}+\lfloor- \frac{ip}{k}\rfloor+\frac12\big)$. Evaluating $\CZ^{(k,p)}_b[\Delta](\mu_1-\nu,m_1-n)$ at the $(s,t)$ pole gives
\be \frac{(qx_1^{-1};q)_\infty}{(\tilde x_1^{-1};\tilde q^{-1})_\infty} (x_1^{-1};q^{-1})_s(\tilde x_1^{-1};\tilde q^{-1})_t\,. \ee
Moreover, setting $\nu=-ibs-ib^{-1}t$ and using the facts that $s-rt\equiv n$ (mod $k$) and $rp\equiv 1$ (mod $2k$), the exponential prefactor in the integrand $(-1)^n e^{\frac{i\pi}{k}(...)}$ rearranges beautifully to become
\be (-1)^{s+t}x_2^{-s}\tilde x_2^{-t}q^{\frac12 s(s+1)}\tilde q^{\frac12 t(t+1)}\,. \ee

Summing all the residues of the integrand (and simplifying a bit), we arrive at
\begin{align*} \label{A2}
\CZ^{(k,p)}_b[\text{bip}_2](\vec\mu,\vec m) &=  -i\frac{(qx_1^{-1};q)_\infty}{(\tilde x_1^{-1};\tilde q^{-1})_\infty}  e^{-\frac{i\pi}{12k}(b^2+b^{-2})+\frac{i\pi}{4}+i\pi S(-p,k)}  \\
&\quad \times \sum_{n=0}^{k-1} \sum_{{s,t\geq 0 \atop s-rt\equiv n}}  (qx_1^{-1}x_2^{-1})^s (\tilde q\tilde x_1^{-1}\tilde x_2^{-1})^t \frac{(x_1;q)_s(\tilde x_1;q)_t}{(q)_s(\tilde q)_t} \\
&\hspace{-.5in} = 
  -i\frac{(qx_1^{-1};q)_\infty}{(\tilde x_1^{-1};\tilde q^{-1})_\infty}  e^{-\frac{i\pi}{12k}(b^2+b^{-2})+\frac{i\pi}{4}+i\pi S(-p,k)}  \bigg(\sum_{s=0}^\infty (qx_1^{-1}x_2^{-1})^s \frac{(x_1;q)_s}{(q)_s}\bigg)\bigg(\sum_{t=0}^\infty  (\tilde q\tilde x_1^{-1}\tilde x_2^{-1})^t \frac{(\tilde x_1;q)_t}{(\tilde q)_t}\bigg) \\
&\hspace{-.5in} = 
  -i e^{-\frac{i\pi}{12k}(b^2+b^{-2})+\frac{i\pi}{4}+i\pi S(-p,k)} 
    \frac{(qx_1^{-1};q)_\infty}{(\tilde x_1^{-1};\tilde q^{-1})_\infty} \frac{(qx_2^{-1};q)_\infty}{(\tilde x_2^{-1};\tilde q^{-1})_\infty} \frac{(\tilde x_1^{-1}\tilde x_2^{-1};\tilde q^{-1})_\infty}{(qx_1^{-1}x_2^{-1};q)_\infty}.
\end{align*}
Here in the second step we used the sum over $n$ (and $n$-independence of the summand) to remove the constraint $s-rt\equiv n$ on $s,t$, obtaining a sum over the full positive lattice $(s,t)\in \Z^2_{\geq 0}$. In the third step, we used the $q$-binomial identity to perform the sums.

The final expression here is very close to the desired result $\CZ_b^{(k,p)}[{\rm bip}_3](\vec\mu,\vec m)$. To finish, we use the formula \eqref{ZZCS-p} to invert the third ratio of $q$-factorials, giving
\begin{align}  \CZ^{(k,p)}_b[\text{bip}_2](\vec\mu,\vec m)  &= e^{-\tfrac{i\pi}{6k}(Q^2-2)-\tfrac{i\pi}{4}+i\pi\big(S(-p,k)-\frac{1}{6k}\kappa(k,r)\big)} \frac{(qx_1^{-1};q)_\infty}{(\tilde x_1^{-1};\tilde q^{-1})_\infty} \frac{(qx_2^{-1};q)_\infty}{(\tilde x_2^{-1};\tilde q^{-1})_\infty} \frac{(x_1x_2;q)_\infty}{(\tilde q^{-1}\tilde x_1\tilde x_2;\tilde q^{-1})_\infty} \notag \\
 &= C_3(k,p,r)\, \CZ_b^{(k,p)}[\text{bip}_3](\vec\mu,\vec m)\,.
\end{align}

In order to reduce these expressions to simpler forms at $p=r=1$, it is convenient to observe that the Dedekind sum $S(-p,k)$ and the function $\kappa(k,r)$ from \eqref{ZZCS-p} simplify to
\be S(-1,k) = -\frac{(k-1)(k-2)}{12k}\,,\qquad \kappa(k,1) = 1-k^2\,.\ee

\vspace{.5in}

\bibliographystyle{JHEP_TD}
\bibliography{toolbox}

\providecommand{\href}[2]{#2}\begingroup\raggedright\begin{thebibliography}{10}

\bibitem{DGH}
T.~Dimofte, S.~Gukov, and L.~Hollands, {\it Vortex Counting and Lagrangian
  3-manifolds},  {\em Lett. Math. Phys.} {\bf 98} (2011) 225--287,
  [\href{http://xxx.lanl.gov/abs/1006.0977}{{\tt arXiv:1006.0977}}].

\bibitem{Yamazaki-3d}
Y.~Terashima and M.~Yamazaki, {\it SL(2,R) Chern-Simons, Liouville, and Gauge
  Theory on Duality Walls},  {\em JHEP} {\bf 1108} (2011) 135,
  [\href{http://xxx.lanl.gov/abs/1103.5748}{{\tt arXiv:1103.5748}}].

\bibitem{DGG}
T.~Dimofte, D.~Gaiotto, and S.~Gukov, {\it Gauge Theories Labelled by
  Three-Manifolds},  {\em Comm. Math. Phys.} {\bf 325} (2014) 367--419,
  [\href{http://xxx.lanl.gov/abs/1108.4389}{{\tt arXiv:1108.4389}}].

\bibitem{CCV}
S.~Cecotti, C.~Cordova, and C.~Vafa, {\it Braids, Walls, and Mirrors},
  \href{http://xxx.lanl.gov/abs/1110.2115}{{\tt arXiv:1110.2115}}.

\bibitem{DGV-hybrid}
T.~Dimofte, D.~Gaiotto, and R.~van~der Veen, {\it RG Domain Walls and Hybrid
  Triangulations},  \href{http://xxx.lanl.gov/abs/1304.6721}{{\tt
  arXiv:1304.6721}}.

\bibitem{CJ-S3}
C.~Cordova and D.~L. Jafferis, {\it Complex Chern-Simons from M5-branes on the
  Squashed Three-Sphere},  \href{http://xxx.lanl.gov/abs/1305.2891}{{\tt
  arXiv:1305.2891}}.

\bibitem{CDGS}
H.-J. Chung, T.~Dimofte, S.~Gukov, and P.~Su{\l}kowski, {\it 3d-3d
  Correspondence Revisited},  \href{http://xxx.lanl.gov/abs/1405.3663}{{\tt
  arXiv:1405.3663}}.

\bibitem{D-volume}
T.~Dimofte, {\it 3d Superconformal Theories from 3-Manifolds},  {\em Lett.
  Math. Phys., special volume (to appear)} (2014).

\bibitem{Witten-GL6d}
E.~Witten, {\it Geometric Langlands From Six Dimensions},  {\em A celebration
  of the mathematical legacy of Raoul Bott, CRM Proc. Lecture Notes 50} (2010)
  281--310, [\href{http://xxx.lanl.gov/abs/0905.2720}{{\tt arXiv:0905.2720}}].

\bibitem{Tachi-discrete}
Y.~Tachikawa, {\it On the 6d origin of discrete additional data of 4d gauge
  theories},  {\em JHEP} {\bf 1405} (2014) 020,
  [\href{http://xxx.lanl.gov/abs/1309.0697}{{\tt arXiv:1309.0697}}].

\bibitem{Witten-gravCS}
E.~Witten, {\it 2+1 Dimensional Gravity as an Exactly Soluble System},  {\em
  Nucl. Phys.} {\bf B311} (1988), no.~1 46--78.

\bibitem{Witten-cx}
E.~Witten, {\it Quantization of Chern-Simons Gauge Theory with Complex Gauge
  Group},  {\em Comm. Math. Phys} {\bf 137} (1991) 29--66.

\bibitem{BW-cx}
D.~Bar-Natan and E.~Witten, {\it Perturbative expansion of Chern-Simons theory
  with non-compact gauge group},  {\em Comm. Math. Phys.} {\bf 141} (1991),
  no.~2 423--440.

\bibitem{gukov-2003}
S.~Gukov, {\it Three-Dimensional Quantum Gravity, Chern-Simons Theory, and the
  A-Polynomial},  {\em Commun. Math. Phys.} {\bf 255} (2005), no.~3 577--627,
  [\href{http://xxx.lanl.gov/abs/hep-th/0306165v1}{{\tt hep-th/0306165v1}}].

\bibitem{hikami-2006}
K.~Hikami, {\it Generalized Volume Conjecture and the A-Polynomials - the
  Neumann-Zagier Potential Function as a Classical Limit of Quantum Invariant},
   {\em J. Geom. Phys.} {\bf 57} (2007), no.~9 1895--1940,
  [\href{http://xxx.lanl.gov/abs/math/0604094v1}{{\tt math/0604094v1}}].

\bibitem{DGLZ}
T.~Dimofte, S.~Gukov, J.~Lenells, and D.~Zagier, {\it Exact Results for
  Perturbative Chern-Simons Theory with Complex Gauge Group},  {\em Comm. Num.
  Thy. and Phys.} {\bf 3} (2009), no.~2 363--443,
  [\href{http://xxx.lanl.gov/abs/0903.2472}{{\tt arXiv:0903.2472}}].

\bibitem{FS-curvedSUSY}
G.~Festuccia and N.~Seiberg, {\it Rigid Supersymmetric Theories in Curved
  Superspace},  {\em JHEP} {\bf 1106} (2011) 114,
  [\href{http://xxx.lanl.gov/abs/1105.0689}{{\tt arXiv:1105.0689}}].

\bibitem{CDFZ-3d}
C.~Closset, T.~T. Dumitrescu, G.~Festuccia, and Z.~Komargodski, {\it
  Supersymmetric Field Theories on Three-Manifolds},  {\em JHEP} {\bf 1305}
  (2013) 017, [\href{http://xxx.lanl.gov/abs/1212.3388}{{\tt
  arXiv:1212.3388}}].

\bibitem{CDFZ-geom}
C.~Closset, T.~T. Dumitrescu, G.~Festuccia, and Z.~Komargodski, {\it The
  Geometry of Supersymmetric Partition Functions},  {\em JHEP} {\bf 1401}
  (2014) 124, [\href{http://xxx.lanl.gov/abs/1309.5876}{{\tt
  arXiv:1309.5876}}].

\bibitem{Kapustin-3dloc}
A.~Kapustin, B.~Willett, and I.~Yaakov, {\it Exact Results for Wilson Loops in
  Superconformal Chern-Simons Theories with Matter},  {\em JHEP} {\bf 1003}
  (2010) 089, [\href{http://xxx.lanl.gov/abs/0909.4559}{{\tt
  arXiv:0909.4559}}]. Published in: JHEP 1003:089,2010 32 pages.

\bibitem{HHL}
N.~Hama, K.~Hosomichi, and S.~Lee, {\it SUSY Gauge Theories on Squashed
  Three-Spheres},  {\em JHEP} {\bf 1105} (2011) 014,
  [\href{http://xxx.lanl.gov/abs/1102.4716}{{\tt arXiv:1102.4716}}].

\bibitem{AGT}
L.~F. Alday, D.~Gaiotto, and Y.~Tachikawa, {\it Liouville Correlation Functions
  from Four-Dimensional Gauge Theories},  {\em Lett. Math. Phys.} {\bf 91}
  (2010), no.~2 167--197, [\href{http://xxx.lanl.gov/abs/0906.3219}{{\tt
  arXiv:0906.3219}}].

\bibitem{HLP-wall}
K.~Hosomichi, S.~Lee, and J.~Park, {\it AGT on the S-duality Wall},  {\em JHEP}
  {\bf 1012} (2010) 079, [\href{http://xxx.lanl.gov/abs/1009.0340}{{\tt
  arXiv:1009.0340}}].

\bibitem{DGG-index}
T.~Dimofte, D.~Gaiotto, and S.~Gukov, {\it 3-Manifolds and 3d Indices},  {\em
  Adv. Theor. Math. Phys.} {\bf 17} (2013) 975--1076,
  [\href{http://xxx.lanl.gov/abs/1112.5179}{{\tt arXiv:1112.5179}}].

\bibitem{Kim-index}
S.~Kim, {\it The complete superconformal index for N=6 Chern-Simons theory},
  {\em Nucl. Phys.} {\bf B821} (2009) 241--284,
  [\href{http://xxx.lanl.gov/abs/0903.4172}{{\tt arXiv:0903.4172}}].

\bibitem{IY-index}
Y.~Imamura and S.~Yokoyama, {\it Index for three dimensional superconformal
  field theories with general R-charge assignments},  {\em JHEP} {\bf 1104}
  (2011) 007, [\href{http://xxx.lanl.gov/abs/1101.0557}{{\tt
  arXiv:1101.0557}}].

\bibitem{KW-index}
A.~Kapustin and B.~Willett, {\it Generalized Superconformal Index for Three
  Dimensional Field Theories},  \href{http://xxx.lanl.gov/abs/1106.2484}{{\tt
  arXiv:1106.2484}}.

\bibitem{LY-S2}
S.~Lee and M.~Yamazaki, {\it 3d Chern-Simons Theory from M5-branes},  {\em
  JHEP} {\bf 1312} (2013) 035, [\href{http://xxx.lanl.gov/abs/1305.2429}{{\tt
  arXiv:1305.2429}}].

\bibitem{Yagi-S2}
J.~Yagi, {\it 3d TQFT from 6d SCFT},  {\em arXiv} (2013)
  [\href{http://xxx.lanl.gov/abs/1305.0291}{{\tt arXiv:1305.0291}}].

\bibitem{BNY-Lens}
F.~Benini, T.~Nishioka, and M.~Yamazaki, {\it 4d Index to 3d Index and 2d
  TQFT},  {\em Phys. Rev.} {\bf D86} (2012) 065015,
  [\href{http://xxx.lanl.gov/abs/1109.0283}{{\tt arXiv:1109.0283}}].

\bibitem{IY-Lens}
Y.~Imamura and D.~Yokoyama, {\it S3/Zn partition function and dualities},  {\em
  JHEP} {\bf 1211} (2012) 122, [\href{http://xxx.lanl.gov/abs/1208.1404}{{\tt
  arXiv:1208.1404}}].

\bibitem{IMY-fact}
Y.~Imamura, H.~Matsuno, and D.~Yokoyama, {\it Factorization of S3/Zn partition
  function},  {\em Phys. Rev.} {\bf D89} (2014) 085003,
  [\href{http://xxx.lanl.gov/abs/1311.2371}{{\tt arXiv:1311.2371}}].

\bibitem{DGG-Kdec}
T.~Dimofte, M.~Gabella, and A.~B. Goncharov, {\it K-Decompositions and 3d Gauge
  Theories},  \href{http://xxx.lanl.gov/abs/1301.0192}{{\tt arXiv:1301.0192}}.

\bibitem{Dimofte-QRS}
T.~Dimofte, {\it Quantum Riemann Surfaces in Chern-Simons Theory},  {\em Adv.
  Theor. Math. Phys.} {\bf 17} (2013) 479--599,
  [\href{http://xxx.lanl.gov/abs/1102.4847}{{\tt arXiv:1102.4847}}].

\bibitem{KashAnd}
J.~E. Andersen and R.~Kashaev, {\it A TQFT from quantum Teichm{\"u}ller
  theory},  \href{http://xxx.lanl.gov/abs/1109.6295}{{\tt arXiv:1109.6295}}.

\bibitem{AK-new}
J.~E. Andersen and R.~Kashaev, {\it A new formulation of the Teichm{\"u}ller
  TQFT},  \href{http://xxx.lanl.gov/abs/1305.4291}{{\tt arXiv:1305.4291}}.

\bibitem{Gar-index}
S.~Garoufalidis, {\it The 3D index of an ideal triangulation and angle
  structures},  \href{http://xxx.lanl.gov/abs/1208.1663}{{\tt
  arXiv:1208.1663}}.

\bibitem{GHRS-index}
S.~Garoufalidis, C.~D. Hodgson, J.~H. Rubinstein, and H.~Segerman, {\it
  1-efficient triangulations and the index of a cusped hyperbolic 3-manifold},
  \href{http://xxx.lanl.gov/abs/1303.5278}{{\tt arXiv:1303.5278}}.

\bibitem{Wit-anal}
E.~Witten, {\it Analytic Continuation of Chern-Simons Theory},  {\em
  Chern-Simons gauge theory: 20 Years After (AMS/IP Stud. Adv. Math.)} (2011)
  347--446, [\href{http://xxx.lanl.gov/abs/1001.2933}{{\tt arXiv:1001.2933}}].

\bibitem{faddeev-1994}
L.~Faddeev, {\it Current-Like Variables in Massive and Massless Integrable
  Models},  {\em in Quantum groups and their applications in physics (Varenna
  1994), Proc. Internat. School Phys. Enrico Fermi, 127, IOS, Amsterdam} (1996)
  117--135.

\bibitem{AHISS}
O.~Aharony, A.~Hanany, K.~Intriligator, N.~Seiberg, and M.~J. Strassler, {\it
  Aspects of N=2 Supersymmetric Gauge Theories in Three Dimensions},  {\em
  Nucl. Phys.} {\bf B499} (1997), no.~1-2 67--99,
  [\href{http://xxx.lanl.gov/abs/hep-th/9703110v1}{{\tt hep-th/9703110v1}}].

\bibitem{FG-Teich}
V.~V. Fock and A.~B. Goncharov, {\it Moduli spaces of local systems and higher
  Teichmuller theory},  {\em Publ. Math. Inst. Hautes Etudes Sci.} {\bf 103}
  (2006) 1--211, [\href{http://xxx.lanl.gov/abs/math/0311149v4}{{\tt
  math/0311149v4}}].

\bibitem{cooper-1994}
D.~Cooper, M.~Culler, H.~Gillet, D.~Long, and P.~Shalen, {\it Plane Curves
  Associated to Character Varieties of 3-Manifolds},  {\em Invent. Math.} {\bf
  118} (1994), no.~1 47--84.

\bibitem{GMNIII}
D.~Gaiotto, G.~W. Moore, and A.~Neitzke, {\it Framed BPS States},  {\em Adv.
  Theor. Math. Phys.} {\bf 17} (2013) 241--397,
  [\href{http://xxx.lanl.gov/abs/1006.0146}{{\tt arXiv:1006.0146}}].

\bibitem{Pasquetti-fact}
S.~Pasquetti, {\it Factorisation of N = 2 theories on the squashed 3-sphere},
  {\em JHEP} {\bf 1204} (2012) 120,
  [\href{http://xxx.lanl.gov/abs/1111.6905}{{\tt arXiv:1111.6905}}].

\bibitem{BDP-blocks}
C.~Beem, T.~Dimofte, and S.~Pasquetti, {\it Holomorphic Blocks in Three
  Dimensions},  \href{http://xxx.lanl.gov/abs/1211.1986}{{\tt
  arXiv:1211.1986}}.

\bibitem{AMRP-fact}
L.~F. Alday, D.~Martelli, P.~Richmond, and J.~Sparks, {\it Localization on
  Three-Manifolds},  {\em JHEP} {\bf 1310} (2013) 095,
  [\href{http://xxx.lanl.gov/abs/1307.6848}{{\tt arXiv:1307.6848}}].

\bibitem{QMF}
D.~Zagier, {\it Quantum Modular Forms},  {\em Quanta of Maths, Clay Math.
  Proc.} {\bf 11} (2010) 659--675. AMS, Prividence, RI.

\bibitem{kashaev-1997}
R.~M. Kashaev, {\it The hyperbolic volume of knots from quantum dilogarithm},
  {\em Lett. Math. Phys.} {\bf 39} (1997) 269--265,
  [\href{http://xxx.lanl.gov/abs/q-alg/9601025v2}{{\tt q-alg/9601025v2}}].

\bibitem{Mur-Mur}
H.~Murakami and J.~Murakami, {\it The colored Jones polynomials and the
  simplicial volume of a knot},  {\em Acta Math.} {\bf 186} (Jan, 2001)
  85--104, [\href{http://xxx.lanl.gov/abs/math/9905075v2}{{\tt
  math/9905075v2}}].

\bibitem{DG-quantumNZ}
T.~D. Dimofte and S.~Garoufalidis, {\it The quantum content of the gluing
  equations},  {\em Geom. Topol.} {\bf 17} (2013), no.~3 1253--1315,
  [\href{http://xxx.lanl.gov/abs/1202.6268}{{\tt arXiv:1202.6268}}].

\bibitem{AG-HW}
J.~E. Andersen and N.~L. Gammelgaard, {\it The Hitchin-Witten connection and
  Complex Chern-Simons Theory}, [\href{http://xxx.lanl.gov/abs/1409.xxxx}{{\tt arXiv:1409.xxxx}}] (2014).

\bibitem{A-MCG}
J.~E. Andersen, {\it The genus one Complex Quantum Chern-Simons representation
  of the Mapping Class Group}, [\href{http://xxx.lanl.gov/abs/1409.xxxx}{{\tt arXiv:1409.xxxx}}] (2014).

\bibitem{AK-complexCS}
J.~E. Andersen and R.~Kashaev, {\it Complex Quantum Chern-Simons}, [\href{http://xxx.lanl.gov/abs/1409.xxxx}{{\tt arXiv:1409.xxxx}}] (2014).

\bibitem{Fock-Teich}
V.~V. Fock, {\it Dual Teichm\"uller spaces},
  \href{http://xxx.lanl.gov/abs/dg-ga/9702018v3}{{\tt dg-ga/9702018v3}}.

\bibitem{AtiyahBott-YM}
M.~F. Atiyah and R.~Bott, {\it The Yang-Mills equations over Riemann surfaces},
   {\em Philos. Trans. Roy. Soc. London Ser. A} {\bf 308} (1983), no.~1505
  523--615.

\bibitem{Kapustin-Witten}
A.~Kapustin and E.~Witten, {\it Electric-Magnetic Duality And The Geometric
  Langlands Program},  {\em Comm. Num. Th. and Phys.} {\bf 1} (2007) 1--236,
  [\href{http://xxx.lanl.gov/abs/hep-th/0604151v3}{{\tt hep-th/0604151v3}}].

\bibitem{Wfiveknots}
E.~Witten, {\it Fivebranes and Knots},  {\em Quantum Topol.} {\bf 3} (2012),
  no.~1 1--137, [\href{http://xxx.lanl.gov/abs/1101.3216}{{\tt
  arXiv:1101.3216}}].

\bibitem{AGGTV}
L.~F. Alday, D.~Gaiotto, S.~Gukov, Y.~Tachikawa, and H.~Verlinde, {\it Loop and
  Surface Operators in N=2 Gauge Theory and Liouville Modular Geometry},  {\em
  JHEP} {\bf 1001} (2010) 113, [\href{http://xxx.lanl.gov/abs/0909.0945}{{\tt
  arXiv:0909.0945}}].

\bibitem{DGOT}
N.~Drukker, J.~Gomis, T.~Okuda, and J.~Teschner, {\it Gauge Theory Loop
  Operators and Liouville Theory},  {\em JHEP} {\bf 1002} (2010) 057,
  [\href{http://xxx.lanl.gov/abs/0909.1105}{{\tt arXiv:0909.1105}}].

\bibitem{RY-Lens}
S.~S. Razamat and M.~Yamazaki, {\it S-duality and the N=2 Lens Space Index},
  {\em JHEP} {\bf 1310} (2013) 048,
  [\href{http://xxx.lanl.gov/abs/1306.1543}{{\tt arXiv:1306.1543}}].

\bibitem{Witten-sl2}
E.~Witten, {\it SL(2,Z) Action On Three-Dimensional Conformal Field Theories
  With Abelian Symmetry},  \href{http://xxx.lanl.gov/abs/hep-th/0307041v3}{{\tt
  hep-th/0307041v3}}.

\bibitem{Shale-rep}
D.~Shale, {\it Linear Symmetries of Free Boson Fields},  {\em Trans. Amer.
  Math. Soc.} {\bf 103} (1962) 149--167.

\bibitem{Weil-rep}
A.~Weil, {\it Sur Certains Groupes d'Op{\'e}rateurs Unitaires},  {\em Acta
  Math.} {\bf 111} (1964) 143--211.

\bibitem{NZ}
W.~D. Neumann and D.~Zagier, {\it Volumes of hyperbolic three-manifolds},  {\em
  Topology} {\bf 24} (1985), no.~3 307--332.

\bibitem{thurston-1980}
W.~Thurston, {\it The Geometry and Topology of Three-Manifolds},  {\em Lecture
  notes at Princeton University} (1980).

\bibitem{BFG-sl3}
N.~Bergeron, E.~Falbel, and A.~Guilloux, {\it Tetrahedra of flags, volume and
  homology of SL(3)},  \href{http://xxx.lanl.gov/abs/1101.2742}{{\tt
  arXiv:1101.2742}}.

\bibitem{GGZ-sln}
S.~Garoufalidis, M.~Goerner, and C.~K. Zickert, {\it Gluing equations for
  PGL(n,C)-representations of 3-manifolds},
  \href{http://xxx.lanl.gov/abs/1207.6711}{{\tt arXiv:1207.6711}}.

\bibitem{neumann-combinatorics}
W.~Neumann, {\it Combinatorics of Triangulations and the Chern-Simons Invariant
  for Hyperbolic 3-Manifolds},  {\em in Topology '90, Ohio State Univ. Math.
  Res. Inst. Publ.} {\bf 1} (1992).

\bibitem{DV-NZ}
T.~Dimofte and R.~van~der Veen, {\it A Spectral Perspective on Neumann-Zagier},
   \href{http://xxx.lanl.gov/abs/1403.5215}{{\tt arXiv:1403.5215}}.

\bibitem{Guilloux-PGL}
A.~Guilloux, {\it Representations of 3-manifold groups in PGL(n,C) and their
  restriction to the boundary},  \href{http://xxx.lanl.gov/abs/1310.2907}{{\tt
  arXiv:1310.2907}}.

\bibitem{GZ-gluing}
S.~Garoufalidis and C.~K. Zickert, {\it The symplectic properties of the
  PGL(n,C)-gluing equations},  \href{http://xxx.lanl.gov/abs/1310.2497}{{\tt
  arXiv:1310.2497}}.

\bibitem{garoufalidis-2004}
S.~Garoufalidis, {\it On the Characteristic and Deformation Varieties of a
  Knot},  {\em Geom. Topol. Monogr.} {\bf 7} (2004) 291--304,
  [\href{http://xxx.lanl.gov/abs/math/0306230v4}{{\tt math/0306230v4}}].

\bibitem{Champ-hypA}
A.~Champanerkar, {\it A-Polynomial and Bloch Invariants of Hyperbolic
  3-Manifolds},  {\em Ph.D. Thesis, Columbia University} (2003).

\bibitem{Dunfield-Mahler}
D.~W. Boyd, F.~Rodriguez-Villegas, and N.~M. Dunfield, {\it Mahler's Measure
  and the Dilogarithm (II)},  {\em Canad. J. Math.} {\bf 54} (2002), no.~3
  468--492, [\href{http://xxx.lanl.gov/abs/math/0308041v2}{{\tt
  math/0308041v2}}].

\bibitem{GS-quant}
S.~Gukov and P.~Su{\l}kowski, {\it A-polynomial, B-model, and Quantization},
  {\em JHEP} {\bf 1202} (2012) 070,
  [\href{http://xxx.lanl.gov/abs/1108.0002}{{\tt arXiv:1108.0002}}].

\bibitem{Zeil-AB}
M.~Petkov{\v s}ek, H.~S. Wilf, and D.~Zeilberger, {\it A=B},  {\em A.K. Peters,
  Wellesley, MA} (1996) xii+212.

\bibitem{Jafferis-Zmin}
D.~L. Jafferis, {\it The Exact Superconformal R-Symmetry Extremizes Z},  {\em
  JHEP} {\bf 1205} (2012) 159, [\href{http://xxx.lanl.gov/abs/1012.3210}{{\tt
  arXiv:1012.3210}}].

\bibitem{snappy}
M.~Culler, N.~Dunfield, and J.~R. Weeks, {\it SnapPy, a computer program for
  studying the geometry and topology of 3-manifolds}, .
  http://snappy.computop.org.

\bibitem{CV-tt*}
S.~Cecotti and C.~Vafa, {\it Topological-anti-topological fusion},  {\em Nucl.
  Phys.} {\bf B367} (1991), no.~2 359--461.

\bibitem{CGV}
S.~Cecotti, D.~Gaiotto, and C.~Vafa, {\it tt* Geometry in 3 and 4 Dimensions},
  {\em JHEP} {\bf 1405} (2014) 055,
  [\href{http://xxx.lanl.gov/abs/1312.1008}{{\tt arXiv:1312.1008}}].

\bibitem{GK-Gg}
S.~Garoufalidis and R.~Kashaev, {\it From state integrals to q-series},
  \href{http://xxx.lanl.gov/abs/1304.2705}{{\tt arXiv:1304.2705}}.

\end{thebibliography}\endgroup

\end{document}